\documentclass[draft]{agujournal2019}
\usepackage{float}

\usepackage{url} %this package should fix any errors with URLs in refs.
\usepackage{lineno}
\usepackage[inline]{trackchanges} %for better track changes. finalnew option will compile document with changes incorporated.
\usepackage{soul}
\usepackage{graphicx}
\usepackage{subfigure}

\draftfalse

\journalname{JGR: Solid Earth}

\begin{document}

%%%%%%%%%%%%%%%%%%%%%%%%%%%%%%%%%%%%%%%%%%%%%%%
%  TITLE
%
% (A title should be specific, informative, and brief. Use
% abbreviations only if they are defined in the abstract. Titles that
% start with general keywords then specific terms are optimized in
% searches)
%
%%%%%%%%%%%%%%%%%%%%%%%%%%%%%%%%%%%%%%%%%%%%%%%

% Example: \title{This is a test title}

\title{Benchmarking seismic phase associators: \\ Insights from synthetic scenarios}

%%%%%%%%%%%%%%%%%%%%%%%%%%%%%%%%%%%%%%%%%%%%%%%
%
%  AUTHORS AND AFFILIATIONS
%
%%%%%%%%%%%%%%%%%%%%%%%%%%%%%%%%%%%%%%%%%%%%%%%

% Authors are individuals who have significantly contributed to the
% research and preparation of the article. Group authors are allowed, if
% each author in the group is separately identified in an appendix.)

% List authors by first name or initial followed by last name and
% separated by commas. Use \affil{} to number affiliations, and
% \thanks{} for author notes.
% Additional author notes should be indicated with \thanks{} (for
% example, for current addresses).

% Example: \authors{A. B. Author\affil{1}\thanks{Current address, Antartica}, B. C. Author\affil{2,3}, and D. E.
% Author\affil{3,4}\thanks{Also funded by Monsanto.}}

\authors{Jorge Puente\affil{1,2}, Christian Sippl\affil{1}, Jannes Münchmeyer\affil{3}, Ian W. McBrearty\affil{4}}

% \affiliation{1}{First Affiliation}
% \affiliation{2}{Second Affiliation}
% \affiliation{3}{Third Affiliation}
% \affiliation{4}{Fourth Affiliation}

\affiliation{1}{Institute of Geophysics, Czech Academy of Sciences, Prague, Czech Republic}
\affiliation{2}{Charles University, Faculty of Mathematics and Physics, Department of Geophysics, Prague, Czech Republic}
\affiliation{3}{Univ. Grenoble Alpes, Univ. Savoie Mont Blanc, CNRS, IRD, Univ. Gustave Eiffel, ISTerre, Grenoble, France.}
\affiliation{4}{Department of Geophysics, Stanford University, Stanford, California, U.S.A.}
%(repeat as many times as is necessary)

% Corresponding author mailing address and e-mail address:

% (include name and email addresses of the corresponding author.  More
% than one corresponding author is allowed in this LaTeX file and for
% publication; but only one corresponding author is allowed in our
% editorial system.)

% Example: \correspondingauthor{First and Last Name}{email@address.edu}

\correspondingauthor{Jorge Puente}{puente@ig.cas.cz}

%%%%%%%%%%%%%%%%%%%%%%%%%%%%%%%%%%%%%%%%%%%%%%%
% KEY POINTS
%%%%%%%%%%%%%%%%%%%%%%%%%%%%%%%%%%%%%%%%%%%%%%%
%  List up to three key points (at least one is required)
%  Key Points summarize the main points and conclusions of the article
%  Each must be 140 characters or fewer with no special characters or punctuation and must be complete sentences

% Example:
% \begin{keypoints}
% \item	List up to three key points (at least one is required)
% \item	Key Points summarize the main points and conclusions of the article
% \item	Each must be 140 characters or fewer with no special characters or punctuation and must be complete sentences
% \end{keypoints}

\begin{keypoints}
\item We compare the performance of five seismic phase associators using synthetic datasets for crustal and subduction zone scenarios.
\item We evaluate the influence of different noise levels, event densities, and the performance under out-of-network conditions.
\item PyOcto and GENIE achieve the best performance in both scenarios, REAL and GaMMA show satisfactory performance for most cases.
\end{keypoints}

%%%%%%%%%%%%%%%%%%%%%%%%%%%%%%%%%%%%%%%%%%%%%%%
%
%  ABSTRACT and PLAIN LANGUAGE SUMMARY
%
% A good Abstract will begin with a short description of the problem
% being addressed, briefly describe the new data or analyses, then
% briefly states the main conclusion(s) and how they are supported and
% uncertainties.

% The Plain Language Summary should be written for a broad audience,
% including journalists and the science-interested public, that will not have 
% a background in your field.
%
% A Plain Language Summary is required in GRL, JGR: Planets, JGR: Biogeosciences,
% JGR: Oceans, G-Cubed, Reviews of Geophysics, and JAMES.
% see http://sharingscience.agu.org/creating-plain-language-summary/)
%
%%%%%%%%%%%%%%%%%%%%%%%%%%%%%%%%%%%%%%%%%%%%%%%

%% \begin{abstract} starts the second page

\begin{abstract}
Reliable seismicity catalogs are fundamental for seismological analysis. Following phase picking, phase association groups arrivals into sets with consistent origins (i.e., events), determines event counts, and identifies outlier picks. To handle the substantial increase in the quantity of seismic phase picks from improved picking methods and larger deployments, several novel phase associators have recently been proposed. This study presents a detailed benchmark analysis of five seismic phase associators, including classical and machine learning-based approaches: PhaseLink, REAL, GaMMA, GENIE, and PyOcto. We use synthetic datasets mimicking real seismicity characteristics in crustal and subduction zone scenarios. We evaluate performance for different conditions, including low- and high- noise environments, out-of-network events, very high event rates, and variable station density. The results reveal notable differences in precision, recall, and computational efficiency. GENIE and PyOcto demonstrate robust performance, with almost perfect perfect performance for most scenarios. Only for the most challenging conditions with high noise levels and event rates performance drops, but still maintains F1 scores above 0.8. PhaseLink's performance declines with noise and event density, particularly in subduction zones, dropping to near zero in the most complex cases. GaMMA outperforms PhaseLink but struggles with accuracy and scalability in high-noise, high-density scenarios. REAL performs reasonably but loses recall under extreme conditions. PyOcto and PhaseLink show the quickest runtimes for smaller-scale datasets, while REAL and GENIE are more than an order of magnitude slower for these cases. At the highest pick rates, GENIE’s runtime disadvantage diminishes, matching PyOcto and scaling effectively. Our results can guide practitioners compiling seismicity catalogs and developers designing novel associators. 

\end{abstract}

\section*{Plain Language Summary}

In order to detect and locate the thousands of small earthquakes that occur in seismically active regions every day, researchers use a variety of algorithms. A "picking algorithm", nowadays often based on deep learning, analyzes the recorded time series data at each station and identifies likely arrivals of seismic waves. Following this, a second algorithm, termed "associator", is tasked with grouping arrivals from different stations into distinct events, after which the earthquakes can be located and characterized.
In this study, we evaluate the performance of five such associator algorithms. This is achieved by feeding them synthetically produced arrival time picks as well as randomly placed false picks, and then evaluating what proportion of events the algorithms successfully retrieve and which picks they correctly associate. We evaluated three classical algorithms and two deep learning based methods, and find significant differences in performance, particularly for the most challenging datasets with high event rates and false picks obscuring true events. The results of our algorithm comparison can be instructive for seismologists who want to use an associator as part of a (semi)automated earthquake detection and location workflow.

%%%%%%%%%%%%%%%%%%%%%%%%%%%%%%%%%%%%%%%%%%%%%%%
%
%  BODY TEXT
%
%%%%%%%%%%%%%%%%%%%%%%%%%%%%%%%%%%%%%%%%%%%%%%%

%%% Suggested section heads:
\section{Introduction}
High-quality and reliable seismicity catalogs are an essential resource in seismology and fundamental for understanding earthquake processes. They form the basis for a wide range of studies in seismology and beyond, including travel time tomography \cite{White2021-mn}, statistical seismology \cite{Hainzl2019-wk,Xiong2023-xm}, hazard assessment \cite{Mancini2022-gf}, as well as research into tectonic processes \cite{Maharaj2023-xv,Sippl2019-gu}. Commonly, earthquake detection is performed with a two-step approach: phase picking and phase association. During phase picking, the task is to identify the onset of seismic phases, usually P- and S- waves, at individual seismic stations. Once phase picking is complete, the next fundamental step is phase association. Phase association involves the process of grouping the seismic phases that were detected at different stations into common seismic events. A group of picks belongs to an event if all of them originate from the same location at the same time, i.e., a distinct hypocenter. The accuracy of phase association is essential for determining earthquake location, depth, and magnitude, and hence forms the backbone of subsequent seismological analyses. In addition, phase association allows discarding spurious phase picks, as these will usually not be consistent across stations.

Historically, both phase picking and phase association were performed manually. However, to keep up with the rapidly growing data availability, automatic methods were developed \cite{Allen1978-uo}. For phase association, early automated approaches were grid-based, involving the creation of a grid over a region of interest and associating phases based on the best-fitting grid points, using travel time tables \cite{Johnson1995-cv, Ringdal1989-bv}. However, the runtime of such approaches becomes prohibitive when faced with a high number of picks. While historically issues were most commonly encountered with dense seismic activity such as aftershock sequences, the growing size of seismic networks and the advent of novel picking methods now routinely leads to vast quantities of picks that produce challenges for association even during background seismicity rates. In particular, the advent of machine learning techniques in phase picking has increased the volume of picks of small earthquakes to an unprecedented level, posing a new challenge to the phase association process \cite{Zhu2018-ki, Ross2018-xf, Zhu2019-lb, Mousavi2019-tf, Mousavi2020-yc, Yang2021-bc, Munchmeyer2022-eh, Woollam2022-ri, Zhu2022-pu}.

Given these developments, the performance of phase associators has become increasingly important in the pursuit of building accurate earthquake catalogs with ever lower magnitudes-of-completeness. Consequently, there are now significant efforts to improve seismic phase association using a range of modern approaches. These approaches build on modified traditional techniques \cite{Zhang2019-qq, Munchmeyer2024-kk}, or use machine learning \cite{Zhu2022-ad} and deep learning \cite{Ross2019-rx, McBrearty2023-th} techniques and represent a significant advancement in the field. In addition to their different conceptual approaches, each algorithm's performance is dependent on the specific configuration of parameters used, and different associators may behave differently under different conditions (e.g., picks, number of stations, and noise density). Here, we conduct an in-depth benchmarking study to understand how different phase associators perform in a range of different scenarios. In addition, we provide insights into effective parameter choices for each associator. As establishing a ''ground truth'' catalog in a real scenario is nearly impossible, we use synthetic scenarios for our benchmark. This allows us to create ''ground truth'' datasets, with which the performance of each associator can be determined based on event and pick-level metrics. In addition, synthetics allow us to evaluate the impact of aspects such as event density or noise levels. Such a controlled environment is an effective way to systematically compare the methods and identify their strengths and limitations.

\section{The algorithms} 

We evaluate five different algorithms for seismic phase association, with each of them taking labeled arrival times of P and S phases as input. 

\textbf{PhaseLink} \cite{Ross2019-rx} is a deep learning (DL) approach for seismic phase association that uses a recurrent neural network with long short-term memory units to process a sliding window of phase picks. The input to PhaseLink is a fixed length sequence of picks from multiple stations, and the network predicts which picks belong to the same source. The predictions are aggregated over time to form clusters, identifying individual earthquakes. The network is trained using a supervised learning approach, with the loss function optimized to minimize the misclassification of picks. PhaseLink requires training that can use real or synthetic data. The use of synthetic training data is crucial, as it allows exposing the network to a large range of seismicity scenarios. For the training step, providing a 1D velocity model of the region of interest is necessary. 

\textbf{REAL} (Rapid Earthquake Association and Location; \citeA{Zhang2019-qq}) is an optimized grid search-based algorithm. It is designed to rapidly and simultaneously associate seismic phases and locate seismic events. REAL performs a grid search in three dimensions around each station, with the earliest P arrival determining potential event locations. This reduces the search space from the entire study area to a smaller volume and eliminates the time dimension from the search, as the approximate origin time for each potential event can be inferred from the initial pick. The theoretical P and S travel-time tables are pre-calculated using a given homogeneous or 1D velocity model. The initial event location is determined at the grid point with the most associated P and S picks. If multiple grid points have the same maximum number of picks, the grid point with the smallest travel-time residuals is selected. REAL implements parallelization to reduce runtime.

\textbf{GaMMA} (Gaussian Mixture Model Association; \citeA{Zhu2022-ad}) treats the phase association problem as an unsupervised clustering problem within a probabilistic framework. It models each seismic event as a mixture component within a Gaussian Mixture Model (GMM) \cite{Bishop2006-ei}. It uses an expectation-maximization algorithm for optimizing the clusters. This iterative process can identify optimal phase associations by maximizing the likelihood of the observed data considering both arrival time and amplitude. DBSCAN \cite{Ester1996-nz} is employed to segment phase picks into sub-windows prior to running the GMM for association. Each cluster can be associated in parallel to maximize CPU usage. GaMMA identifies ``core points'' based on the density of neighboring points to form clusters around them. This preprocessing step helps to manage the computational complexity and increase the scalability and efficiency by dividing the data into smaller, manageable segments, making the subsequent Expectation-Maximization algorithm more efficient. GaMMA can model travel-times with homogeneous and 1D models. In addition, it can incorporate amplitude decay relationships. We do not use amplitude information in this study for consistency with the other methods.

\textbf{GENIE} (Graph Earthquake Neural Interpretation Engine; \citeA{McBrearty2023-th}) employs a graph neural network (GNN) to predict earthquake source locations and the likelihood of phase associations. GENIE constructs two graphs: one representing the seismic stations (station graph) and another representing the potential source locations (source graph). The source graph's nodes span the source region of interest, with edges connecting nearby spatial elements. Similarly, the station graph links nearby stations. Both graphs enable transfer and sharing of information between the connected elements to help the GNN identify likely source hypocenters and association assignments. Training GENIE involves generating synthetic data that covers a wide range of station configurations, source distributions, and pick sets. Synthetic catalogs are created by sampling network realizations, computing arrival times, corrupting data with noise by a certain percentage, and adding a percentage of false picks to the dataset. The generation of training data can make use of homogeneous or 1D velocity models. This diverse approach to training ensures that the model is exposed to a wide range of scenarios. GENIE supports both CPU and GPU processing.

\textbf{PyOcto} \cite{Munchmeyer2024-kk} employs a 4D space-time partitioning strategy inspired by the Oct(o)tree data structure. This way, PyOcto focuses computational resources on promising origin regions and reduces complexity. To minimize runtime, PyOcto discards event-free nodes early and uses a priority queue to scan promising nodes first. Once a node has reached a critically small size, PyOcto locates and outputs the event. Picks associated with the event are removed from the input set, to avoid duplicate associations. To model travel times, PyOcto supports homogeneous and 1D velocity models. PyOcto uses parallelization across different time blocks, to optimize CPU usage.

\section{Benchmarking approach}

\subsection{Event-station scenarios}

\begin{figure}[H]
\centering
\includegraphics[scale=.29]
{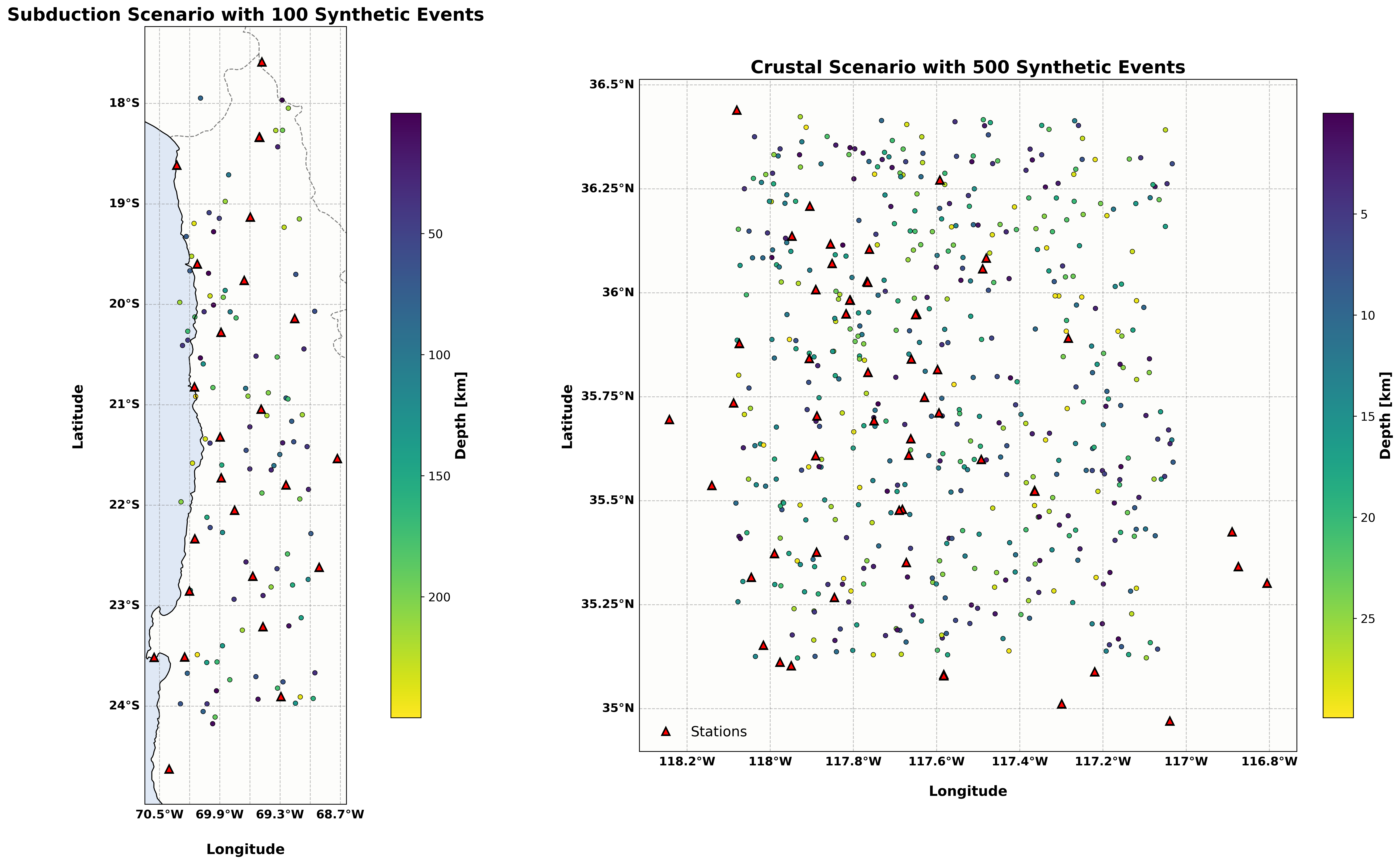}
\caption{\textbf{Left}: Station configuration and synthetic seismic event distribution example for the subduction zone scenario. Red triangles represent the IPOC network's seismic stations, colored dots an example set of 100 synthetic events. The fixed station layout, combined with variable event densities and noise levels, forms the basis of our associator evaluation.
\textbf{Right}: Station configuration and synthetic seismic event distribution example for the shallow seismicity scenario. We show an example realization with 500 events.}
\label{fig: combined_synthetic_scenarios}
\end{figure}

We conduct our benchmark study with two typical examples of seismic network geometry and seismicity depth range: a crustal seismicity scenario and a subduction zone scenario. Both scenarios are designed to replicate real-world conditions in terms of station density and distribution as well as the range of hypocentral depths. Note that we do not use real seismicity distributions but prefer events that are randomly distributed in space to test the algorithms' ability to detect arbitrarily located events (see Section 3.2). The station distributions and 1D velocity models for both scenarios are based on existing seismic network deployments (Figure \ref{fig: combined_synthetic_scenarios}) and geological settings.

For the crustal seismicity scenario, we use a set of stations from the Southern California Seismic Network \cite{California-Institute-of-Technology-and-United-States-Geological-Survey-Pasadena1926-jy} and the 1D velocity model of \citeA{Hadley1977-ko}. Seismic events are randomly generated within the region depicted in Figure \ref{fig: combined_synthetic_scenarios} (right) following a uniform distribution and covering the depth range of 0-30 km.

The subduction zone scenario employs the station distribution of the IPOC (Integrated Plate Boundary Observatory Chile; \citeA{GFZ-German-Research-Centre-for-Geosciences2006-yf}) CX seismic network in Northern Chile and the 1D velocity model of \citeA{Graeber1999-yc}. As for the crustal scenario, we generate seismic events randomly distributed in space and time in a uniform way, but cover a much larger range of hypocentral depths, from shallow crustal to intermediate-depth intraslab earthquakes (0-250 km; see Figure \ref{fig: combined_synthetic_scenarios}, left). 

\subsection{Synthetic pick/event generation}
\label{sec:synthetic_data_generation}

We create our synthetic benchmark datasets by closely following the approach outlined by \citeA{McBrearty2023-th}. From randomly generated origin times and hypocentral locations, we generate labeled P and S arrival times at the different stations with the respective 1D velocity models and station distributions. The dataset construction process comprises the following steps:

\begin{enumerate}
    \item \textbf{Event Location and Timing Selection}: Event locations are randomly generated within the station network of the scenario being simulated. Origin times are arbitrarily assigned within a 24-hour time span.
    \item \textbf{Arrival Time Calculation}: For each event, at all stations, P- and S-wave arrival times are computed using the NonLinLoc raytracer (http://alomax.free.fr/nlloc/) and a 1D velocity model
    \item \textbf{Arrival-Time Data Corruption}: To simulate real-world arrival time heterogeneity due to 3D velocity structure as well as picking errors, arrival times are perturbed by adding random noise. While picking errors are often modeled as Gaussian \cite{Diehl2009-px}, velocity model uncertainties can lead to higher-tailed distributions of error, so we perturb the arrival time data by uniform random noise proportional to travel time (-1 to 1\% of travel time).
    \item \textbf{Application of Distance Threshold}: A randomly determined cutoff distance threshold (uniformly between 70 km and 150 km for the crustal scenario and between 160 km and 500 km for the subduction scenario) is assigned to each event. All arrivals from source-station paths exceeding this limit are deleted. This can be seen as a rough approximation of event magnitude.    
    \item \textbf{Station Dropout}: A percentage of the stations (20\%) are randomly deleted for each event to introduce operational variability.
    \item \textbf{False Pick Integration}: The process concludes with the incorporation of a predefined percentage (30\%, 100\% and 300\%) of additional false picks (or ``noise picks''), effectively simulating automatic picker outputs that often contain many picks that do not belong to actual earthquakes. These false picks are randomly uniformly distributed over time, stations, and phase type.
\end{enumerate}

Details of the resulting event and pick distributions are provided in Tables S1 and S2. The distance threshold ensures the generation of a diverse set of events, including ``large-moveout'' events, that are detected across the majority of the seismic network, as well as ``small-moveout'' events, that are detected by a limited amount of stations (see pick count distributions in Figure S1). The synthetic seismicity we use is randomly distributed across the regions, different from real-world patterns where seismicity is concentrated near active faults, or inside the downgoing slab in subduction zones. However, for purposes of performance evaluation, the approach of evaluating all possible event locations, whether they are tectonically likely or not, has the advantage that it ensures the associators can also detect events in areas that have not previously had seismicity.

Although we attempt to design our synthetic scenarios in a realistic way, a number of complications that exist in real-world applications are still neglected. For instance, a real-world subduction zone dataset will most likely contain out-of-network events offshore. The level of arrival time noise we assume may easily be exceeded in real applications, and we unrealistically assumed that a station always has both a P- and S-pick. We thus perform our synthetic experiments in two main steps. The main set of experiments is performed with the above approach for creating synthetic datasets, and performance is evaluated for different amounts of events within 24 hours as well as different proportions of noise picks. After this evaluation, we perform a suite of tests where we introduce additional real-world problems such as having different proportions of out-of-network events, higher travel-time noise levels, and increased rates of missed picks. We evaluate the effect on performance each of these complications has one-by-one (see Section 4.4).   

\subsection{Performance evaluation approach}

To assess the performance of the seismic phase associators, we employ a set of evaluation metrics at both the event level and the pick level. This means that we first check how many events were correctly retrieved, how many were missed and how many false events were created from noise picks. We consider an event correctly retrieved if the associator yields an event that contains at least 50\% of the picks originally created for the synthetic event. In this way, we ensure that the original set of picks can not create more than one real event, and the loss of a fraction of real picks does not affect whether or not the event is correctly retrieved.

On the pick level, we then evaluate how many picks are correctly associated to an event (commonly associated picks), how many are missed (missed picks), how many are wrongly associated (i.e. picks from one event that get associated to a different one) and how many false picks are added to an event. Ground truth picks refers to the picks that the synthetic event actually has, predicted picks are the picks retrieved for this event (may contain correctly associated, wrongly associated and false picks). 

We employ the following set of metrics: 

\begin{itemize}
    \item {\textbf{Precision}: Measures the proportion of true positives (TP) in the entire output. High precision indicates few false positives (FP).
    
     \begin{equation}
     Precision = TP / (TP + FP)
     \end{equation}
     On the event level, TP corresponds to correctly identified events, FP to false or additional events that were associated from ground truth or noise picks. For the pick level analysis, TP marks the amount of picks correctly associated to an event, whereas FP is the sum of the number of noise picks added to the event and the number of wrongly associated picks that stem from other events.
     }
    
    \item {\textbf{Recall}: Measures the proportion of true positives compared to all ground truth correct associations. High recall indicates few false negatives (i.e., most actual events or picks were detected).

    \begin{equation}
     Recall = TP / (TP + FN)
    \end{equation}
    On the event level, TP again corresponds to the correctly identified events, while FN are the ground truth events that are missed. At pick level, TP are the picks correctly associated to an event, and FN are the ground truth picks from that event that are missing in the associated event.
    
    }

    \item {\textbf{F1 Score}: The harmonic mean of precision and recall, providing a balance between sensitivity and accuracy. A high F1 score indicates strong overall performance in correctly determined detections while minimizing false detections.

    \begin{equation}
    F1 = 2 \times \left( \frac{Precision \times Recall}{Precision + Recall} \right)
    \end{equation}}

    \item{\textbf{Runtimes}: Runtime is a crucial metric for practical implementations. Here, we evaluate the processing speed of each associator. We measure the time from the initiation of the model to the generation of its outputs. It is important to note that our measurement does not take into account any preprocessing steps such as the construction of the velocity model and travel time tables, or the model training for the DL-based associators, because these are processes usually executed only once within a given application framework. Our experiments are conducted using a consistent computational environment: all associators are run on systems utilizing 25 CPU threads, with access to 200 GB of RAM. For DL-based associators that leverage GPU acceleration, such as GENIE and PhaseLink, we use an NVIDIA A40 GPU for both training and inference.}
\end{itemize}

\subsection{Parameter optimization approach}

The performance of each association algorithm is heavily dependent on the choice of tuning parameters. Except for the association threshold, which defines the minimum number of picks needed to define an event, the different algorithms have very different parameter sets, a consequence of their quite different approaches. In order to provide a fair comparison between the different algorithms, we have to optimize the parameter choices for each of them, which is a time-consuming activity. For the sake of comparability and also to mimic real-world applications, we chose an association threshold of 10 picks for declaring an event (without specification how many of them have to be P or S) for all associators.

\begin{figure}[H]
    \includegraphics[width=1\textwidth]{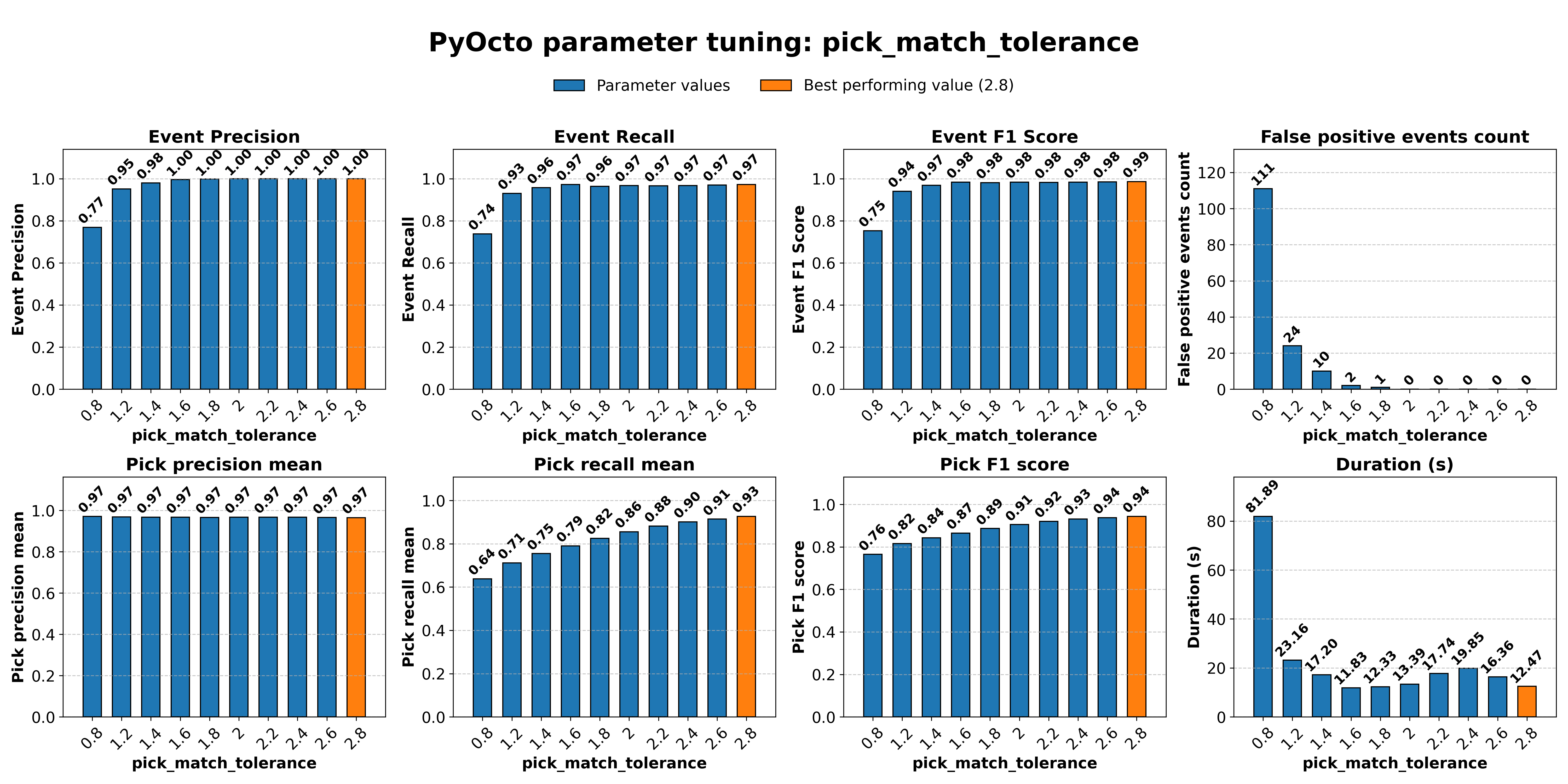}
    \caption{Example of our parameter optimization approach, here for parameter \textit{pick match tolerance} of PyOcto. The metrics event- and pick-level precision, recall and F1 score, as well as runtime and false positive count, are monitored against a systematic change of this parameter. For the run shown here, the choice marked in orange is evaluated to perform best. Note that we do not show the entire extent of the utilized search space here, values $>$2.8 were also tested. The finally chosen optimum parameters are determined by comparing performance for all nine runs (with 100, 500 and 2000 events as well as 30, 100 and 300$\%$ of noise picks) that we evaluate in Section 4.}
    \label{tdx}
\end{figure}

For REAL, GaMMA and PyOcto, we then conduct a large series of runs, changing parameters one by one and evaluating the change in performance metrics in response to these changes. While varying parameters individually is not the optimal approach, conducting a complete grid search would be computationally prohibitive. Where available, we used published parameter choices from an earlier associator comparison \cite{Munchmeyer2024-kk} or application studies \cite{Becker2024-oj} as an initial parameter guess. An example of optimizing a single parameter for PyOcto is shown in Figure \ref{tdx}: we systematically vary the parameter \textit{pick match tolerance}, and determine metrics like precision, recall, F1 score (on event and pick level) as well as runtime for each of these trial runs. The parameter choice with the overall best performance, as indicated by these different metrics, is chosen (here highlighted in orange color). We optimized two separate sets of parameter choices for the crustal and subduction zone scenario. For each of these sets, the final parameter choice is a compromise between the optimizations on all nine different runs (all combinations of 100, 500 and 2000 events as well as 30, 100 and 300$\%$ noise picks). That is, once selected, the same set of parameters is used for all tests, regardless of the number of noise picks and event rates.

The neural network based algorithms, PhaseLink and GENIE, require a training step before application, in which the majority of parameter optimization occurs. Because this step is time-consuming, the iterative tuning strategy as used for the traditional associators is not possible, and only a minimal amount of parameter tuning was possible for these methods. To create the training datasets for these algorithms we used the codes available with each method, which follow a similar approach of synthetic pick and event creation as outlined in Section 3.2. For details of the training process for PhaseLink and GENIE, please refer to Text S2 in the Supplementary Material and descriptions supplied in the original publications. All our final parameter choices for each associator and scenario are listed in Tables S3-S7 in the Supplementary Material.

\section{Results}
We evaluate the performance of the five seismic phase associators — PhaseLink, REAL, GaMMA, GENIE, and PyOcto — in the two different event-station scenarios introduced in Section 3.1. GaMMA, REAL and PyOcto offer the possibility of using either a homogeneous seismic velocity (0D model) or a 1D velocity model for the association process. We tested the different configurations and here only use their best-performing configurations as identified by our analysis (see Text S1, Figures S2 and S3, and Table S8 in the Supplementary Material). For each of the two event-station scenarios, we performed a total of 9 different runs, which feature different event numbers (100, 500 and 2000 events within 24 hours) as well as different proportions of randomly distributed ``noise picks'' (30, 100 and 300\% of the true picks). 

\subsection{Event-Level Performance Metrics}

\begin{figure}[H]
\centering
\includegraphics[scale=.6]
{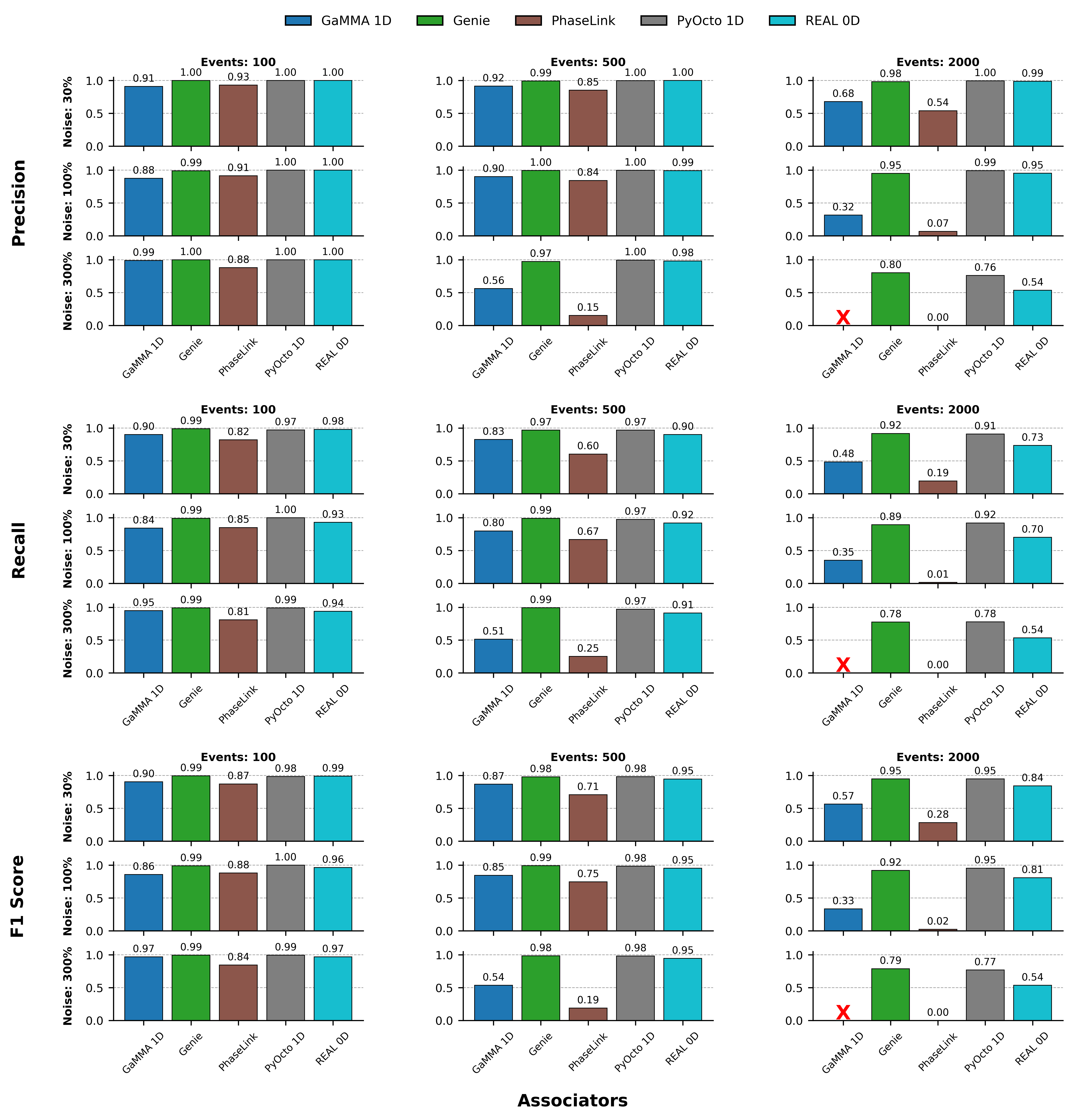}
\caption{Comparing precision, recall and F1 score across associators under different noise levels and event density, for the subduction scenario. A red X mark is indicated where an associator did not complete the run, thus did not obtain any results.}
\label{fig:subduction_event_metrics}
\end{figure}

\begin{figure}[H]
\centering
\includegraphics[scale=.6]
{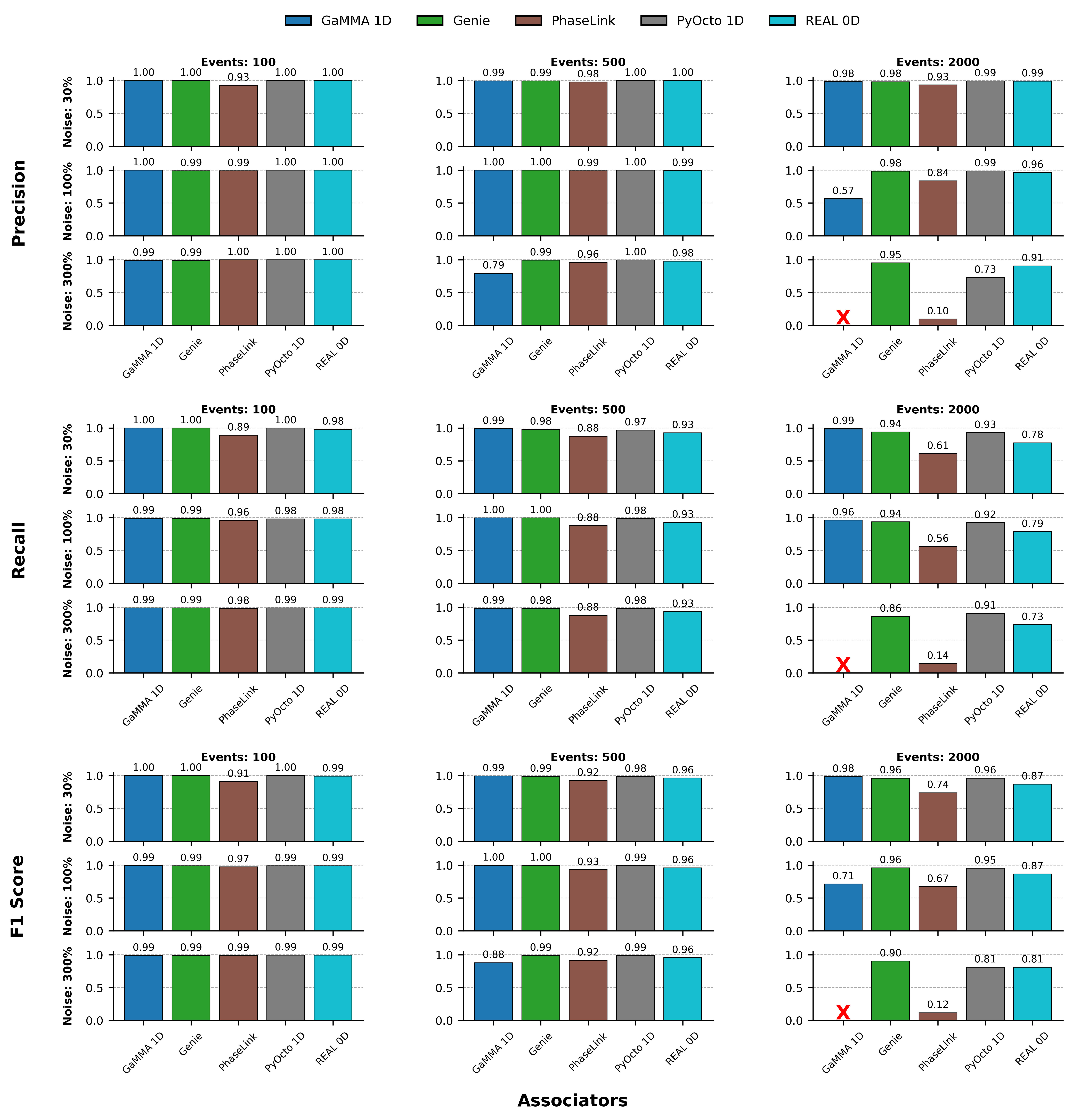}
\caption{Comparing precision, recall and F1 score across associators under different noise levels and event density, for the crustal scenario. A red X mark is indicated where an associator did not complete the run, thus did not obtain any results.}
\label{fig:crustal_event_metrics}
\end{figure}

The event-level results are presented in Figure~\ref{fig:subduction_event_metrics} for the subduction scenario and Figure~\ref{fig:crustal_event_metrics} for the crustal scenario.
The full numerical results are available in Tables S8 and S9.
At low noise levels (30\% noise) and small event counts (100 events), all associators demonstrate high event-level precision and recall, with relatively minor differences between different scenarios and associators. While GENIE, PyOcto and REAL show values above 0.97 for precision, recall and F1 score in both scenarios, GaMMA obtains lower scores around 0.9 for the subduction zone scenario, and PhaseLink scores around or even below 0.9 for both scenarios. With higher noise levels and event counts, the performance of the different associators diverges significantly. This is especially true for the subduction zone scenario, where the performance drops for the more difficult runs are more pronounced than for the crustal scenario. 

Adding more noise picks to the smallest run with only 100 events has no major impact on performance, whereas increasing event numbers deteriorates performance values more clearly. Of all associators, PhaseLink exhibits the most drastic performance drops with increasing event numbers and noise percentages. While it still performs reasonably (metrics largely above 0.8) in most crustal scenario runs except for the most difficult case of 2000 events and 300\% noise, its accuracy deteriorates significantly in the subduction zone scenario. There, it already has low precision, recall and F1 score values below 0.3 for the run with 500 events and 300\% noise as well as for all runs with 2000 events. For 2000 events and more than 100\% noise, its F1 score is nearly zero. Overall, PhaseLink's precision results are higher than its recall values. GaMMA performs markedly better than PhaseLink overall, but also exhibits a performance drop of F1 values to between 0.5 and 0.55 already in the high-noise case of 500 events for the subduction zone scenario. In the crustal scenario, it achieves clearly better results than in the subduction zone scenario, with a clear performance drop only for the case with 2000 events. For the most complex runs (2000 events with 300\% noise), GaMMA does not complete the processing due to memory allocation issues. The high computational demands of clustering large volumes of data with significant noise leads to excessive memory usage for GaMMA, exceeding the available RAM (200 GB). There is a clear tendency of reduced precision with more stable recall for GaMMA when moving to the more challenging runs in the crustal scenario, while no such systematic tendency can be seen for the subduction zone scenario.

REAL achieves overall good results in the crustal scenario, with metrics above 0.9 everywhere except for the runs with 2000 events. There, its recall drops more significantly (to values around 0.75) than its precision (still above 0.9) for the most challenging runs. This constitutes a significantly better performance than GaMMA. In the subduction zone scenario, REAL likewise only experiences a significant performance drop for the runs with 2000 events, but here it performs worse than in the crustal scenario, with recall dropping to around 0.7 already for low-noise conditions. Again, REAL's precision is generally higher than recall, but both decrease to 0.54 for the most challenging run. Finally, GENIE and PyOcto achieve the highest scores throughout the different runs, with only very minor differences between the two algorithms. Their metrics are above 0.97 for all runs with 100 or 500 events, in both scenarios. For the runs with 2000 events, precision and recall stay above 0.9 except for the very last run with 300\% noise. There, they both drop just under 0.8 for precision and recall in the subduction scenario, whereas GENIE obtains a higher precision than PyOcto (0.95 vs. 0.73) at similar recall (0.86 vs. 0.91) in the crustal scenario.

\subsection{Pick-Level Performance Metrics}

Due to the previously used definition of an event being correctly identified if it contains at least 50\% of the original (ground truth) picks, event-level metrics do not fully indicate which associator has a tendency to miss picks or to incorporate ``noise picks'' into correctly retrieved events. Such information becomes apparent when analyzing the performance on the pick level. Here, we perform this pick-level analysis. Picks are classified as correctly associated (CA), wrongly associated (WAP) or false as described in Section 3.3, and precision, recall and F1 scores are calculated similar to the event-level metrics. While the presented differences in pick-level performance do not have direct consequences on event retrieval (only correctly retrieved events were evaluated here in Section 4.1), missing picks and especially the incorporation of erroneous picks can have a large impact on the quality of the final seismicity catalog, leading to wrong and more uncertain hypocentral locations and magnitudes if no additional post-processing is applied. It should be noted that these pick-level results are derived only from the events that are successfully retrieved, i.e. that exceed the threshold of 50\% matching picks to the ground truth event. This criterion ensures that only events with a significant overlap between the predicted dataset and the ground truth synthetic dataset are considered. This implies that the set of events considered differs between the different associators, and it also means that additional false events that may be created from the remainder of ground truth picks, noise picks or a mixture of the two, do not impact the pick-level metrics. Hence, these pick-level metrics do not take into account event-level precision, which decreases proportional to the extent that false events are created, and which can be highly variable between different algorithms, as shown in Figures ~\ref{fig:subduction_event_metrics} and ~\ref{fig:crustal_event_metrics}.

\begin{figure}[H]
\centering
\includegraphics[scale=.30]{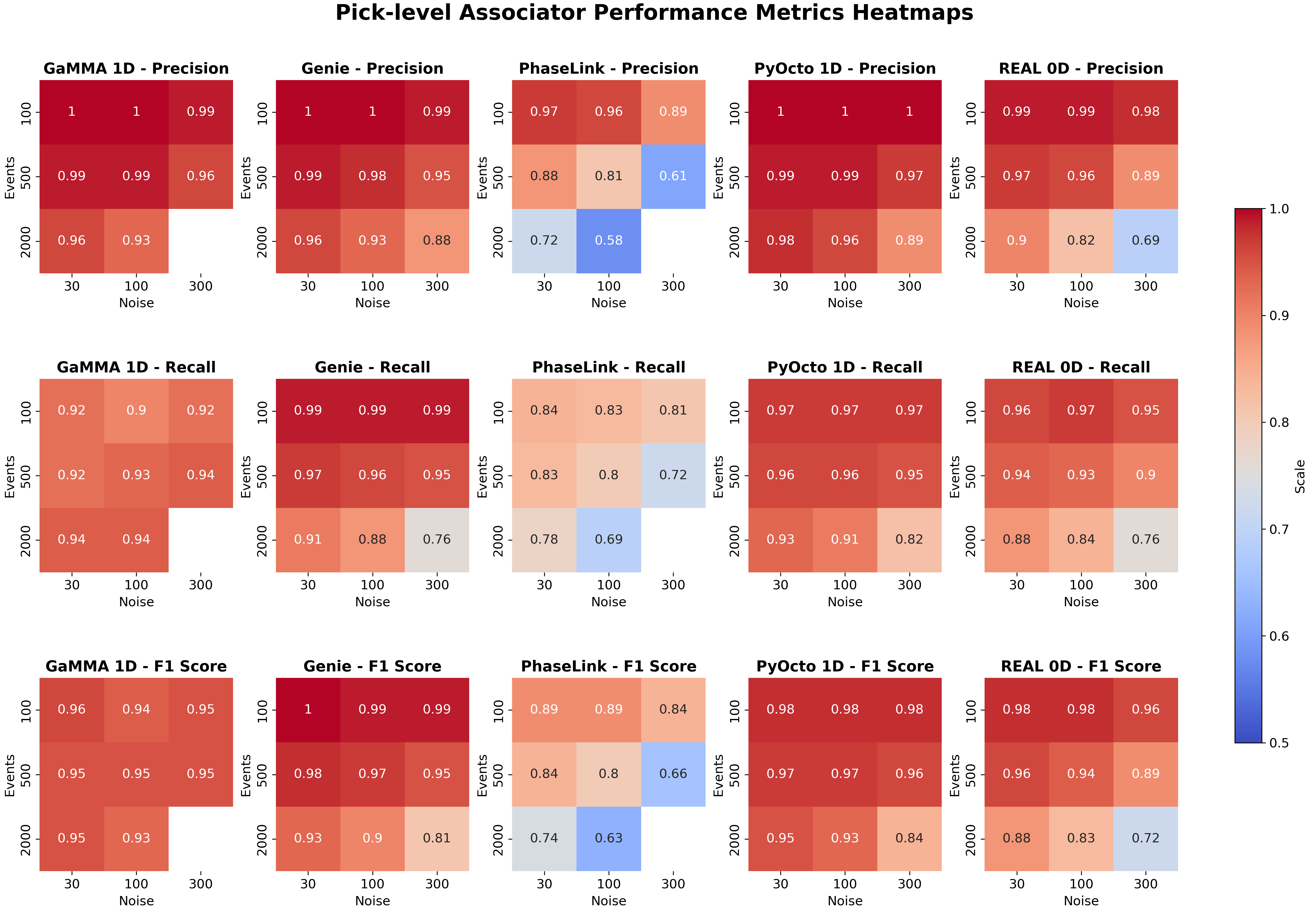}
\caption{Heatmap depiction of performance metrics (precision, recall, F1 score) across different noise and event densities for the subduction scenario. Each panel shows the mean performance derived from events that matched the synthetic ground truth.}
\label{fig:pick_lvl_heatmaps_sz}
\end{figure}

\begin{figure}[H]
\includegraphics[width=1\textwidth]{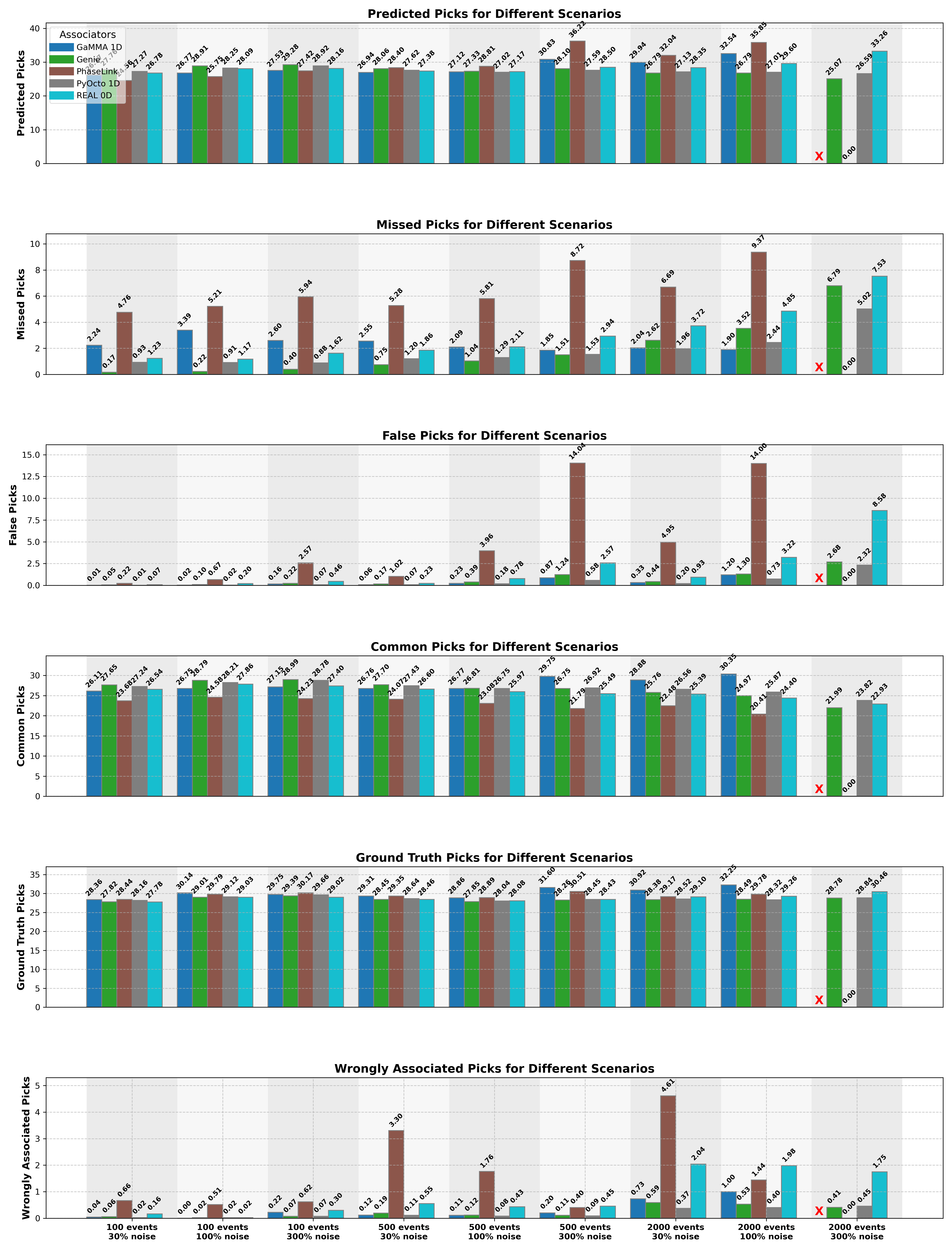}
\caption{Mean values of six pivotal metrics for each associator, set against the backdrop of different noise levels and event densities, of the subduction zone scenario. Metrics reflect only those seismic events that meet or exceede a 50\% matching threshold with the ground truth synthetic dataset, thereby focusing on matched events to gauge associator performance accurately.}
\label{fig:extra_pick_stats_counts_sz}
\end{figure}

\begin{figure}[H]
\centering
\includegraphics[scale=.30]{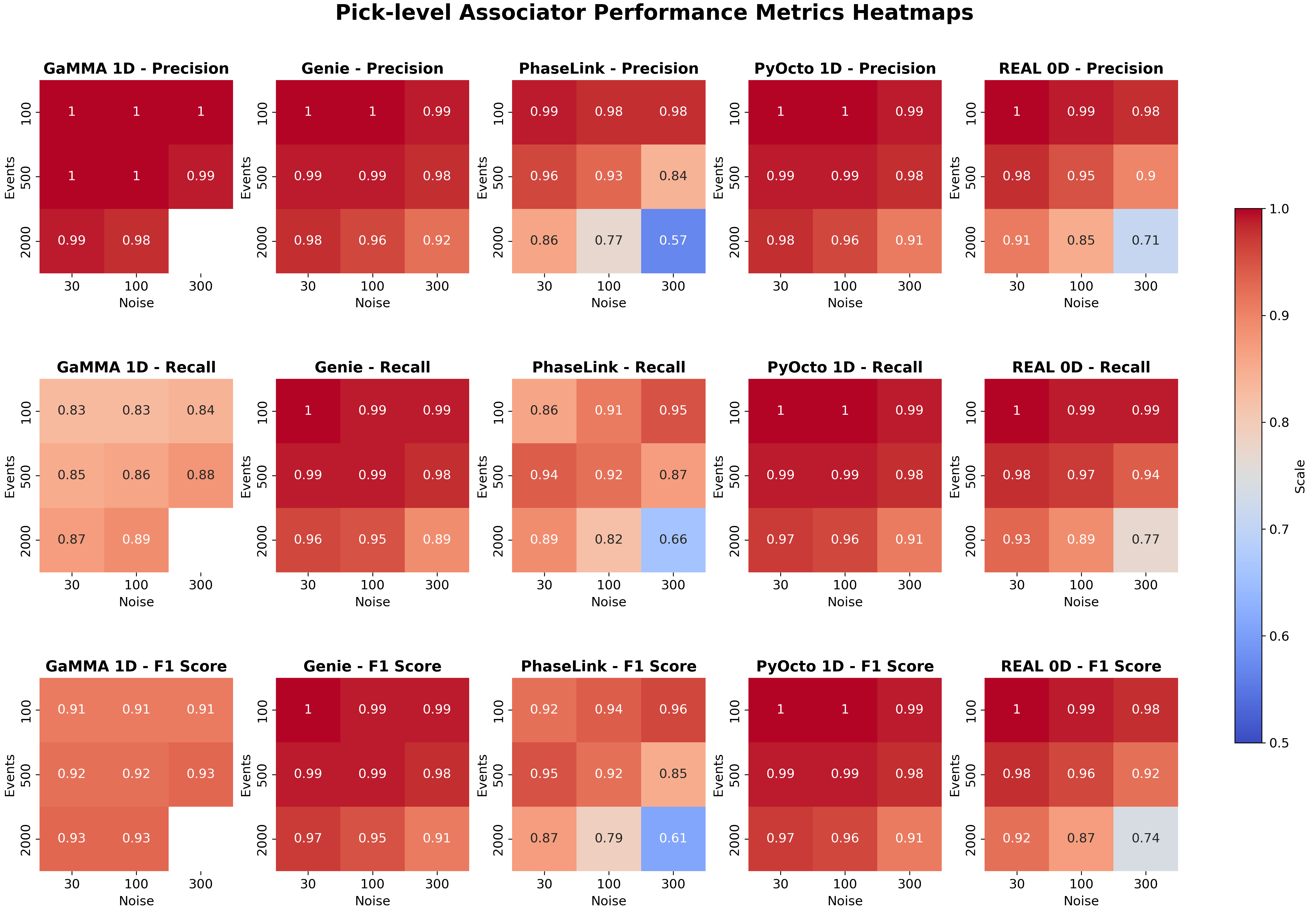}
\caption{Heatmap depiction of performance metrics (precision, recall, F1 score) across different noise and event densities for the crustal scenario. Each panel shows the mean performance derived from events that matched the synthetic ground truth.}
\label{fig:pick_lvl_heatmaps_cr}
\end{figure}

\begin{figure}[H]
\includegraphics[width=1\textwidth]{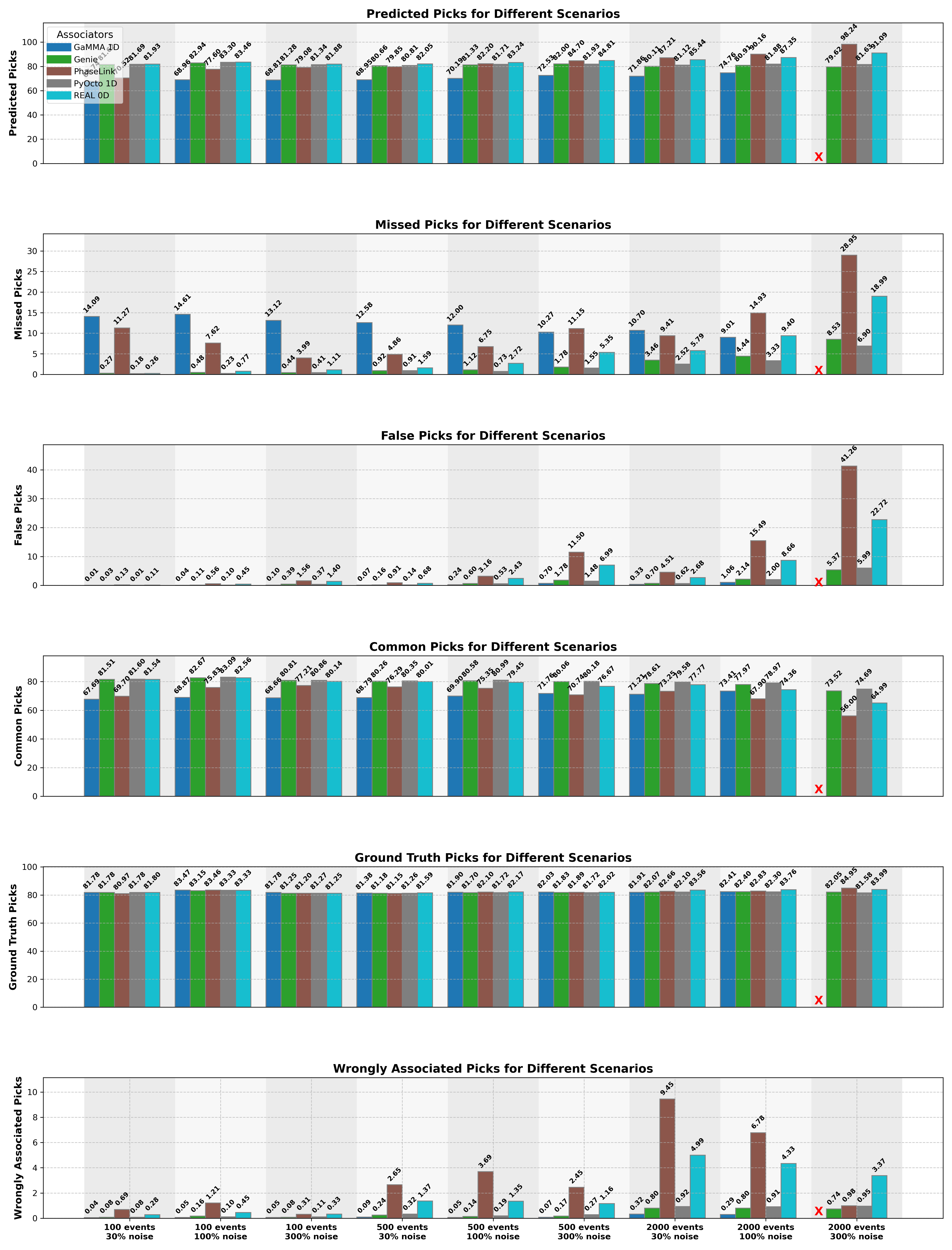}
\caption{Mean values of six pivotal metrics for each associator, set against the backdrop of different noise levels and event densities, of the crustal scenario. Metrics reflect only those seismic events that meet or exceed a 50\% matching threshold with the ground truth synthetic dataset, thereby focusing on matched events to gauge associator performance accurately.}
\label{fig:extra_pick_stats_counts_cr}
\end{figure}

Heat maps in Figures \ref{fig:pick_lvl_heatmaps_sz} and \ref{fig:pick_lvl_heatmaps_cr} show the mean values of precision, recall, and F1 score at pick level. Figures \ref{fig:extra_pick_stats_counts_sz} and \ref{fig:extra_pick_stats_counts_cr} show the mean values for the pick-level results per event (ground truth picks, predicted picks, commonly associated picks, missed picks, false picks, and wrongly associated picks) across the different associators and runs. All values shown in these figures are also provided numerically in Tables S11 and S12 in the Supplementary Materials. The observed general performance trends are largely similar to the event level ones. At low noise levels and smaller event counts, all associators demonstrate high pick-level accuracy, which deteriorates with increasing event numbers and noise picks. GENIE and PyOcto again show the highest accuracy, with performance metrics above 0.9 in nearly all cases, and retaining values above 0.8 even under the most adverse conditions. REAL nearly matches their performance in the smaller-scale runs, but performs worse in the runs with 2000 events, where it obtains values around 0.7 for the most challenging run with 2000 events and 300\% noise. GaMMA features high precision, but recall does not exceed 0.92 even for the smallest and simplest runs, and it fails to finish the hardest case due to memory issues. Moreover, low event-level precision for GaMMA, especially in the subduction zone scenario, implies that it creates many secondary events with falsely associated picks. Lastly, PhaseLink has the weakest overall results, with performance deteriorating (values below 0.8) already at the intermediate-difficulty runs, and nearly zero performance for the hardest runs.

The detailed pick statistics (Figures \ref{fig:extra_pick_stats_counts_sz} and \ref{fig:extra_pick_stats_counts_cr}) show that most associators tend to miss an average of one or two picks per event even for the easiest runs, whereas the incorporation of false or wrongly associated picks is virtually zero there. As the runs become more demanding, more picks are missed, but this is largely compensated by also incorporating more false or wrongly associated picks, so that the average total number of picks per event does not change significantly. PhaseLink starts to miss large amounts of picks already in the intermediate difficulty scenarios and at the same time incorporates many false or wrongly associated picks. For the hardest test case, PhaseLink does not retrieve any correct events in the subduction scenario, which is why missed, false, and wrongly associated picks for PhaseLink are zero for this case. For the other associators, the tendency to miss or wrongly incorporate picks is less strong than for PhaseLink. GaMMA misses a substantial amount of picks (an average of 14 per event in the crustal scenario) in the easier runs, and this proportion of missed picks stays relatively stable across the different runs. REAL's performance is close to the level of GENIE and PyOcto throughout most of the runs but deteriorates faster for the highest event rates, where it misses more picks and incorporates more noise or wrongly associated picks than these algorithms.

\subsection{Processing runtimes}

\begin{figure}[H]
\includegraphics[width=1\textwidth]{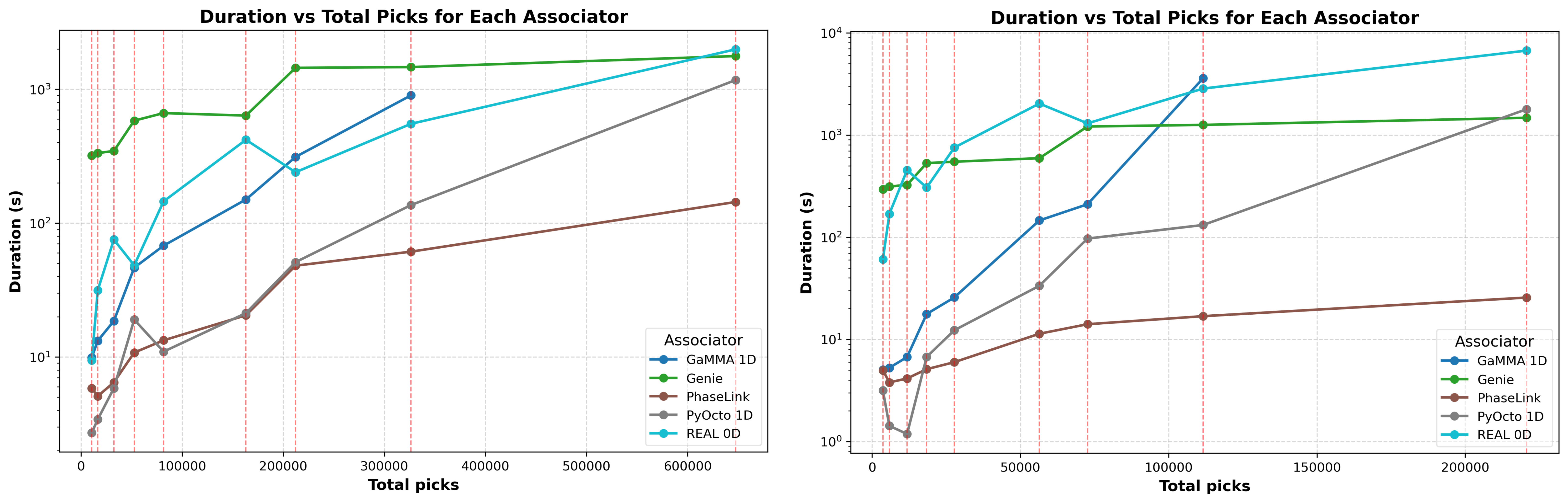}
\caption{Left: Logarithmic scale comparison of processing duration against the total number of picks for various seismic phase associators for the crustal seismicity scenario. Right: Logarithmic scale comparison of processing duration against the total number of picks for various seismic phase associators for the subduction scenario. Note that GaMMA 1D did not complete processing for the most complex case in both scenarios due to memory allocation issues, and thus its runtime is not plotted for those cases.}
\label{fig:runtime_both}
\end{figure}

The last performance metric we evaluate is processing runtime, as introduced in Section 3.3. Figure \ref{fig:runtime_both} shows summaries of runtimes for all different associators and evaluated runs (values are also listed in Tables S9 and S10 in the Supplementary Material), which are represented by the total number of picks (ground truth plus noise). Although the crustal scenario features a larger amount of stations (Figure \ref{fig: combined_synthetic_scenarios}) and thus more picks by a factor of 3-4, runtimes are generally slower for the subduction scenario. This is likely a consequence of events being distributed over a larger spatial region, as well as extending to much deeper depths. This increases the search space for potential sources and may also necessitate more complex travel time calculations. Processing times generally increase with scenario size, but the different associators show very different scaling behavior. While PyOcto, PhaseLink and partially also GaMMA finish the smaller scenarios in less than or around 10 seconds, REAL and especially GENIE are slower by an order of magnitude or more. For higher pick rates, it is apparent that the neural network-based associators (PhaseLink and GENIE) have better scalability than the other methods, in that their runtimes grow less strongly with an increasing number of picks. PyOcto, REAL and GaMMA have more significant processing time growth with total pick numbers, with GaMMA's curve being the steepest. However, since GENIE is quite slow for small scenarios, this flatter curve only means that its processing time is similar to PyOcto and somewhat faster than REAL for the largest scenarios we evaluate. PhaseLink, on the other hand, clearly processes large-scale problems fastest, but due to its near-zero performance for such cases (Section 4.1) it is still not an effective choice for processing such datasets. One can also observe that for GENIE and PhaseLink, which are based on neural networks, the amount of noise picks does not influence the total processing time significantly, whereas it plays a major role for the other, more classical associators.

\subsection{Further tests}

Although we took care to design our main synthetic experiments in a way that resembles natural use cases in many ways, there are still a few additional sources of complexity that we did not address in those tests. Out-of-network events are a common occurrence in most settings, especially in subduction zones where most of the plate interface as well as the often seismically active outer rise are located offshore \cite{Stern2002-ym}. Moreover, our synthetic scenarios have rather low levels of travel time noise ($\pm$1\% of travel time), all stations have either paired P+S picks or no picks at all for each event, and events reflect relatively large magnitude cases, with high numbers of constituent picks across the network over large geographic regions. All of these conditions are typically not present in real-world applications; instead, travel time noise levels may be higher, and most events will be small magnitude and hence only observed on a small fraction of the network. Thus, we also evaluate the deterioration in accuracy that occurs when these complications are increased to more challenging real-world levels. We conduct these additional tests on the intermediate subduction zone scenario with 500 events and 100\% noise picks. In a first set of runs, we systematically increase travel time noise levels, then move on to introduce different proportions of out-of-network events to the west of the station network (see Figure \ref{fig:oon_analysis}), and finally remove a higher proportion of P- or S-phases, which emulates the creation of smaller magnitude events. The evaluation of these additional complications complements the main analysis presented in Sections 4.1 through 4.3.

\subsubsection{Travel time noise}

\begin{figure}[H]
\includegraphics[width=\textwidth]{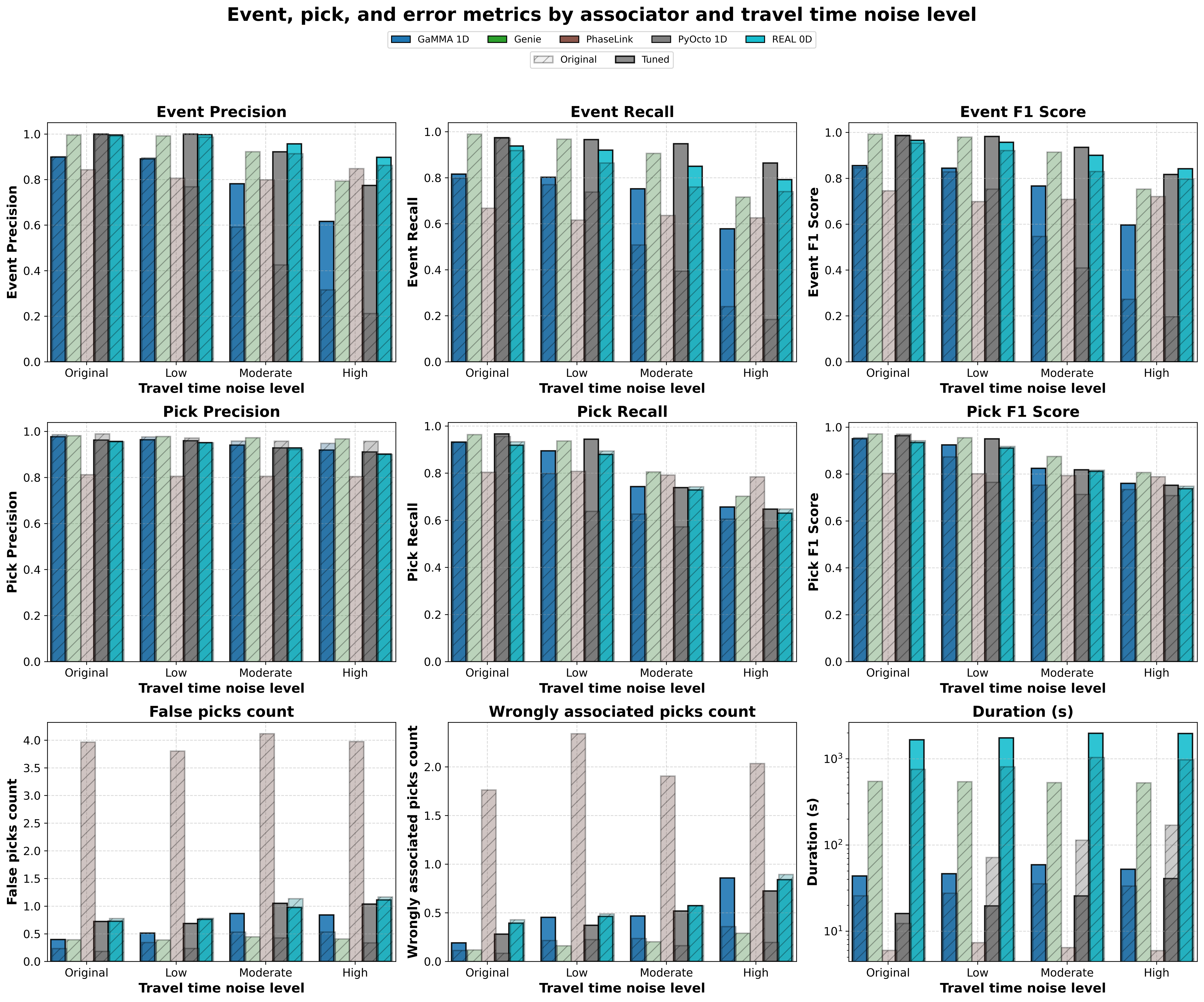}
\caption{Event-level (upper row), pick-level (middle row) and other (lower row) metrics for runs with systematically changed levels of travel time noise. The original configuration corresponds to the run with 500 events and 100\% noise picks of the subduction zone scenario. To simulate low, moderate and high noise conditions, $\pm$1-5\%, $\pm$5-10\% and $\pm$10-15\% of the travel time are added as noise to the picks. Note that for REAL, GaMMA and PyOcto, two different runs are shown, one with the original parameter optimization (hatched bars) and one with their tolerance parameters re-optimized for each noise level (solid bars).}
\label{fig:noise}
\end{figure}

In order to gauge the effect of adding more noise onto the utilized picks, we conducted three additional runs with noise added from random uniform distributions of $\pm$1-5\%, $\pm$5-10\% and $\pm$10-15\% of travel time. Results from these runs, in addition to the one with the original $\pm$0-1\% noise, are shown in Figure \ref{fig:noise}. REAL, GaMMA and PyOcto have tolerance-type parameters (REAL: \textit{nrt}; GaMMA: \textit{max\_sigma11}; PyOcto: \textit{pick\_match\_tolerance}) that put an upper bound on what misfit between predicted and observed arrival times is permissible. In a first series of runs, we kept the tolerance parameters fixed at the same values as determined in our previous optimizations. We then re-optimized these single parameters for each of these associators and runs, which in some cases yield significantly better results (see hatched and filled bars in Figure \ref{fig:noise}). The neural network based associators, PhaseLink and GENIE, were kept with their originally chosen parameters, as these methods appear less sensitive to travel time noise levels. However, if these models were re-trained for the higher expected noise levels this would likely increase performance further.

We observe a general performance decay of all associators with increasing travel time noise level, with a clearer decrease of recall values compared to precision. PhaseLink appears to be least affected by increasing travel time noise levels, but since its performance is already relatively low for the low noise levels of the original application, it is still not among the best performing algorithms for the highest noise levels. PyOcto and GaMMA show a large dependence on the re-optimization, with the sometimes very low event recall levels of the original tolerance parameter choices (below 0.25 for the high-noise case) significantly improving to around 0.6 (GaMMA) or 0.85 (PyOcto) with new, more adequate choice of tolerance parameters. For REAL, in contrast, re-optimizing the tolerance parameter only brings a subtle performance increase even with high noise, as it already performs quite well (recall above 0.75) with the original setting. While increasing the tolerance parameter leads to better metrics for GaMMA and PyOcto, it also causes a notable increase in false and wrongly associated picks that are incorporated into the retrieved events. GENIE maintains $>$0.75 F1 scores for both event-level and pick-level metrics even at the highest travel time noise levels, despite not being re-optimized for these higher noise level cases.

These results underscore the importance of parameter optimization, and illustrate the inherent tradeoff between robustness against travel time noise and the incorporation of noise picks that REAL, GaMMA and PyOcto exhibit. The two associators using a neural network approach, PhaseLink and GENIE, are more flexible with respect to travel time noise and largely do not need to be re-optimized once they are properly trained.

\subsubsection{Out-of-network events}

The correct identification and accurate location of out-of-network events represents a major challenge in seismology \cite{Williamson2023-aq}. We thus conduct three additional runs where we add an additional 150, 300 and 450 out-of-network events to the subduction zone scenario run with 500 events and 100\% noise picks. These events are randomly placed up to 200 km west of the network and at depths of 0-40 km. We do not preform any additional parameter optimization for these runs, but use the previously determined optimal parameters. 

\begin{figure}[H]
\includegraphics[width=\textwidth]{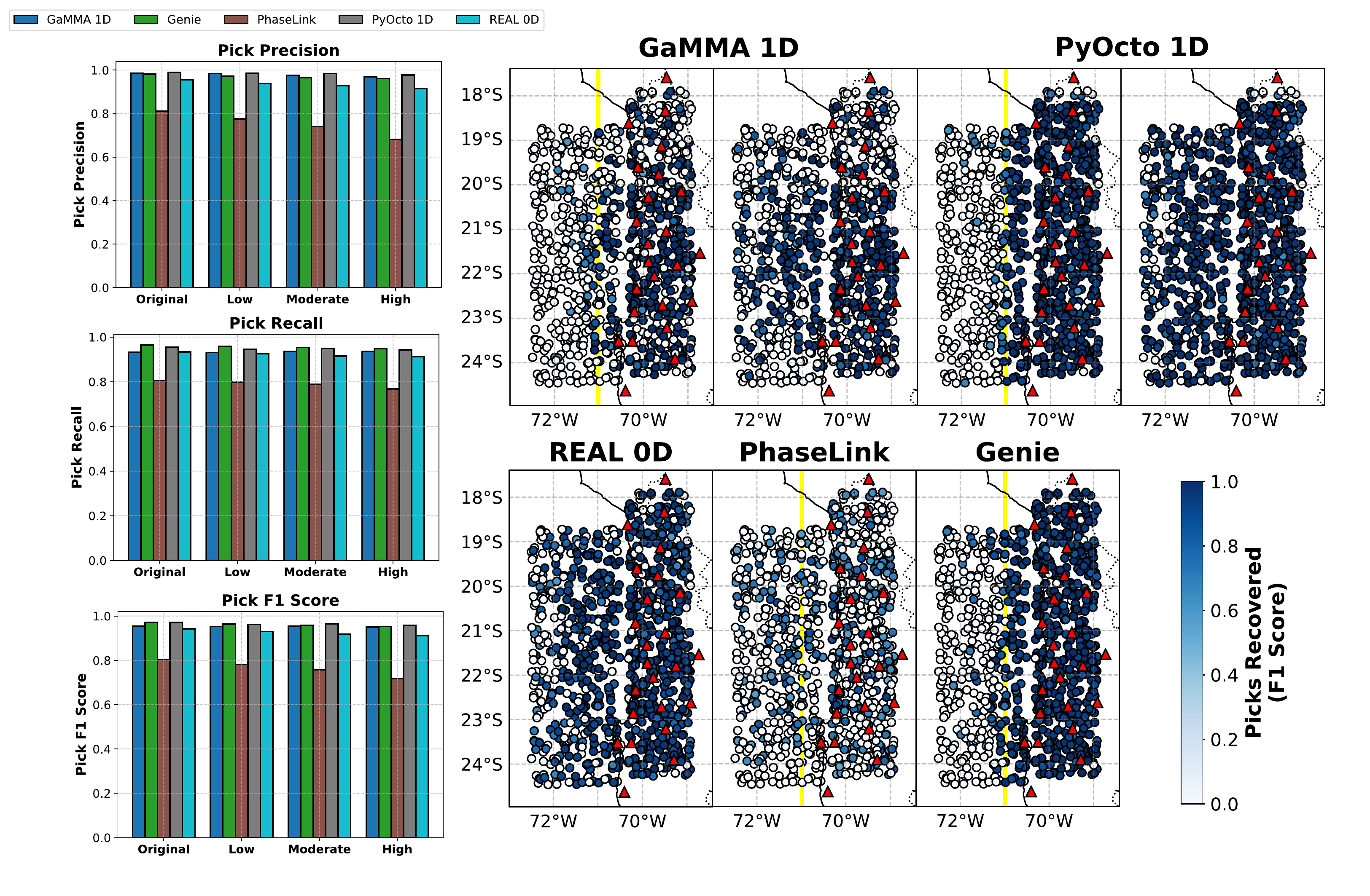}
\caption{Test results for including 150, 300 and 450 out-of-network events placed up to 200 km west of the network, at depths between 0 and 40 km. The left panel shows performance metrics for the in-network events for the different runs, on the right map view plots for the run with 450 out-of-network events are shown that indicate the performance of the different associators for the single events that are colored by pick-level F1 score. A yellow line shows the western edge of the search area in those runs where it falls within the out-of-network events.}
\label{fig:oon_analysis}
\end{figure}

In Figure \ref{fig:oon_analysis}, we show the performance for the in-network events with bar charts in the left panel, and the retrieval of out-of-network events (only for the case with 450 such events) in the map plots on the right. The bar charts reveal that the influence of out-of-network events on the correct association of in-network events is small, with only a slight decrease in performance for the run with the highest amount of out-of-network events being apparent for most associators. However, the algorithms differ markedly in how well they retrieve out-of-network events. In all cases, the event retrieval rate declines with distance from the network, but the nature of this decline is different among all associators. In our original parameter optimization, out-of-network events were not expected, so that the permissible search area for all algorithms but REAL (which does not feature such a parameter) was set to 71$^{\circ}$W. When keeping this choice, PyOcto and GENIE perform very well for closeby out-of-network events, but then show a sharp decline in retrieval rate in the close vicinity of this boundary. This means that events only slightly outside this search space limit will be missed, highlighting the importance of choosing it large enough. For GaMMA and PhaseLink, a substantial amount of closeby out-of-network events is missed, but a small proportion of events beyond the search area limit are retrieved as well. For REAL, the search space does not have to be defined by the user. Our results show that it has a high event retrieval rate that declines with distance from the network. At distances that roughly correspond to the location of the seismically active outer rise in a subduction zone, REAL still retrieves a reasonable proportion of events. 

When re-configuring GaMMA and PyOcto to include all out-of-network events in the search space, PyOcto's performance surpasses the one of REAL, although it also starts to miss events towards the western edge of the event cloud. GaMMA's results are less convincing, and it only retrieves a significant proportion of events west of the network center, whereas very few events are found at the northern and southern end of the out-of-network event cloud.

%Other bits and pieces:\\
%- REAL runs into algorithm design trouble when events are more than one network width away\\
%- mention additional test with small-aperture array surrounded by events up to large distances (supplement)?

\subsubsection{Removal of P- or S-phases}

In a last additional test, we remove different amounts (20, 40 and 60\%) of each event's picks, randomly between stations and P or S phases. This is meant to investigate the associators' performance in case many stations only have one pick, and not both paired P and S picks. At the same time, this run modifies the original distribution of pick numbers (Figure S1) in which only a relatively small proportion of events have pick numbers close to the association threshold of 10, creating a more realistic configuration in which most events emulate small magnitude events and only slightly exceed the association threshold.

\begin{figure}[H]
    \includegraphics[width=\textwidth]{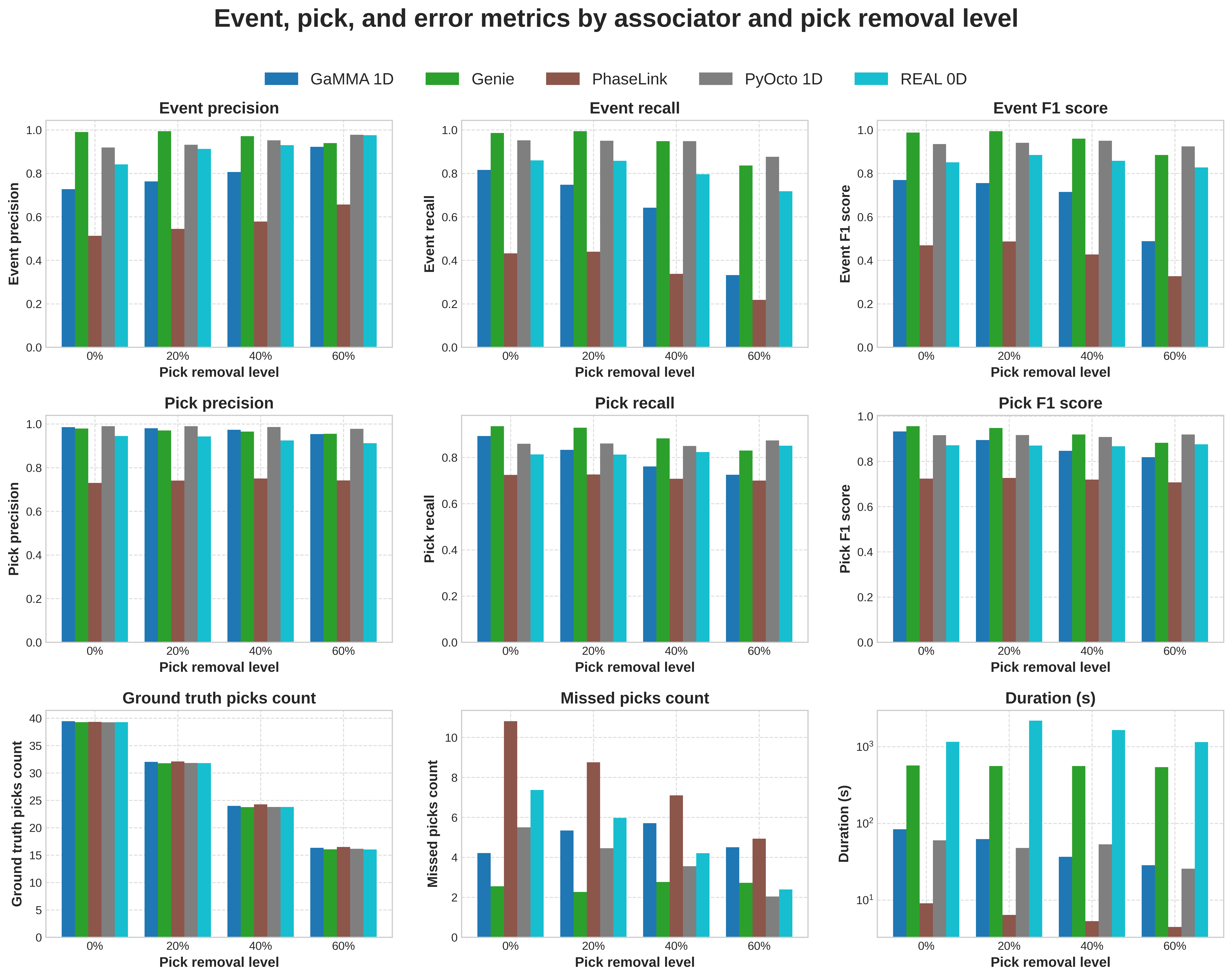}
    \caption{Results of the random pick removal test. Starting from the subduction zone scenario with 500 events and 100\% noise picks, we remove 20, 40 and 60\% of picks (randomly P or S) to simulate more sparsely detected events. Results are provided in the same form as in previous figures.}
    \label{fig:pick_removal}
\end{figure}

Results from this test are shown in Figure \ref{fig:pick_removal}. The average number of ground truth picks per event, shown in the lower left panel, demonstrates the decrease of total picks for the different runs, where the last run of 60\% removed picks only has about 15 picks per event on average, which is close to the association threshold of 10 picks. The scores for event precision and recall show that event precision actually increases with pick removal for most associators, most clearly for GaMMA and PhaseLink. At the same time, event recall deteriorates for all associators. As in most other runs, GENIE and PyOcto show the best overall performance, with scores $>$0.8 throughout all runs, and only minor differences between them. Here, PyOcto is slightly better for the run with the highest pick removal rate, whereas GENIE has minimally higher scores for the runs with fewer picks removed. REAL shows effective results, but its performance is systematically inferior to GENIE and PyOcto by about 0.1 in precision and recall. While PhaseLink performs poorly throughout all runs, GaMMA's recall also decreases significantly (to only about 0.3) for the highest rate of pick removal. For this run, the average event only comprises 15 picks, and as GaMMA misses more than 4 picks per event on average, this leads to many events moving below the association threshold and thus not being detected.

\section{Discussion}

\subsection{Associator configuration or training}

As briefly outlined in Section 3.4, each of the used associators requires the tuning of a number of parameters. Parameter choices are specific to the setting, to conditions such as station distribution or the amount and quality of input picks. This means that in all cases, a certain amount of tuning to the setting at hand is required, and none of the algorithms can be generalized to perform well ``out-of-the-box''. In real-world use cases, ground truth catalogs in the form of verified picks and events may not exist, so users must rely on experience and conducting and analyzing test runs to configure parameters. The amount of effort and expertise that is required to properly configure and apply the different algorithms also differs widely. In this Section, we discuss some of the tradeoffs that are inherent to the parameter optimization and comment on the practical use of the different algorithms.

For the backpropagation-based associators (REAL, PyOcto) as well as GaMMA, a tolerance-type parameter defines the allowed arrival time misfit for picks. This parameter has to be adapted to the expected travel time noise level (see Section 4.4.1) due to pick uncertainties or the deviation of the used velocity model from reality, and a suboptimal choice can have severe consequences for associator performance (e.g., Figure \ref{fig:noise}). A too high value will lead to the incorporation of noise picks and thus decreased precision, whereas a too small value will lead to many true picks being missed (lower pick-level and event-level recall). While such a tolerance-type parameter also exists implicitly in the training process of the DL algorithms and the level of travel time noise added to training picks, its role in defining the quality of achievable phase associations is less prominent. Secondly, a choice of grid density or refinement level is required, which leads to a second substantial tradeoff. A very fine parameterization will typically lead to improved performance, though it can severely increase runtimes, while a too coarse parameterization will lead to quick runtimes but inferior results.  

While the previous parameter tradeoffs have to be addressed when configuring PyOcto, REAL or GaMMA, these algorithms nevertheless feature a relatively limited set of parameters that need tuning, which means that finding a suitable (while maybe not optimal) configuration is not very time-intensive. Training the neural networks for PhaseLink and GENIE needs a higher amount of effort and expertise, and it could take a substantial amount of time to find a well-working setup for new users. In particular, the choice of the parameters used for the generated synthetic picks in training (e.g. proportion of noise picks, event density, levels of travel time noise, etc.) are important, yet may be hard to tune. The choice of the level of noise and event rates during training the DL associators implicitly affects the precision and recall tradeoffs, however directly assessing this tradeoff is difficult as it requires multiple rounds of training. For GENIE, it is required to set a few scale-dependent parameters such as the maximum moveout distance of sources, and the label kernel widths. The level of travel time noise and event rates can also be chosen to roughly reflect realistic conditions. In case of a real-world application, it is important to use the real data characteristics to guide the choice of training data parameters, but this process may necessitate some trial-and-error until a working configuration is found.

The advantage offered by the DL algorithms is flexibility, which is illustrated in the test runs with different noise levels (Figure \ref{fig:noise}). Once properly trained, PhaseLink and GENIE generally do not require parameter adaptation to perform well in a wide range of settings, while REAL, GaMMA or PyOcto have to be adjusted in case different conditions are encountered. Thus, the higher amount of initial investment in training the network can result in increased flexibility. For GENIE, since it relies on graph neural networks, this flexibility also extends to handling significantly different station configurations, e.g., if a seismic network is heavily modified over time by adding or removing stations, robustness can be maintained without requiring re-training. For example, GENIE performs well when trained on a dense network of 91 stations and then applied to a much smaller subnetwork of 21 stations, while PhaseLink has to be re-trained for such an application (see Text S3 and Figures S5 and S6 in the Supplementary Material).

\subsection{Event duplicates and multiplets}

An issue we did not analyze in detail is the possible creation of duplicate or multiplet events by phase associators. As none of the associators allows a single pick to be used by more than one event, our event definition of $\geq$50\% of ground truth picks ensures that only one output event per ground truth event is analyzed. Whether additional false events with smaller amounts of ground truth picks, possibly mixed with noise picks, are created was not evaluated independently. However, this effect is encoded in the statistics for event-level precision, pick-level recall as well as missed picks (Figures 3, 4, 6 and 8).

In the subduction zone scenario, both GaMMA and PhaseLink have decreased event-level precision even for the simplest runs, which does not occur in the crustal scenario. In both cases, both algorithms also show lower values for pick-level recall, which is due to missed picks. This likely implies that while picks are simply missed in the crustal scenario, they are at least sometimes combined to secondary events in the subduction case (thus the lower event-level precision). This may be due to the larger spatial search space in this scenario, which allows more possibilities for a secondary event to achieve a consistent source location with several picks ``by chance''. 
Interestingly, decreasing the number of constituent picks per event (Section 4.3 and Figure \ref{fig:pick_removal}) increases the event-level precision of GaMMA and PhaseLink substantially. We interpret that this observation implies that if a smaller total number of picks are available, producing a false secondary event with more than 10 arrivals is less likely.

\subsection{Runtime trends and applicability}

The runtime trends observed across the different associators (Section 4.3) show significant differences in scalability and computational efficiency. The DL-based associators GENIE and PhaseLink here demonstrate superior scalability compared to more traditional methods, but in the case of GENIE this is coupled with much slower runtimes especially in smaller scenarios. 
Runtimes of PhaseLink and GENIE mostly scale with the number of identified events and are largely independent of the number of noise picks, whereas an increase of the total number of picks drastically increases runtimes of REAL, GaMMA and PyOcto.

Comparing directly between the best-performing algorithms PyOcto and GENIE, PyOcto is substantially faster (factor of 100 or more) for the smaller-scale applications we tested, while for the largest runs that encompass $>$100k picks, runtimes are roughly similar between these two algorithms. While the largest scenario we tested, which contains 2000 events within 24 hours on a network of $\sim$50 stations, is already quite extreme in terms of event rate (likely corresponding to the aftershock series of a large earthquake), many current (or future) seismic networks can include 100s or even 1000s of stations. In such cases, an algorithm such as GENIE may be beneficial, and the advantage in runtime for such large datasets as well as its flexibility towards network geometry changes over time may easily outweigh the larger effort in initially training the model. For seismic networks of more limited scale, i.e., many regional and local as well as temporary deployments of $\sim$dozens of stations, PyOcto is potentially the most appropriate choice of associator, as it achieves similar performance as GENIE, is much faster, and requires relatively limited parameter configuration before application. 

\section{Conclusions}

We evaluated five phase association algorithms with scenarios of synthetic picks and events that were designed to approximate real-world conditions. We find that GENIE and PyOcto show the overall best performance across all tested scenarios and runs. These two algorithms are the most recently published algorithms, and are also based on very different techniques: one uses neural networks, while the other uses an efficient back-projection based search scheme.  Our results indicate distinct advantages and tradeoffs of each algorithm and do not allow a decision of which fundamental phase association approach (classical or DL) is superior.

While GaMMA and especially PhaseLink showed significant problems in more challenging conditions, REAL exhibited robust performance overall, but has slow runtimes due to its grid search-like approach. PyOcto and GENIE clearly performed best, with only small differences between the two algorithms except for runtimes. There, PyOcto is substantially faster (factor of $\sim$100) for smaller-scale problems, whereas GENIE catches up for larger problems due to better scalability. For the largest problems we tested, their runtimes were comparable. However, greater differences between the two algorithms may appear for larger seismic network applications, and are also indicated by the need for different levels of re-tuning based on observed seismicity characteristics and noise levels.

Taking into account additional considerations such as parameter tradeoffs and ease of configuration, we conclude that PyOcto is well suited for most phase association problems today, unless they feature very high seismicity rates coupled with more than hundreds or thousands of seismic stations. In this latter case, the better runtime scaling as well as the flexibility towards network geometry changes can be significant assets for GENIE. Such applications may become more commonplace in the future, as instrumentation is ever increasing globally.

\section*{Open Research Section}

No actual data was used in this article, only synthetic experiments were conducted, which can be repeated based on the information given in the paper. The five tested phase association algorithms, PhaseLink (https://github.com/interseismic/PhaseLink), REAL (https://github.com/Dal-mzhang/REAL), GaMMA (https://github.com/AI4EPS/GaMMA), GENIE (https://github.com/imcbrearty/GENIE) and PyOcto (https://github.com/yetinam/pyocto), are all freely available for download under the provided links, and installation instruction as well as documentations are provided. The optimal sets of tuning parameters we derived are given in the Supplementary Material (Tables S3-S7).   

The utilized raytracer is contained in the NonLinLoc software package (http://alomax.free.fr/nlloc/), the 1D velocity models can be found in the respective publications \cite{Graeber1999-yc,Hadley1977-ko}. For our different station configuration scenarios, we used real station locations from the networks CX in Chile \cite{GFZ-German-Research-Centre-for-Geosciences2006-yf}, and networks CE \cite{California-Geological-Survey1972-wy}, CI \cite{California-Institute-of-Technology-and-United-States-Geological-Survey-Pasadena1926-jy}, GS \cite{Albuquerque-Seismological-Laboratory-ASL-/USGS1980-ye}, NN \cite{University-of-Nevada-Reno1971-ea}, NP \cite{US-Geological-Survey1931-yn}, PB (https://www.fdsn.org/networks/detail/PB/) and ZY (https://www.fdsn.org/networks/detail/ZY\_1990/) in California.

\acknowledgments

C.S. received funding from the European Research Council (ERC) through the Horizon 2020 program (ERC Starting Grant MILESTONE; StG2020-947856). J.P. and C.S. were supported by the Czech Academy of Sciences through a Mobility Plus (MPP) Grant, grant number ANID-23-08. I.M. was supported by AFRL under contract PA04-S-D00243-12-TO-03-Stanford. J.M. was supported by the European Union under the grant agreement n°101104996 (“DECODE”). This work was supported by the Ministry of Education, Youth and Sports of the Czech Republic through the e-INFRA CZ (ID:90254).

%%%%%%%%%%%%%%%%%%%%%%%%%%%%%%%%%%%%%%%%%%%%%%%
% REFERENCES and BIBLIOGRAPHY
%
% \bibliography{<name of your .bib file>} don't specify the file extension
% don't specify bibliographystyle
%
%%%%%%%%%%%%%%%%%%%%%%%%%%%%%%%%%%%%%%%%%%%%%%%

\bibliography{assoc_benchmark}

\begin{thebibliography}{}

\bibitem [\protect \citeauthoryear {%
{Albuquerque Seismological Laboratory (ASL)/USGS}%
}{%
{Albuquerque Seismological Laboratory (ASL)/USGS}%
}{%
{\protect \APACyear {1980}}%
}]{%
Albuquerque-Seismological-Laboratory-ASL-/USGS1980-ye}
\APACinsertmetastar {%
Albuquerque-Seismological-Laboratory-ASL-/USGS1980-ye}%
\begin{APACrefauthors}%
{Albuquerque Seismological Laboratory (ASL)/USGS}.%
\end{APACrefauthors}%
\unskip\
\newblock
\APACrefYearMonthDay{1980}{}{}.
\newblock
\APACrefbtitle {{US} geological survey networks.} {{US} geological survey networks.}
\newblock
\APACaddressPublisher{}{International Federation of Digital Seismograph Networks}.
\PrintBackRefs{\CurrentBib}

\bibitem [\protect \citeauthoryear {%
Allen%
}{%
Allen%
}{%
{\protect \APACyear {1978}}%
}]{%
Allen1978-uo}
\APACinsertmetastar {%
Allen1978-uo}%
\begin{APACrefauthors}%
Allen, R\BPBI V.%
\end{APACrefauthors}%
\unskip\
\newblock
\APACrefYearMonthDay{1978}{{\APACmonth{10}}}{}.
\newblock
{\BBOQ}\APACrefatitle {Automatic earthquake recognition and timing from single traces} {Automatic earthquake recognition and timing from single traces}.{\BBCQ}
\newblock
\APACjournalVolNumPages{Bull. Seismol. Soc. Am.}{68}{5}{1521--1532}.
\PrintBackRefs{\CurrentBib}

\bibitem [\protect \citeauthoryear {%
Becker%
, McBrearty%
, Beroza%
\BCBL {}\ \BBA {} Martínez-Garzón%
}{%
Becker%
\ \protect \BOthers {.}}{%
{\protect \APACyear {2024}}%
}]{%
Becker2024-oj}
\APACinsertmetastar {%
Becker2024-oj}%
\begin{APACrefauthors}%
Becker, D.%
, McBrearty, I\BPBI W.%
, Beroza, G\BPBI C.%
\BCBL {}\ \BBA {} Martínez-Garzón, P.%
\end{APACrefauthors}%
\unskip\
\newblock
\APACrefYearMonthDay{2024}{{\APACmonth{05}}}{}.
\newblock
{\BBOQ}\APACrefatitle {Performance of {AI}-based phase picking and event association methods after the large 2023 {MW} 7.8 and 7.6 Türkiye doublet} {Performance of {AI}-based phase picking and event association methods after the large 2023 {MW} 7.8 and 7.6 türkiye doublet}.{\BBCQ}
\newblock
\APACjournalVolNumPages{Bull. Seismol. Soc. Am.}{}{}{}.
\PrintBackRefs{\CurrentBib}

\bibitem [\protect \citeauthoryear {%
Bishop%
}{%
Bishop%
}{%
{\protect \APACyear {2006}}%
}]{%
Bishop2006-ei}
\APACinsertmetastar {%
Bishop2006-ei}%
\begin{APACrefauthors}%
Bishop, C\BPBI M.%
\end{APACrefauthors}%
\unskip\
\newblock
\APACrefYear{2006}.
\newblock
\APACrefbtitle {Pattern Recognition and Machine Learning} {Pattern recognition and machine learning}\ (\PrintOrdinal{1}\ \BEd).
\newblock
\APACaddressPublisher{New York, NY}{Springer}.
\PrintBackRefs{\CurrentBib}

\bibitem [\protect \citeauthoryear {%
{California Geological Survey}%
}{%
{California Geological Survey}%
}{%
{\protect \APACyear {1972}}%
}]{%
California-Geological-Survey1972-wy}
\APACinsertmetastar {%
California-Geological-Survey1972-wy}%
\begin{APACrefauthors}%
{California Geological Survey}.%
\end{APACrefauthors}%
\unskip\
\newblock
\APACrefYearMonthDay{1972}{}{}.
\newblock
\APACrefbtitle {California strong motion instrumentation program.} {California strong motion instrumentation program.}
\newblock
\APACaddressPublisher{}{International Federation of Digital Seismograph Networks}.
\PrintBackRefs{\CurrentBib}

\bibitem [\protect \citeauthoryear {%
{California Institute of Technology and United States Geological Survey Pasadena}%
}{%
{California Institute of Technology and United States Geological Survey Pasadena}%
}{%
{\protect \APACyear {1926}}%
}]{%
California-Institute-of-Technology-and-United-States-Geological-Survey-Pasadena1926-jy}
\APACinsertmetastar {%
California-Institute-of-Technology-and-United-States-Geological-Survey-Pasadena1926-jy}%
\begin{APACrefauthors}%
{California Institute of Technology and United States Geological Survey Pasadena}.%
\end{APACrefauthors}%
\unskip\
\newblock
\APACrefYearMonthDay{1926}{}{}.
\newblock
\APACrefbtitle {Southern California seismic network.} {Southern california seismic network.}
\newblock
\APACaddressPublisher{}{International Federation of Digital Seismograph Networks}.
\PrintBackRefs{\CurrentBib}

\bibitem [\protect \citeauthoryear {%
Diehl%
, Kissling%
, Husen%
\BCBL {}\ \BBA {} Aldersons%
}{%
Diehl%
\ \protect \BOthers {.}}{%
{\protect \APACyear {2009}}%
}]{%
Diehl2009-px}
\APACinsertmetastar {%
Diehl2009-px}%
\begin{APACrefauthors}%
Diehl, T.%
, Kissling, E.%
, Husen, S.%
\BCBL {}\ \BBA {} Aldersons, F.%
\end{APACrefauthors}%
\unskip\
\newblock
\APACrefYearMonthDay{2009}{{\APACmonth{02}}}{}.
\newblock
{\BBOQ}\APACrefatitle {Consistent phase picking for regional tomography models: application to the greater Alpine region} {Consistent phase picking for regional tomography models: application to the greater alpine region}.{\BBCQ}
\newblock
\APACjournalVolNumPages{Geophys J Int}{176}{2}{542--554}.
\PrintBackRefs{\CurrentBib}

\bibitem [\protect \citeauthoryear {%
Ester%
, Kriegel%
, Sander%
\BCBL {}\ \BBA {} Xu%
}{%
Ester%
\ \protect \BOthers {.}}{%
{\protect \APACyear {1996}}%
}]{%
Ester1996-nz}
\APACinsertmetastar {%
Ester1996-nz}%
\begin{APACrefauthors}%
Ester, M.%
, Kriegel, H.%
, Sander, J.%
\BCBL {}\ \BBA {} Xu, X.%
\end{APACrefauthors}%
\unskip\
\newblock
\APACrefYearMonthDay{1996}{{\APACmonth{08}}}{}.
\newblock
{\BBOQ}\APACrefatitle {A density-based algorithm for discovering clusters in large spatial databases with noise} {A density-based algorithm for discovering clusters in large spatial databases with noise}.{\BBCQ}
\newblock
\APACjournalVolNumPages{KDD}{}{}{226--231}.
\PrintBackRefs{\CurrentBib}

\bibitem [\protect \citeauthoryear {%
{GFZ German Research Centre for Geosciences}%
\ \BBA {} {Institut des Sciences de l'Univers-Centre National de la Recherche CNRS-INSU}%
}{%
{GFZ German Research Centre for Geosciences}%
\ \BBA {} {Institut des Sciences de l'Univers-Centre National de la Recherche CNRS-INSU}%
}{%
{\protect \APACyear {2006}}%
}]{%
GFZ-German-Research-Centre-for-Geosciences2006-yf}
\APACinsertmetastar {%
GFZ-German-Research-Centre-for-Geosciences2006-yf}%
\begin{APACrefauthors}%
{GFZ German Research Centre for Geosciences}%
\BCBT {}\ \BBA {} {Institut des Sciences de l'Univers-Centre National de la Recherche CNRS-INSU}.%
\end{APACrefauthors}%
\unskip\
\newblock
\APACrefYearMonthDay{2006}{}{}.
\newblock
\APACrefbtitle {{IPOC} Seismic Network.} {{IPOC} seismic network.}
\newblock
\APACaddressPublisher{}{Integrated Plate boundary Observatory Chile - IPOC}.
\PrintBackRefs{\CurrentBib}

\bibitem [\protect \citeauthoryear {%
Graeber%
\ \BBA {} Asch%
}{%
Graeber%
\ \BBA {} Asch%
}{%
{\protect \APACyear {1999}}%
}]{%
Graeber1999-yc}
\APACinsertmetastar {%
Graeber1999-yc}%
\begin{APACrefauthors}%
Graeber, F\BPBI M.%
\BCBT {}\ \BBA {} Asch, G.%
\end{APACrefauthors}%
\unskip\
\newblock
\APACrefYearMonthDay{1999}{{\APACmonth{09}}}{}.
\newblock
{\BBOQ}\APACrefatitle {Three-dimensional models {ofPwave} velocity {andP}-to-Svelocity ratio in the southern central Andes by simultaneous inversion of local earthquake data} {Three-dimensional models {ofPwave} velocity {andP}-to-svelocity ratio in the southern central andes by simultaneous inversion of local earthquake data}.{\BBCQ}
\newblock
\APACjournalVolNumPages{J. Geophys. Res.}{104}{B9}{20237--20256}.
\PrintBackRefs{\CurrentBib}

\bibitem [\protect \citeauthoryear {%
Hadley%
\ \BBA {} Kanamori%
}{%
Hadley%
\ \BBA {} Kanamori%
}{%
{\protect \APACyear {1977}}%
}]{%
Hadley1977-ko}
\APACinsertmetastar {%
Hadley1977-ko}%
\begin{APACrefauthors}%
Hadley, D.%
\BCBT {}\ \BBA {} Kanamori, H.%
\end{APACrefauthors}%
\unskip\
\newblock
\APACrefYearMonthDay{1977}{{\APACmonth{10}}}{}.
\newblock
{\BBOQ}\APACrefatitle {Seismic structure of the Transverse Ranges, California} {Seismic structure of the transverse ranges, california}.{\BBCQ}
\newblock
\APACjournalVolNumPages{GSA Bulletin}{88}{10}{1469--1478}.
\PrintBackRefs{\CurrentBib}

\bibitem [\protect \citeauthoryear {%
Hainzl%
, Sippl%
\BCBL {}\ \BBA {} Schurr%
}{%
Hainzl%
\ \protect \BOthers {.}}{%
{\protect \APACyear {2019}}%
}]{%
Hainzl2019-wk}
\APACinsertmetastar {%
Hainzl2019-wk}%
\begin{APACrefauthors}%
Hainzl, S.%
, Sippl, C.%
\BCBL {}\ \BBA {} Schurr, B.%
\end{APACrefauthors}%
\unskip\
\newblock
\APACrefYearMonthDay{2019}{{\APACmonth{08}}}{}.
\newblock
{\BBOQ}\APACrefatitle {Linear relationship between aftershock productivity and seismic coupling in the northern Chile subduction zone} {Linear relationship between aftershock productivity and seismic coupling in the northern chile subduction zone}.{\BBCQ}
\newblock
\APACjournalVolNumPages{J. Geophys. Res. Solid Earth}{124}{8}{8726--8738}.
\PrintBackRefs{\CurrentBib}

\bibitem [\protect \citeauthoryear {%
Johnson%
, Bittenbinder%
, Bogaert%
, Dietz%
\BCBL {}\ \BBA {} Kohler%
}{%
Johnson%
\ \protect \BOthers {.}}{%
{\protect \APACyear {1995}}%
}]{%
Johnson1995-cv}
\APACinsertmetastar {%
Johnson1995-cv}%
\begin{APACrefauthors}%
Johnson, C\BPBI E.%
, Bittenbinder, A.%
, Bogaert, B.%
, Dietz, L.%
\BCBL {}\ \BBA {} Kohler, W.%
\end{APACrefauthors}%
\unskip\
\newblock
\APACrefYearMonthDay{1995}{}{}.
\newblock
{\BBOQ}\APACrefatitle {Earthworm: A flexible approach to seismic network processing} {Earthworm: A flexible approach to seismic network processing}.{\BBCQ}
\newblock
\APACjournalVolNumPages{Iris newsletter}{14}{2}{1--4}.
\PrintBackRefs{\CurrentBib}

\bibitem [\protect \citeauthoryear {%
Maharaj%
\ \protect \BOthers {.}}{%
Maharaj%
\ \protect \BOthers {.}}{%
{\protect \APACyear {2023}}%
}]{%
Maharaj2023-xv}
\APACinsertmetastar {%
Maharaj2023-xv}%
\begin{APACrefauthors}%
Maharaj, A.%
, Roecker, S.%
, Alvarado, P.%
, Trad, S.%
, Beck, S.%
\BCBL {}\ \BBA {} Comte, D.%
\end{APACrefauthors}%
\unskip\
\newblock
\APACrefYearMonthDay{2023}{{\APACmonth{10}}}{}.
\newblock
{\BBOQ}\APACrefatitle {Are volatiles from subducted ridges on the Pampean flat slab fracking the crust? Evidence from an enhanced seismicity catalog} {Are volatiles from subducted ridges on the pampean flat slab fracking the crust? evidence from an enhanced seismicity catalog}.{\BBCQ}
\newblock
\APACjournalVolNumPages{Geochem. Geophys. Geosyst.}{24}{10}{}.
\PrintBackRefs{\CurrentBib}

\bibitem [\protect \citeauthoryear {%
Mancini%
\ \protect \BOthers {.}}{%
Mancini%
\ \protect \BOthers {.}}{%
{\protect \APACyear {2022}}%
}]{%
Mancini2022-gf}
\APACinsertmetastar {%
Mancini2022-gf}%
\begin{APACrefauthors}%
Mancini, S.%
, Segou, M.%
, Werner, M\BPBI J.%
, Parsons, T.%
, Beroza, G.%
\BCBL {}\ \BBA {} Chiaraluce, L.%
\end{APACrefauthors}%
\unskip\
\newblock
\APACrefYearMonthDay{2022}{{\APACmonth{11}}}{}.
\newblock
{\BBOQ}\APACrefatitle {On the use of high-resolution and deep-learning seismic catalogs for short-term earthquake forecasts: Potential benefits and current limitations} {On the use of high-resolution and deep-learning seismic catalogs for short-term earthquake forecasts: Potential benefits and current limitations}.{\BBCQ}
\newblock
\APACjournalVolNumPages{J. Geophys. Res. Solid Earth}{127}{11}{e2022JB025202}.
\PrintBackRefs{\CurrentBib}

\bibitem [\protect \citeauthoryear {%
McBrearty%
\ \BBA {} Beroza%
}{%
McBrearty%
\ \BBA {} Beroza%
}{%
{\protect \APACyear {2023}}%
}]{%
McBrearty2023-th}
\APACinsertmetastar {%
McBrearty2023-th}%
\begin{APACrefauthors}%
McBrearty, I\BPBI W.%
\BCBT {}\ \BBA {} Beroza, G\BPBI C.%
\end{APACrefauthors}%
\unskip\
\newblock
\APACrefYearMonthDay{2023}{{\APACmonth{04}}}{}.
\newblock
{\BBOQ}\APACrefatitle {Earthquake Phase Association with Graph Neural Networks} {Earthquake phase association with graph neural networks}.{\BBCQ}
\newblock
\APACjournalVolNumPages{Bull. Seismol. Soc. Am.}{113}{2}{524--547}.
\PrintBackRefs{\CurrentBib}

\bibitem [\protect \citeauthoryear {%
Mousavi%
, Ellsworth%
, Zhu%
, Chuang%
\BCBL {}\ \BBA {} Beroza%
}{%
Mousavi%
\ \protect \BOthers {.}}{%
{\protect \APACyear {2020}}%
}]{%
Mousavi2020-yc}
\APACinsertmetastar {%
Mousavi2020-yc}%
\begin{APACrefauthors}%
Mousavi, S\BPBI M.%
, Ellsworth, W\BPBI L.%
, Zhu, W.%
, Chuang, L\BPBI Y.%
\BCBL {}\ \BBA {} Beroza, G\BPBI C.%
\end{APACrefauthors}%
\unskip\
\newblock
\APACrefYearMonthDay{2020}{{\APACmonth{08}}}{}.
\newblock
{\BBOQ}\APACrefatitle {Earthquake transformer-an attentive deep-learning model for simultaneous earthquake detection and phase picking} {Earthquake transformer-an attentive deep-learning model for simultaneous earthquake detection and phase picking}.{\BBCQ}
\newblock
\APACjournalVolNumPages{Nat. Commun.}{11}{1}{3952}.
\PrintBackRefs{\CurrentBib}

\bibitem [\protect \citeauthoryear {%
Mousavi%
, Zhu%
, Sheng%
\BCBL {}\ \BBA {} Beroza%
}{%
Mousavi%
\ \protect \BOthers {.}}{%
{\protect \APACyear {2019}}%
}]{%
Mousavi2019-tf}
\APACinsertmetastar {%
Mousavi2019-tf}%
\begin{APACrefauthors}%
Mousavi, S\BPBI M.%
, Zhu, W.%
, Sheng, Y.%
\BCBL {}\ \BBA {} Beroza, G\BPBI C.%
\end{APACrefauthors}%
\unskip\
\newblock
\APACrefYearMonthDay{2019}{{\APACmonth{07}}}{}.
\newblock
{\BBOQ}\APACrefatitle {{CRED}: A deep residual network of convolutional and recurrent units for earthquake signal detection} {{CRED}: A deep residual network of convolutional and recurrent units for earthquake signal detection}.{\BBCQ}
\newblock
\APACjournalVolNumPages{Sci. Rep.}{9}{1}{1--14}.
\PrintBackRefs{\CurrentBib}

\bibitem [\protect \citeauthoryear {%
Münchmeyer%
}{%
Münchmeyer%
}{%
{\protect \APACyear {2024}}%
}]{%
Munchmeyer2024-kk}
\APACinsertmetastar {%
Munchmeyer2024-kk}%
\begin{APACrefauthors}%
Münchmeyer, J.%
\end{APACrefauthors}%
\unskip\
\newblock
\APACrefYearMonthDay{2024}{{\APACmonth{01}}}{}.
\newblock
{\BBOQ}\APACrefatitle {{PyOcto}: A high-throughput seismic phase associator} {{PyOcto}: A high-throughput seismic phase associator}.{\BBCQ}
\newblock
\APACjournalVolNumPages{Seismica}{3}{1}{}.
\PrintBackRefs{\CurrentBib}

\bibitem [\protect \citeauthoryear {%
Münchmeyer%
\ \protect \BOthers {.}}{%
Münchmeyer%
\ \protect \BOthers {.}}{%
{\protect \APACyear {2022}}%
}]{%
Munchmeyer2022-eh}
\APACinsertmetastar {%
Munchmeyer2022-eh}%
\begin{APACrefauthors}%
Münchmeyer, J.%
, Woollam, J.%
, Rietbrock, A.%
, Tilmann, F.%
, Lange, D.%
, Bornstein, T.%
\BDBL {}Soto, H.%
\end{APACrefauthors}%
\unskip\
\newblock
\APACrefYearMonthDay{2022}{{\APACmonth{01}}}{}.
\newblock
{\BBOQ}\APACrefatitle {Which picker fits my data? A quantitative evaluation of deep learning based seismic pickers} {Which picker fits my data? a quantitative evaluation of deep learning based seismic pickers}.{\BBCQ}
\newblock
\APACjournalVolNumPages{J. Geophys. Res. [Solid Earth]}{127}{1}{}.
\PrintBackRefs{\CurrentBib}

\bibitem [\protect \citeauthoryear {%
Ringdal%
\ \BBA {} Kværna%
}{%
Ringdal%
\ \BBA {} Kværna%
}{%
{\protect \APACyear {1989}}%
}]{%
Ringdal1989-bv}
\APACinsertmetastar {%
Ringdal1989-bv}%
\begin{APACrefauthors}%
Ringdal, F.%
\BCBT {}\ \BBA {} Kværna, T.%
\end{APACrefauthors}%
\unskip\
\newblock
\APACrefYearMonthDay{1989}{{\APACmonth{12}}}{}.
\newblock
{\BBOQ}\APACrefatitle {A multi-channel processing approach to real time network detection, phase association, and threshold monitoring} {A multi-channel processing approach to real time network detection, phase association, and threshold monitoring}.{\BBCQ}
\newblock
\APACjournalVolNumPages{Bulletin of the Seismological Society of America}{79}{6}{1927--1940}.
\PrintBackRefs{\CurrentBib}

\bibitem [\protect \citeauthoryear {%
Ross%
, Meier%
, Hauksson%
\BCBL {}\ \BBA {} Heaton%
}{%
Ross%
\ \protect \BOthers {.}}{%
{\protect \APACyear {2018}}%
}]{%
Ross2018-xf}
\APACinsertmetastar {%
Ross2018-xf}%
\begin{APACrefauthors}%
Ross, Z\BPBI E.%
, Meier, M.%
, Hauksson, E.%
\BCBL {}\ \BBA {} Heaton, T\BPBI H.%
\end{APACrefauthors}%
\unskip\
\newblock
\APACrefYearMonthDay{2018}{{\APACmonth{10}}}{}.
\newblock
{\BBOQ}\APACrefatitle {Generalized Seismic Phase Detection with Deep Learning} {Generalized seismic phase detection with deep learning}.{\BBCQ}
\newblock
\APACjournalVolNumPages{Bull. Seismol. Soc. Am.}{108}{5A}{2894--2901}.
\PrintBackRefs{\CurrentBib}

\bibitem [\protect \citeauthoryear {%
Ross%
, Yue%
, Meier%
, Hauksson%
\BCBL {}\ \BBA {} Heaton%
}{%
Ross%
\ \protect \BOthers {.}}{%
{\protect \APACyear {2019}}%
}]{%
Ross2019-rx}
\APACinsertmetastar {%
Ross2019-rx}%
\begin{APACrefauthors}%
Ross, Z\BPBI E.%
, Yue, Y.%
, Meier, M\BHBI A.%
, Hauksson, E.%
\BCBL {}\ \BBA {} Heaton, T\BPBI H.%
\end{APACrefauthors}%
\unskip\
\newblock
\APACrefYearMonthDay{2019}{{\APACmonth{01}}}{}.
\newblock
{\BBOQ}\APACrefatitle {{PhaseLink}: A deep learning approach to seismic phase association} {{PhaseLink}: A deep learning approach to seismic phase association}.{\BBCQ}
\newblock
\APACjournalVolNumPages{J. Geophys. Res. [Solid Earth]}{124}{1}{856--869}.
\PrintBackRefs{\CurrentBib}

\bibitem [\protect \citeauthoryear {%
Sippl%
, Schurr%
, John%
\BCBL {}\ \BBA {} Hainzl%
}{%
Sippl%
\ \protect \BOthers {.}}{%
{\protect \APACyear {2019}}%
}]{%
Sippl2019-gu}
\APACinsertmetastar {%
Sippl2019-gu}%
\begin{APACrefauthors}%
Sippl, C.%
, Schurr, B.%
, John, T.%
\BCBL {}\ \BBA {} Hainzl, S.%
\end{APACrefauthors}%
\unskip\
\newblock
\APACrefYearMonthDay{2019}{{\APACmonth{11}}}{}.
\newblock
{\BBOQ}\APACrefatitle {Filling the gap in a double seismic zone: Intraslab seismicity in Northern Chile} {Filling the gap in a double seismic zone: Intraslab seismicity in northern chile}.{\BBCQ}
\newblock
\APACjournalVolNumPages{Lithos}{346-347}{105155}{105155}.
\PrintBackRefs{\CurrentBib}

\bibitem [\protect \citeauthoryear {%
Stern%
}{%
Stern%
}{%
{\protect \APACyear {2002}}%
}]{%
Stern2002-ym}
\APACinsertmetastar {%
Stern2002-ym}%
\begin{APACrefauthors}%
Stern, R\BPBI J.%
\end{APACrefauthors}%
\unskip\
\newblock
\APACrefYearMonthDay{2002}{{\APACmonth{12}}}{}.
\newblock
{\BBOQ}\APACrefatitle {Subduction zones} {Subduction zones}.{\BBCQ}
\newblock
\APACjournalVolNumPages{Rev. Geophys.}{40}{4}{3--1--3--38}.
\PrintBackRefs{\CurrentBib}

\bibitem [\protect \citeauthoryear {%
{University of Nevada, Reno}%
}{%
{University of Nevada, Reno}%
}{%
{\protect \APACyear {1971}}%
}]{%
University-of-Nevada-Reno1971-ea}
\APACinsertmetastar {%
University-of-Nevada-Reno1971-ea}%
\begin{APACrefauthors}%
{University of Nevada, Reno}.%
\end{APACrefauthors}%
\unskip\
\newblock
\APACrefYearMonthDay{1971}{}{}.
\newblock
\APACrefbtitle {Nevada Seismic Network.} {Nevada seismic network.}
\newblock
\APACaddressPublisher{}{International Federation of Digital Seismograph Networks}.
\PrintBackRefs{\CurrentBib}

\bibitem [\protect \citeauthoryear {%
{U.S. Geological Survey}%
}{%
{U.S. Geological Survey}%
}{%
{\protect \APACyear {1931}}%
}]{%
US-Geological-Survey1931-yn}
\APACinsertmetastar {%
US-Geological-Survey1931-yn}%
\begin{APACrefauthors}%
{U.S. Geological Survey}.%
\end{APACrefauthors}%
\unskip\
\newblock
\APACrefYearMonthDay{1931}{}{}.
\newblock
\APACrefbtitle {United States national strong-motion network.} {United states national strong-motion network.}
\newblock
\APACaddressPublisher{}{International Federation of Digital Seismograph Networks}.
\PrintBackRefs{\CurrentBib}

\bibitem [\protect \citeauthoryear {%
White%
\ \protect \BOthers {.}}{%
White%
\ \protect \BOthers {.}}{%
{\protect \APACyear {2021}}%
}]{%
White2021-mn}
\APACinsertmetastar {%
White2021-mn}%
\begin{APACrefauthors}%
White, M\BPBI C\BPBI A.%
, Fang, H.%
, Catchings, R\BPBI D.%
, Goldman, M\BPBI R.%
, Steidl, J\BPBI H.%
\BCBL {}\ \BBA {} Ben-Zion, Y.%
\end{APACrefauthors}%
\unskip\
\newblock
\APACrefYearMonthDay{2021}{{\APACmonth{06}}}{}.
\newblock
{\BBOQ}\APACrefatitle {Detailed traveltime tomography and seismic catalogue around the 2019 \textit{M}w7.1 Ridgecrest, California, earthquake using dense rapid-response seismic data} {Detailed traveltime tomography and seismic catalogue around the 2019 \textit{M}w7.1 ridgecrest, california, earthquake using dense rapid-response seismic data}.{\BBCQ}
\newblock
\APACjournalVolNumPages{Geophys. J. Int.}{227}{1}{204--227}.
\PrintBackRefs{\CurrentBib}

\bibitem [\protect \citeauthoryear {%
Williamson%
, Lux%
\BCBL {}\ \BBA {} Allen%
}{%
Williamson%
\ \protect \BOthers {.}}{%
{\protect \APACyear {2023}}%
}]{%
Williamson2023-aq}
\APACinsertmetastar {%
Williamson2023-aq}%
\begin{APACrefauthors}%
Williamson, A.%
, Lux, A.%
\BCBL {}\ \BBA {} Allen, R.%
\end{APACrefauthors}%
\unskip\
\newblock
\APACrefYearMonthDay{2023}{{\APACmonth{01}}}{}.
\newblock
{\BBOQ}\APACrefatitle {Improving out of network earthquake locations using prior seismicity for use in earthquake early warning} {Improving out of network earthquake locations using prior seismicity for use in earthquake early warning}.{\BBCQ}
\newblock
\APACjournalVolNumPages{Bull. Seismol. Soc. Am.}{}{}{}.
\PrintBackRefs{\CurrentBib}

\bibitem [\protect \citeauthoryear {%
Woollam%
\ \protect \BOthers {.}}{%
Woollam%
\ \protect \BOthers {.}}{%
{\protect \APACyear {2022}}%
}]{%
Woollam2022-ri}
\APACinsertmetastar {%
Woollam2022-ri}%
\begin{APACrefauthors}%
Woollam, J.%
, Münchmeyer, J.%
, Tilmann, F.%
, Rietbrock, A.%
, Lange, D.%
, Bornstein, T.%
\BDBL {}Soto, H.%
\end{APACrefauthors}%
\unskip\
\newblock
\APACrefYearMonthDay{2022}{{\APACmonth{05}}}{}.
\newblock
{\BBOQ}\APACrefatitle {{SeisBench—A} Toolbox for Machine Learning in Seismology} {{SeisBench—A} toolbox for machine learning in seismology}.{\BBCQ}
\newblock
\APACjournalVolNumPages{Seismol. Res. Lett.}{93}{3}{1695--1709}.
\PrintBackRefs{\CurrentBib}

\bibitem [\protect \citeauthoryear {%
Xiong%
, Brudzinski%
, Gossett%
, Lin%
\BCBL {}\ \BBA {} Hampton%
}{%
Xiong%
\ \protect \BOthers {.}}{%
{\protect \APACyear {2023}}%
}]{%
Xiong2023-xm}
\APACinsertmetastar {%
Xiong2023-xm}%
\begin{APACrefauthors}%
Xiong, Q.%
, Brudzinski, M\BPBI R.%
, Gossett, D.%
, Lin, Q.%
\BCBL {}\ \BBA {} Hampton, J\BPBI C.%
\end{APACrefauthors}%
\unskip\
\newblock
\APACrefYearMonthDay{2023}{{\APACmonth{04}}}{}.
\newblock
{\BBOQ}\APACrefatitle {Seismic magnitude clustering is prevalent in field and laboratory catalogs} {Seismic magnitude clustering is prevalent in field and laboratory catalogs}.{\BBCQ}
\newblock
\APACjournalVolNumPages{Nat. Commun.}{14}{1}{2056}.
\PrintBackRefs{\CurrentBib}

\bibitem [\protect \citeauthoryear {%
Yang%
, Hu%
, Zhang%
\BCBL {}\ \BBA {} Liu%
}{%
Yang%
\ \protect \BOthers {.}}{%
{\protect \APACyear {2021}}%
}]{%
Yang2021-bc}
\APACinsertmetastar {%
Yang2021-bc}%
\begin{APACrefauthors}%
Yang, S.%
, Hu, J.%
, Zhang, H.%
\BCBL {}\ \BBA {} Liu, G.%
\end{APACrefauthors}%
\unskip\
\newblock
\APACrefYearMonthDay{2021}{{\APACmonth{01}}}{}.
\newblock
{\BBOQ}\APACrefatitle {Simultaneous earthquake detection on multiple stations via a convolutional neural network} {Simultaneous earthquake detection on multiple stations via a convolutional neural network}.{\BBCQ}
\newblock
\APACjournalVolNumPages{Seismol. Res. Lett.}{92}{1}{246--260}.
\PrintBackRefs{\CurrentBib}

\bibitem [\protect \citeauthoryear {%
Zhang%
, Ellsworth%
\BCBL {}\ \BBA {} Beroza%
}{%
Zhang%
\ \protect \BOthers {.}}{%
{\protect \APACyear {2019}}%
}]{%
Zhang2019-qq}
\APACinsertmetastar {%
Zhang2019-qq}%
\begin{APACrefauthors}%
Zhang, M.%
, Ellsworth, W\BPBI L.%
\BCBL {}\ \BBA {} Beroza, G\BPBI C.%
\end{APACrefauthors}%
\unskip\
\newblock
\APACrefYearMonthDay{2019}{{\APACmonth{11}}}{}.
\newblock
{\BBOQ}\APACrefatitle {Rapid Earthquake Association and Location} {Rapid earthquake association and location}.{\BBCQ}
\newblock
\APACjournalVolNumPages{Seismol. Res. Lett.}{90}{6}{2276--2284}.
\PrintBackRefs{\CurrentBib}

\bibitem [\protect \citeauthoryear {%
L.~Zhu%
\ \protect \BOthers {.}}{%
L.~Zhu%
\ \protect \BOthers {.}}{%
{\protect \APACyear {2019}}%
}]{%
Zhu2019-lb}
\APACinsertmetastar {%
Zhu2019-lb}%
\begin{APACrefauthors}%
Zhu, L.%
, Peng, Z.%
, McClellan, J.%
, Li, C.%
, Yao, D.%
, Li, Z.%
\BCBL {}\ \BBA {} Fang, L.%
\end{APACrefauthors}%
\unskip\
\newblock
\APACrefYearMonthDay{2019}{{\APACmonth{08}}}{}.
\newblock
{\BBOQ}\APACrefatitle {Deep learning for seismic phase detection and picking in the aftershock zone of 2008 {M7}.9 Wenchuan Earthquake} {Deep learning for seismic phase detection and picking in the aftershock zone of 2008 {M7}.9 wenchuan earthquake}.{\BBCQ}
\newblock
\APACjournalVolNumPages{Phys. Earth Planet. Inter.}{293}{106261}{106261}.
\PrintBackRefs{\CurrentBib}

\bibitem [\protect \citeauthoryear {%
W.~Zhu%
\ \BBA {} Beroza%
}{%
W.~Zhu%
\ \BBA {} Beroza%
}{%
{\protect \APACyear {2018}}%
}]{%
Zhu2018-ki}
\APACinsertmetastar {%
Zhu2018-ki}%
\begin{APACrefauthors}%
Zhu, W.%
\BCBT {}\ \BBA {} Beroza, G\BPBI C.%
\end{APACrefauthors}%
\unskip\
\newblock
\APACrefYearMonthDay{2018}{{\APACmonth{10}}}{}.
\newblock
{\BBOQ}\APACrefatitle {{PhaseNet}: a deep-neural-network-based seismic arrival-time picking method} {{PhaseNet}: a deep-neural-network-based seismic arrival-time picking method}.{\BBCQ}
\newblock
\APACjournalVolNumPages{Geophys. J. Int.}{216}{1}{261--273}.
\PrintBackRefs{\CurrentBib}

\bibitem [\protect \citeauthoryear {%
W.~Zhu%
, McBrearty%
, Mousavi%
, Ellsworth%
\BCBL {}\ \BBA {} Beroza%
}{%
W.~Zhu%
, McBrearty%
\BCBL {}\ \protect \BOthers {.}}{%
{\protect \APACyear {2022}}%
}]{%
Zhu2022-ad}
\APACinsertmetastar {%
Zhu2022-ad}%
\begin{APACrefauthors}%
Zhu, W.%
, McBrearty, I\BPBI W.%
, Mousavi, S\BPBI M.%
, Ellsworth, W\BPBI L.%
\BCBL {}\ \BBA {} Beroza, G\BPBI C.%
\end{APACrefauthors}%
\unskip\
\newblock
\APACrefYearMonthDay{2022}{{\APACmonth{05}}}{}.
\newblock
{\BBOQ}\APACrefatitle {Earthquake phase association using a Bayesian Gaussian mixture model} {Earthquake phase association using a bayesian gaussian mixture model}.{\BBCQ}
\newblock
\APACjournalVolNumPages{J. Geophys. Res. [Solid Earth]}{127}{5}{}.
\PrintBackRefs{\CurrentBib}

\bibitem [\protect \citeauthoryear {%
W.~Zhu%
, Tai%
, Mousavi%
, Bailis%
\BCBL {}\ \BBA {} Beroza%
}{%
W.~Zhu%
, Tai%
\BCBL {}\ \protect \BOthers {.}}{%
{\protect \APACyear {2022}}%
}]{%
Zhu2022-pu}
\APACinsertmetastar {%
Zhu2022-pu}%
\begin{APACrefauthors}%
Zhu, W.%
, Tai, K\BPBI S.%
, Mousavi, S\BPBI M.%
, Bailis, P.%
\BCBL {}\ \BBA {} Beroza, G\BPBI C.%
\end{APACrefauthors}%
\unskip\
\newblock
\APACrefYearMonthDay{2022}{{\APACmonth{03}}}{}.
\newblock
{\BBOQ}\APACrefatitle {An end‐to‐end earthquake detection method for joint phase picking and association using deep learning} {An end‐to‐end earthquake detection method for joint phase picking and association using deep learning}.{\BBCQ}
\newblock
\APACjournalVolNumPages{J. Geophys. Res. Solid Earth}{127}{3}{}.
\PrintBackRefs{\CurrentBib}

\end{thebibliography}

%Reference citation instructions and examples:
%
% Please use ONLY \cite and \citeA for reference citations.
% \cite for parenthetical references
% ...as shown in recent studies (Simpson et al., 2019)
% \citeA for in-text citations
% ...Simpson et al. (2019) have shown...
%
%
%...as shown by \citeA{jskilby}.
%...as shown by \citeA{lewin76}, \citeA{carson86}, \citeA{bartoldy02}, and \citeA{rinaldi03}.
%...has been shown \cite{jskilbye}.
%...has been shown \cite{lewin76,carson86,bartoldy02,rinaldi03}.
%... \cite <i.e.>[]{lewin76,carson86,bartoldy02,rinaldi03}.
%...has been shown by \cite <e.g.,>[and others]{lewin76}.
%
% apacite uses < > for prenotes and [ ] for postnotes
% DO NOT use other cite commands (e.g., \citet, \citep, \citeyear, \nocite, \citealp, etc.).
%

\end{document}

\title{Supporting Information for ``Benchmarking seismic phase associators: Insights from synthetic scenarios"}
\authors{Jorge Puente\affil{1,2}, Christian Sippl\affil{1}, Jannes Münchmeyer\affil{3}, Ian W. McBrearty\affil{4}}

\affiliation{1}{ Institute of Geophysics, Czech Academy of Sciences, Prague, Czech Republic}
\affiliation{2}{Charles University, Faculty of Mathematics and Physics, Department of Geophysics, Prague, Czech Republic}
\affiliation{3}{Univ. Grenoble Alpes, Univ. Savoie Mont Blanc, CNRS, IRD, Univ. Gustave Eiffel, ISTerre, Grenoble, France}
\affiliation{4}{Department of Geophysics, Stanford University, Stanford, California, U.S.A.}

\begin{article}

\noindent\textbf{Contents of this file}
%%%Remove or add items as needed%%%
\begin{enumerate}
\item Texts S1 and S2 
\item Figures S1 to S6
\item Tables S1 to S12
%if Tables are larger than 1 page, upload as separate excel file
\end{enumerate}

\noindent\textbf{Text S1: Evaluating the use of 0D vs. 1D velocity models}

In evaluating seismic phase associators, the choice between a simple homogeneous (0D) velocity model and a more detailed 1D velocity model can significantly impact performance. While 0D models offer simplicity and computational efficiency, they may overlook critical depth-dependent variations in seismic wave propagation that 1D models capture and introduce systematic errors when predicting traveltimes at larger distances. We thus compare the performance of REAL, GaMMA and PyOcto with 0D and 1D velocity models, for the crustal as well as the subduction zone scenario. Results from these runs are shown in Figures S2 and S3, and raw results are provided in Table S8. In the crustal scenario, where seismicity is shallow and thus occurs in a region of relatively uniform velocities, the differences between 0D and 1D models are generally modest. GaMMA and PyOcto showed slight improvements in precision, recall, and runtime efficiency when using the 1D model, particularly at higher event counts and noise levels. For REAL, the 1D model reduced runtime but led to slightly lower performance scores overall. In the subduction zone scenario, characterized by a greater hypocentral depth range and longer raypaths, i.e. higher expected model errors, we encounter larger differences between 0D and 1D model versions. For GaMMA and especially PyOcto, the 1D version significantly outperforms the 0D one, whereas REAL once again shows better performance with the 0D model, suggesting that the simpler homogeneous model is more effective here. Based on these findings, we selected the 0D version of REAL, and 1D versions of GaMMA and PyOcto for comparison against the deep learning-based associators, as shown in Section 4.

\noindent\textbf{Text S2: Training the deep learning based algorithms}

The deep learning-based algorithms need to be trained prior to application, which is usually done with synthetic data. The algorithms' performance critically depends on how large and realistic the training datasets are, as well as on the adequate choice of a number of parameters that steer the training process. We here outline the training approaches for PhaseLink and GENIE.
These associators require extensive synthetic data generation to expose the model to a wide range of event-station geometries. We used an approach highly similar to the one previously outlined (Section 3.2) to create synthetic arrival time data from 1D velocity models of the regions of interest.

PhaseLink is trained with a supervised learning approach, where the ground truth associations (labels) are known. We train PhaseLink for 100 epochs, saving model checkpoints at each epoch, and select the checkpoint with the lowest validation loss. During training, the selection of training parameters for PhaseLink is conducted through an iterative process, similar to how we optimized parameters for the other associators (Section 3.4). For instance, a batch size of 64 is found to be optimal for the subduction zone scenario, whereas the higher station density of the crustal scenario necessitates a higher value of 300. Likewise, we vary the number of fake picks (n\_fake) to simulate different noise levels in the training data. Higher values of fake picks were tested for the crustal scenario to reflect its higher noise environment, ultimately selecting 400 fake picks per batch. For the subduction scenario, we find that 25 fake picks provided a good model performance. Lastly, we generate 1,000,000 synthetic training samples for each scenario (for all parameter choices, refer to Table S6), ensuring that the model is exposed to a wide variety of event locations and noise conditions. The model's performance is evaluated by monitoring the validation loss and assessing the quality of the associations in preliminary runs. To illustrate the model's convergence during training, Figure S4 shows the evolution of validation loss through the 100 epochs, with the best model chosen at epoch 61.

The input of GENIE consists of any number of phase picks over an arbitrary station network, and the model is trained to predict source space-time likelihoods and source-arrival association assignments for the set of input picks. Internally, the model uses two graphs: one for the stations, and another for the source region. For each pair of source and station nodes, the misfit between observed arrivals and the theoretical arrivals is measured, and this information is then shared and transformed between both neighboring stations and source nodes with graph convolutions to detect when and where earthquakes have occurred, and the likely association assignments to these events. Through the training process the model can learn to detect subtle signatures of moveout patterns over seismic networks for both small and large events, and learn to account for the heterogeneous station distribution, noise level, and monitoring conditions. Similar to PhaseLink, GENIE is trained using supervised learning. To train the model, a diverse suite of synthetic training data is generated, which includes sources with arbitrary positions and highly variable levels of noise and observational characteristics. Key training parameters include the maximum moveout distances of sources, the level of travel time noise, the amount of false and corrupted picks, and the maximum rate of events (Table S7). Additionally, users must set the target source region, velocity model, and choose kernel sizes for the space-time Gaussian labels. Hence, while the model can handle changing station distributions between training and future applications, for applying the model to entirely new regions it is helpful to retrain the model so that the chosen kernel sizes, velocity model, and spatial extent of the source graphs are all well calibrated to the study region of interest. The number of epochs, learning rate, and batch size can also be varied, however these are typically set to nominal values. 

\noindent\textbf{Text S3: Testing different station densities} 

\begin{figure}
    \centering
    \includegraphics[scale=0.25]{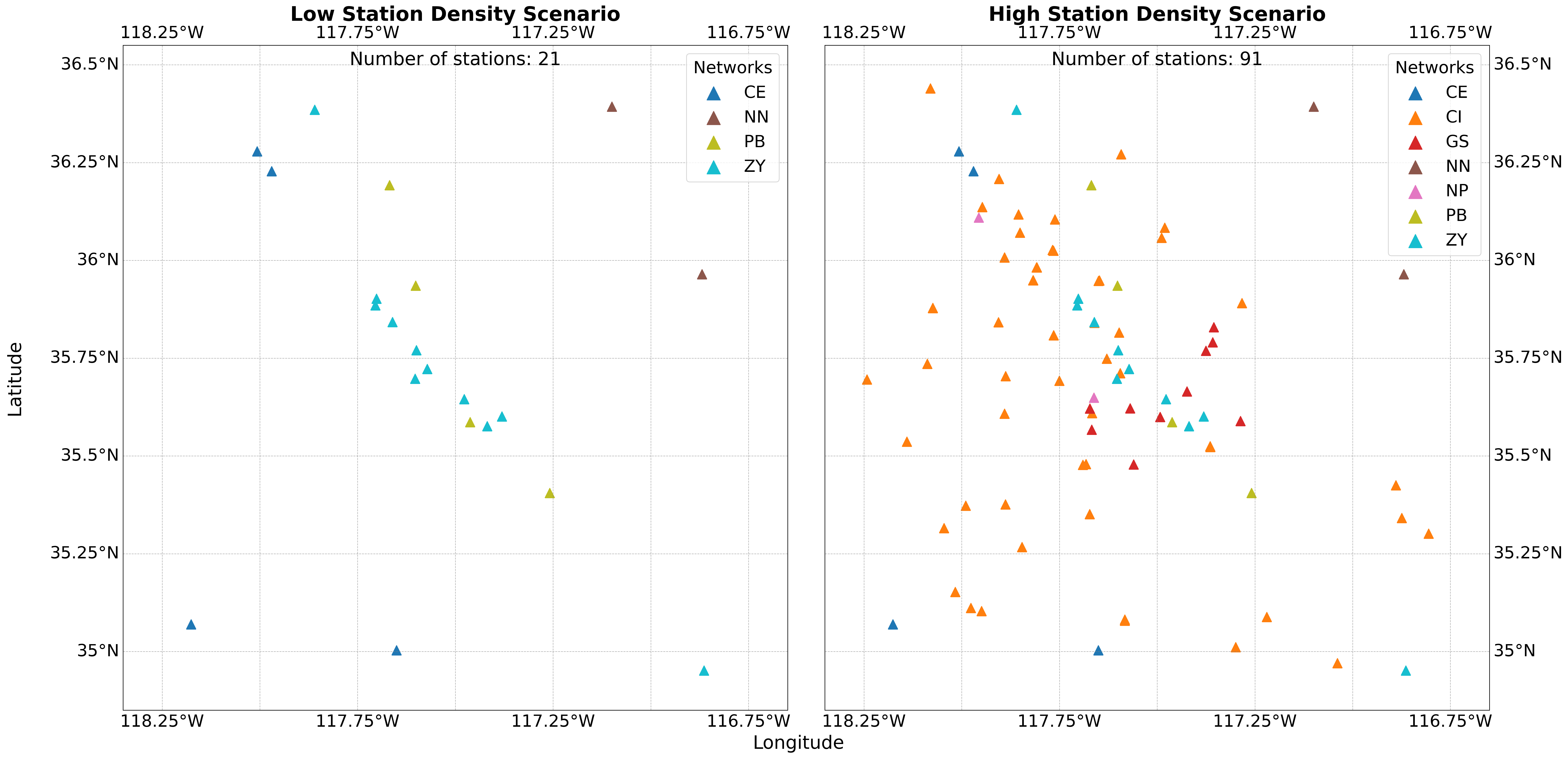}
    \caption{Station density configurations within the 1.5$^\circ$ x 1.5$^\circ$ area of the crustal scenario. \textbf{Left:} Low station density configuration (21 stations). \textbf{Right:} High station density configuration (91 stations). The seismic networks (CE, PB, ZY, NN, CI, GS, NP) are indicated in the legend by colors.}
    \label{fig:station_density_maps}
\end{figure}

To test the effect of different seismic network densities on associator performance, we conducted an additional test by modifying the California-based crustal scenario. While a higher station density can enable more accurate event detection, having more closely-spaced stations also increases the possibility of cross-associating phases to the wrong event in the case of dense seismicity such as aftershock sequences. We created two distinct station configurations derived from the crustal scenario within the same geographic area of 1.5$^\circ$×1.5$^\circ$ (Figure \ref{fig:station_density_maps}), using real-world seismic networks from the Southern California Seismic Network (SCSN). The low-density configuration comprises a total of 21 stations, the high-density configuration has 91 stations.

\begin{figure}
    \centering
    \includegraphics[scale=0.39]{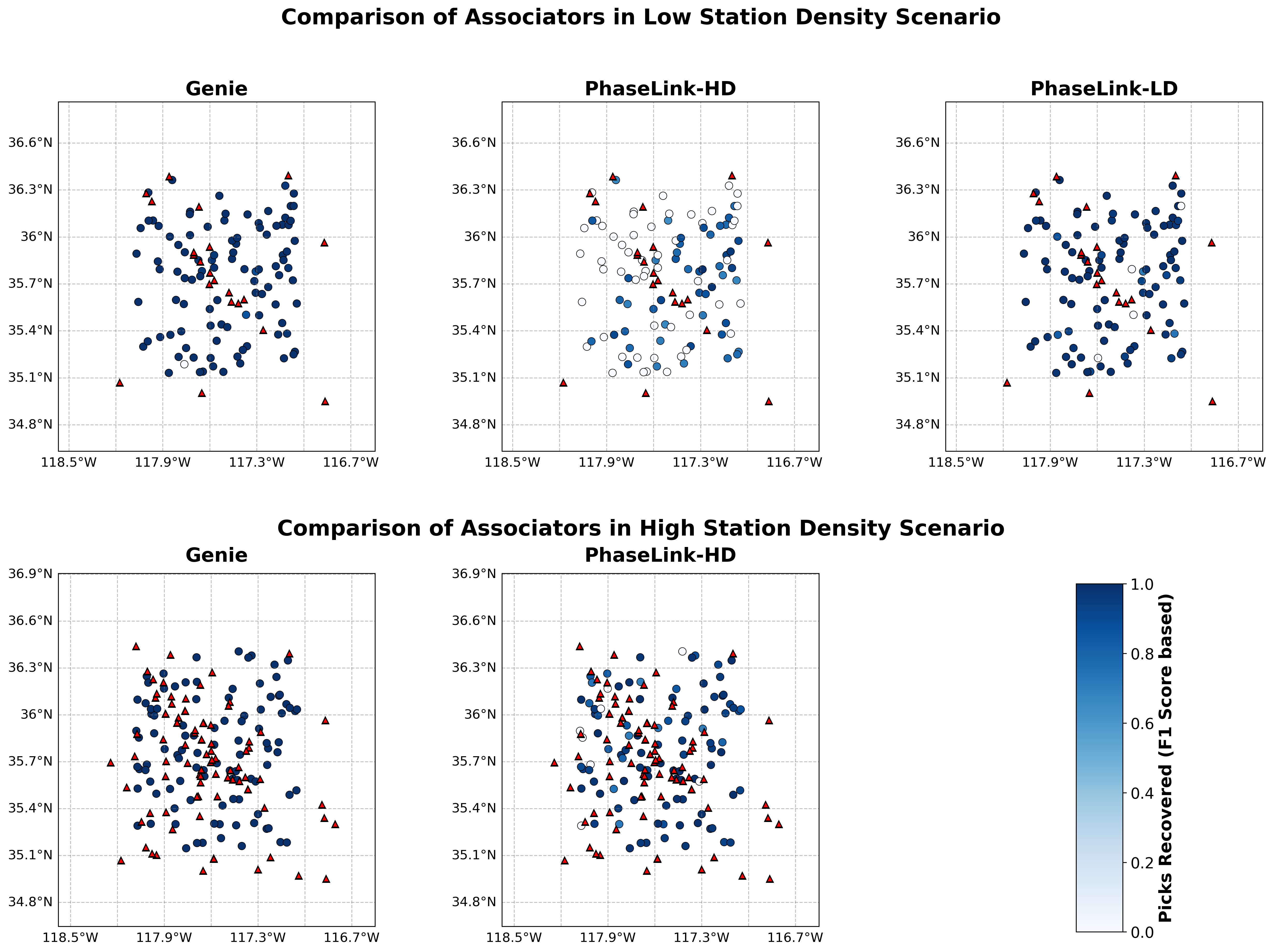}
    \caption{Comparison of associators for low (top) and high station density (bottom) scenarios. Triangles indicate station locations and circles represent events, colored by the achieved F1 score on pick level by each associator. The increase in station density (bottom row) generally improves event association and pick recovery, as shown by the more densely populated and darker-colored events in those subplots. PhaseLink-HD and PhaseLink-LD refer to the PhaseLink associator that was trained with the high-density and low-density scenario, respectively. GENIE was only trained on the high-density scenario here.}
    \label{fig:station_density_genie_pl}
\end{figure}

When repeating the different runs from the crustal scenario (see Section 4) with the modified station sets, we find that the precision of most associators decreases significantly, which is mainly due to our choice of the same association threshold (10 picks) for all runs. What we find is an inherent trade-off between event detection sensitivity and precision. In high-density networks, a low association threshold enhances sensitivity to smaller events but increases the risk of false associations due to random noise picks. Conversely, increasing the threshold improves precision by filtering out false associations, but will reduce sensitivity. Notably, GENIE is less affected by this issue. It consistently maintains high precision and recall across both scenarios without the need to adjust the association threshold or other parameters, and even in the high-density crustal case worked well with 10 required picks while maintaining a low rate of false positives. This independence of parameter optimization appears to be an important advantage of neural network based methods. While GENIE could be applied in all three cases (high density, low density, original) with the same training, we found that PhaseLink needs to be re-trained in order to perform well across different station densities (see Figure \ref{fig:station_density_genie_pl}). 

\end{article}

\clearpage

\begin{figure}
    \centering
    \includegraphics[width=\linewidth]{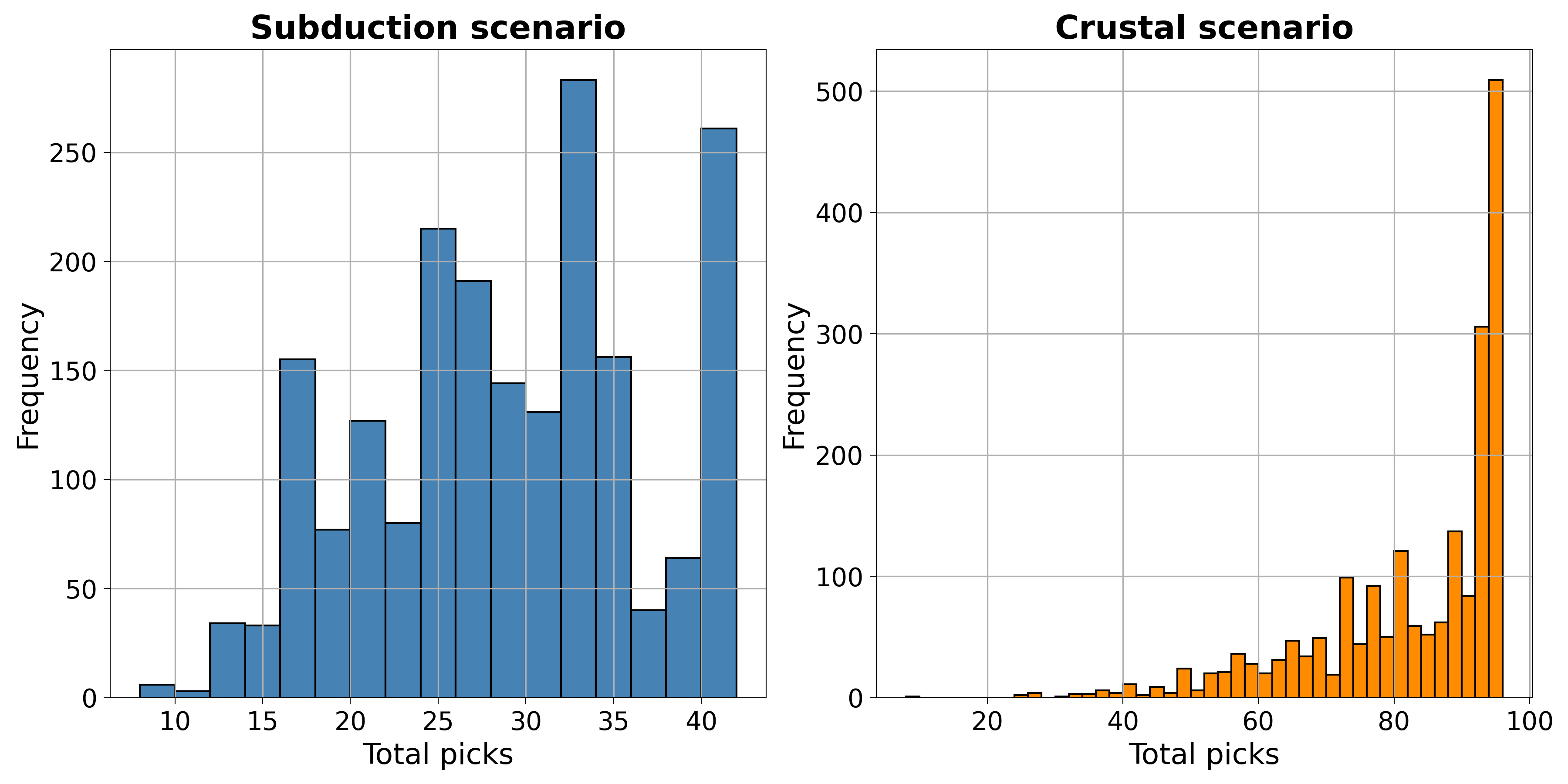}
    \caption{Distributions of event pick count for 2000 events in subduction zone (left) and crustal (right) scenario.}
    \label{fig:hist}
\end{figure}

\begin{figure}
\centering
\includegraphics[scale=.45]{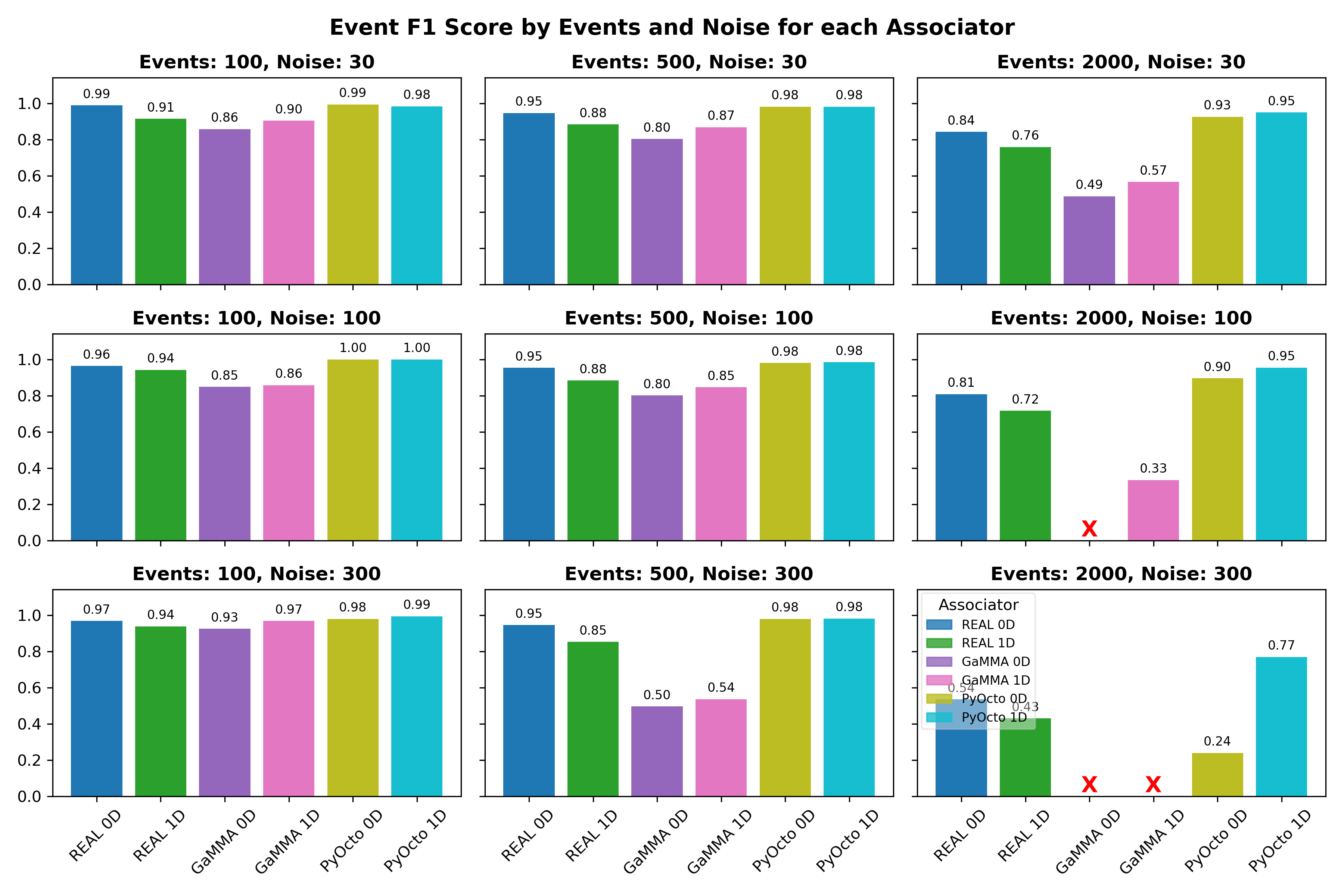}
\caption{Comparison of Event F1 Score for GaMMA, REAL, and PyOcto using 0D (homogeneous) and 1D velocity models across different noise levels and event densities in subduction zone scenario. For the processing times of the different runs, please refer to Table S8.}
\label{fig: subduction_f1score_velocity_models} 
\end{figure}

\begin{figure}
\centering
\includegraphics[scale=.45]{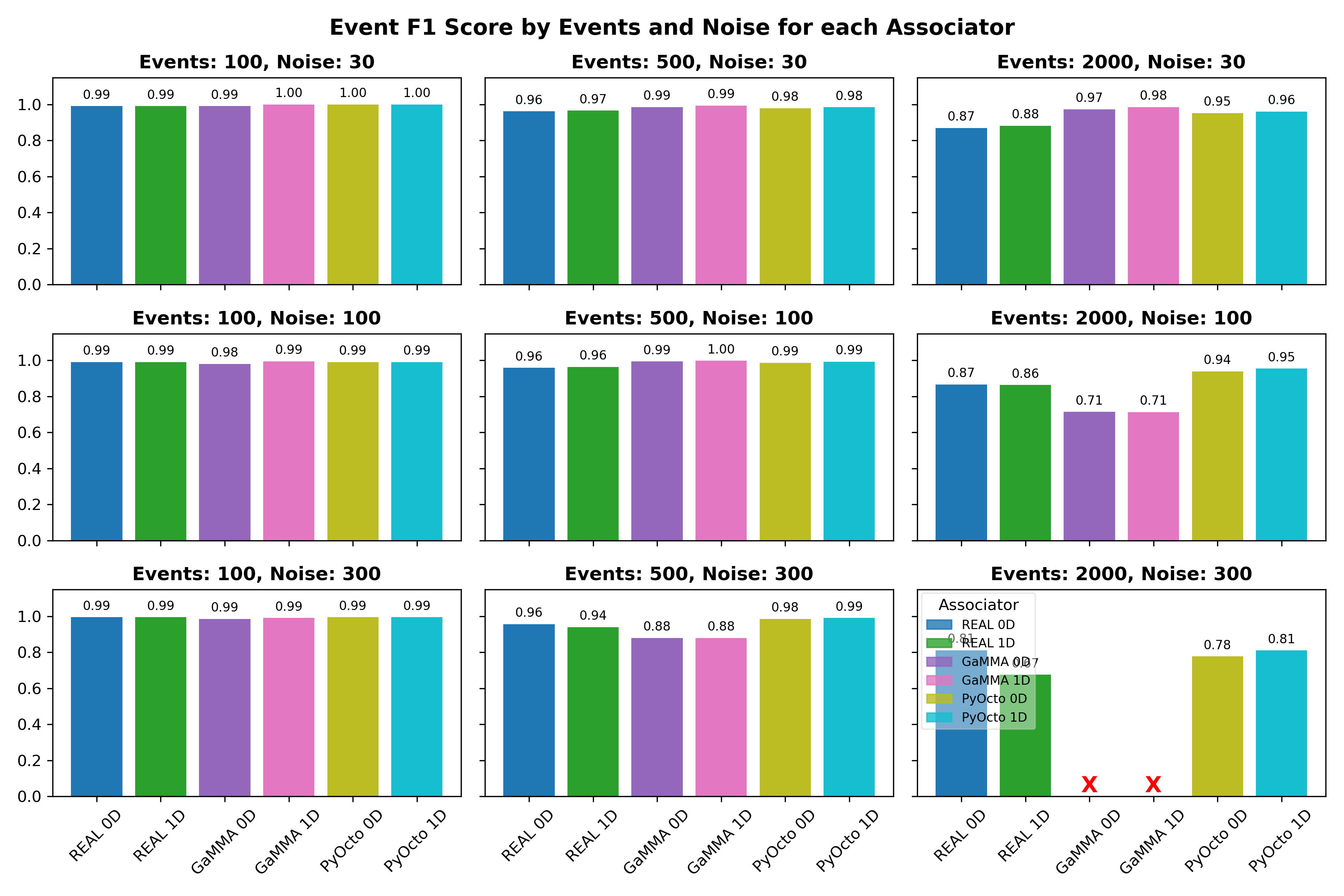}
\caption{Comparison of Event F1 Score for GaMMA, REAL, and PyOcto using 0D (homogeneous) and 1D velocity models across different noise levels and event densities in the crustal scenario. For the processing times of the different runs, please refer to Table S8.}
\label{fig: subduction_f1score_velocity_models} 
\end{figure}

\begin{figure}
\centering
\includegraphics[scale=.42]
{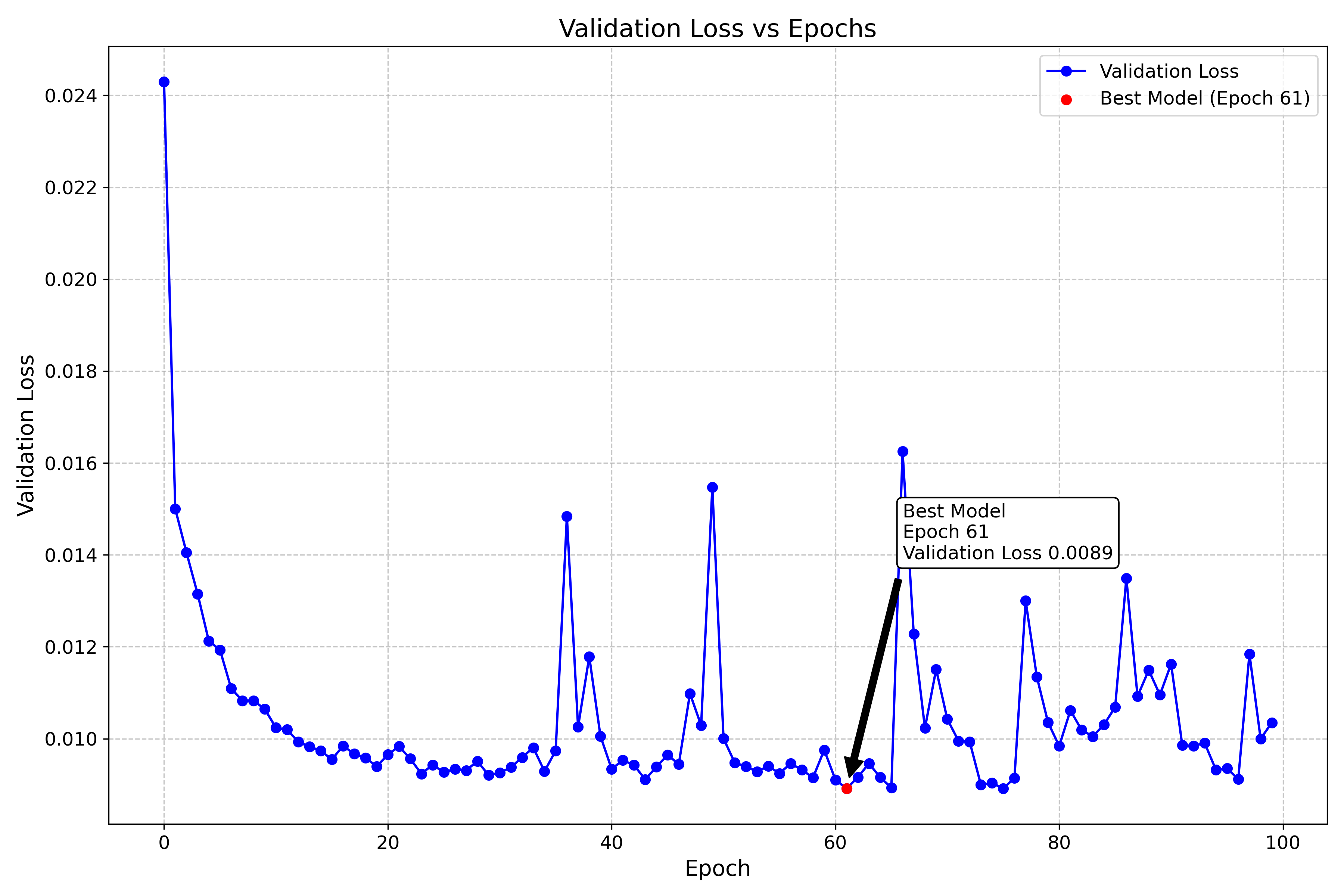}
\caption{Validation loss vs. epochs for an example training of PhaseLink. The plot shows the validation loss at each epoch during the training process. The best model, indicated by the red marker, was here achieved at epoch 61 with a validation loss of 0.0089. The plot shows a general trend of decreasing validation loss as the training progresses, demonstrating the model's improvement over time.}
\label{fig:validation_epochs_PL}
\end{figure}

%\begin{figure} \includegraphics[width=1\textwidth]{figures/oon/multipanel_oon_plot.png} \caption{Comparison of associators in an out-of-network scenario, illustrating their performance across a region. Each subplot represents a different associator, showing stations as triangles and events as circles. The color intensity of the circles represents the picks recovered by each associator, based on the F1 score, with darker shades indicating higher recovery. The final subplot (bottom-right) shows the event retrieval counts, visualizing the number of associators that successfully retrieved each event. This comparison highlights variability in associator performance in an out-of-network context, with denser and more accurate event recovery around the station network's central area.} \label{fig:multipanel_oon_plot}
%\end{figure}

\clearpage

\begin{table}

 \caption{Dataset statistics of subduction scenario.}
 \centering
 \begin{tabular}{lccccc}
\hline
 Events &  Noise (\%) &  Event picks &  False picks &  Total picks &  Picks per event \\
\hline
    100 &         30 &         2794 &          838 &         3632 &           27.940 \\
    100 &        100 &         2912 &         2912 &         5824 &           29.120 \\
    100 &        300 &         2946 &         8838 &        11784 &           29.460 \\
    500 &         30 &        14150 &         4244 &        18394 &           28.300 \\
    500 &        100 &        13864 &        13864 &        27728 &           27.728 \\
    500 &        300 &        14100 &        42300 &        56400 &           28.200 \\
   2000 &         30 &        55874 &        16762 &        72636 &           27.937 \\
   2000 &        100 &        55822 &        55822 &       111644 &           27.911 \\
   2000 &        300 &        55190 &       165570 &       220760 &           27.595 \\
\hline
\end{tabular}
\end{table}

\begin{table}
\caption{Dataset statistics of shallow seismicity scenario.}
\centering
\begin{tabular}{lccccc}
\hline
 Events &  Noise (\%) &  Event picks &  False picks &  Total picks &  Picks per event \\
\hline
    100 &         30 &         8178 &         2452 &        10630 &           81.780 \\
    100 &        100 &         8298 &         8298 &        16596 &           82.980 \\
    100 &        300 &         8124 &        24372 &        32496 &           81.240 \\
    500 &         30 &        40490 &        12146 &        52636 &           80.980 \\
    500 &        100 &        40842 &        40842 &        81684 &           81.684 \\
    500 &        300 &        40788 &       122364 &       163152 &           81.576 \\
   2000 &         30 &       163270 &        48980 &       212250 &           81.635 \\
   2000 &        100 &       163240 &       163240 &       326480 &           81.620 \\
   2000 &        300 &       161874 &       485622 &       647496 &           80.937 \\
\hline
\end{tabular}
\end{table}

\begin{table}
\centering
\caption{Parameters for GaMMA 1D in the Crustal and Subduction Scenarios. Italicized values represent parameters that were adjusted during the tuning process, while non-italicized values indicate parameters that were kept fixed.}
\label{tab:gamma1D_tuned_parameters}
\begin{tabular}{lll}
\hline
         Parameter &              Crustal Scenario &            Subduction Scenario \\
\hline
     \texttt{use amplitude} &                         False &                          False \\
             \texttt{vel p} &                           \textit{6.2} &                            \textit{7.0} \\
             \texttt{vel s} &                           \textit{3.4} &                            \textit{4.0} \\
            \texttt{method} &                          BGMM &                           BGMM \\
        \texttt{use dbscan} &                          True &                           True \\
 \texttt{oversample factor} &                             \textit{3} &                              \textit{2} \\
        \texttt{dbscan eps} &                             \textit{7} &                             \textit{20} \\
\texttt{dbscan min samples} &                            \textit{20} &                              \textit{5} \\
  \texttt{min picks per eq} &                            10 &                             10 \\
       \texttt{max sigma11} &                           \textit{2.0} &                            \textit{2.0} \\
       \texttt{max sigma22} &                           \textit{1.0} &                            \textit{1.0} \\
       \texttt{max sigma12} &                           \textit{1.0} &                            \textit{1.0} \\
              \texttt{ncpu} &                            25 &                             25 \\
  \texttt{1D velocity model} &                          True &                           True \\
             \texttt{x(km)} &                    [385, 520] &                     [250, 600] \\
             \texttt{y(km)} &                  [3860, 4040] &                   [7200, 8000] \\
             \texttt{z(km)} &                     [0, 30.0] &                       [0, 250] \\
             \texttt{local crs} &                          32611 &                            9155 \\
\hline

\end{tabular}
\end{table}

\begin{table}
\centering
\caption{Parameters for PyOcto 1D in the Crustal and Subduction Scenarios. Italicized values represent parameters that were adjusted during the tuning process, while non-italicized values indicate parameters that were kept fixed.}
\label{tab:pyocto1D_tuned_parameters}
\begin{tabular}{lll}
\hline
                                 Parameter &               Crustal Scenario &             Subduction Scenario \\
\hline
                       \texttt{spatial limits xlim} &                     [385, 520] &                  [250.0, 600.0] \\
                       \texttt{spatial limits ylim} &                   [3860, 4040] &                [7200.0, 8000.0] \\
                       \texttt{spatial limits zlim} &                        [0, 30] &                      [0, 250.0] \\
                    \texttt{association cutoff distance} &                      \textit{200} &                             \textit{350} \\
                             \texttt{time before} &                          \textit{100.0} &                           \textit{300.0} \\
                           \texttt{min node size} &                             \textit{10} &                              \textit{10} \\
                  \texttt{min node size location} &                            \textit{2.5} &                             \textit{1.5} \\
                    \texttt{pick match tolerance} &                            \textit{0.8} &                             \textit{0.8} \\
                     \texttt{min interevent time} &                            \textit{3.0} &                             \textit{3.0} \\
                        \texttt{max pick overlap} &                              \textit{4} &                               \textit{4} \\
                                 \texttt{n picks} &                             10 &                              10 \\
                               \texttt{n p picks} &                              5 &                               5 \\
                               \texttt{n s picks} &                              5 &                               5 \\
                         \texttt{n p and s picks} &                              4 &                               4 \\
                   \texttt{refinement iterations} &                              \textit{3} &                               \textit{3} \\
                            \texttt{time slicing} &                         \textit{1200.0} &                          \textit{1200.0} \\
                    \texttt{location split depth} &                              \textit{6} &                               \textit{6} \\
                   \texttt{location split return} &                              \textit{4} &                               \textit{4} \\
                       \texttt{min pick fraction} &                            \textit{0.0} &                            \textit{0.25} \\
                               \texttt{n threads} &                             25 &                              25 \\
                                \texttt{VelMod1D} &                           True &                            True \\
                  \texttt{velocity model tolerance} &                            \textit{1.0} &                             \textit{1.0} \\
                                 \texttt{local crs} &                          32611 &                            9155 \\
                     \texttt{tt table grid spacing} &                            \textit{1.0} &                             \textit{0.5} \\
                         \texttt{tt table x extent} &                            \textit{300} &                             \textit{500} \\
                         \texttt{tt table y extent} &                            \textit{300} &                             \textit{800} \\
\hline

\end{tabular}

\end{table}

%REAL parameters

\begin{table}
\centering
\caption{Parameters for REAL 0D in the Crustal and Subduction Scenarios. Italicized values represent parameters that were adjusted during the tuning process, while non-italicized values indicate parameters that were kept fixed.}
\label{tab:REAL0D_tuned_parameters}
\begin{tabular}{lll}
\hline
                                 Parameter &               Crustal Scenario &             Subduction Scenario \\
\hline
          \texttt{tt config dist} &                           \textit{4} &                           \textit{9} \\
           \texttt{tt config dep} &                          \textit{30} &                         \textit{250} \\
         \texttt{tt config ddist} &                         \textit{0.6} &                         \textit{1.0} \\
          \texttt{tt config ddep} &                           \textit{1} &                           \textit{8} \\
        \texttt{1D velocity model} &                       False &                       False \\
                \texttt{latitude} &                        35.0 &                   -21.18148 \\
         \texttt{R rx} &                           \textit{1} &                           \textit{1} \\
         \texttt{R rh} &                          \textit{30} &                         \textit{250} \\
         \texttt{R tdx} &                         \textit{0.1} &                         \textit{0.1} \\
         \texttt{R tdh} &                           \textit{8} &                          \textit{10} \\
         \texttt{R tint} &                         \textit{0.1} &                         \textit{0.1} \\
         \texttt{V vp0} &                         \textit{6.2} &                         \textit{6.8} \\
         \texttt{V vs0} &                         \textit{3.3} &                         \textit{4.0} \\
         \texttt{V s\_vp0} &                         \textit{5.4} &                         \textit{5.3} \\
         \texttt{V s\_vs0} &                         \textit{3.3} &                         \textit{3.1} \\
         \texttt{V ielev} &                           1 &                           1 \\
         \texttt{S np0} &                           4 &                           4 \\
         \texttt{S ns0} &                           \textit{4} &                           \textit{4} \\
         \texttt{S nps0} &                          10 &                          10 \\
         \texttt{S npsboth0} &                           4 &                           4 \\
         \texttt{S std0} &                         \textit{2.0} &                         \textit{2.0} \\
         \texttt{S dtps} &                           \textit{0} &                           \textit{0} \\
         \texttt{S nrt} &                           \textit{2} &                           \textit{2} \\
         \texttt{S drt} &                           \textit{2} &                           \textit{2} \\
         \texttt{S nxd} &                         \textit{0.4} &                         \textit{0.4} \\
         \texttt{S rsel} &                           \textit{6} &                           \textit{6} \\
\hline

\end{tabular}

\end{table}

\begin{table}
\centering
\caption{Parameters for PhaseLink in the Crustal and Subduction Scenarios. Italicized values represent parameters that were adjusted during the tuning process, while non-italicized values indicate parameters that were kept fixed.}
\label{tab:phaselink_tuned_parameters}
\begin{tabular}{lll}
\hline
                   Parameter &                                   Crustal Scenario &                                Subduction Scenario \\
\hline
             \texttt{t win} &                                                \textit{250} &                                                \textit{120} \\
          \texttt{n epochs} &                                                \textit{100} &                                                \textit{100} \\
       \texttt{n max picks} &                                                \textit{300} &                                                \textit{120} \\
        \texttt{batch size} &                                                \textit{64} &                                                \textit{64} \\
        \texttt{n min nucl} &                                                \textit{12} &                                                 \textit{6} \\
       \texttt{n min merge} &                                                \textit{2} &                                                 \textit{2} \\
         \texttt{n min det} &                                                 10 &                                                 10 \\
       \texttt{avg eve sep} &                                                \textit{20} &                                                \textit{12} \\
            \texttt{pr min} &                                                \textit{0.5} &                                                \textit{0.5} \\
      \texttt{n train samp} &                                            1000000 &                                            1000000 \\
      \texttt{n min radius} &                                                \textit{8} &                                                 \textit{8} \\
            \texttt{n fake} &                                                \textit{400} &                                                \textit{25} \\
   \texttt{max event depth} &                                                 30 &                                                250 \\
     \texttt{min hypo dist} &                                               \textit{50.0} &                                              \textit{80.0} \\
     \texttt{max hypo dist} &                                               \textit{80.0} &                                             \textit{450.0} \\
    \texttt{max pick error} &                                                1.0 &                                                1.0 \\
     \texttt{min pick dist} &                                                0.5 &                                                0.5 \\
           \texttt{min sep} &                                                \textit{0.6} &                                               \textit{0.6} \\
           \texttt{lat min} &                                              34.87 &                                              -25.0 \\
           \texttt{lat max} &                                               36.5 &                                              -17.0 \\
           \texttt{lon min} &                                             118.28 &                                              -71.0 \\
           \texttt{lon max} &                                              116.7 &                                              -66.0 \\
\hline

\end{tabular}

\end{table}

\begin{table}
\centering
\caption{Chosen parameters for GENIE 1D in the Crustal and Subduction Scenarios. Italicized entries were partially tuned with 2$-$3 rounds of re-training, all other values were chosen to reflect the characteristic spatial scale and expected event rates of either scenario. \vspace{0.5 cm}}
\label{tab:genie_tuned_parameters}
\begin{tabular}{lll}
\hline
         Parameter &              Crustal Scenario &            Subduction Scenario \\
\hline
             \texttt{k\_sta\_edges} & 8 & 8 \\
             \texttt{k\_spc\_edges} & 15 & 15 \\
             \texttt{n\_of\_spatial\_nodes} & 1000 & 1500 \\
             \texttt{kernel\_sig\_t} & 3.0 & 8.0 \\
             \texttt{src\_t\_kernel} & 3.0 & 8.0 \\
             \texttt{src\_x\_kernel} & 15000 & 45000 \\
             \texttt{spc\_random} & 15000 & 10000 \\
             \texttt{spc\_thresh\_ran} & 15000 & 135000 \\
             \texttt{sig\_t} & 0.01 & \textit{0.0075} \\
             \texttt{min\_sta\_arrival} & 12 & 8 \\
             \texttt{thresh\_noise\_max} & 2.25 & \textit{0.75} \\
             \texttt{total\_bias} & 0.01 & \textit{0.0075} \\
             \texttt{dist\_range} & [5000, 250000] & [100000, 1490000] \\ 
             \texttt{max\_rate\_events} & \textit{225} & \textit{280} \\ 
             \texttt{max\_false\_events} & \textit{650} & \textit{650} \\ 
             \texttt{miss\_pick\_fraction} & [0.05, 0.35] & [0.05, 0.35] \\ 
             \texttt{thresh} & 0.6 & 0.6  \\ 
             \texttt{thresh\_assoc} & 0.6 & 0.6 \\ 
             \texttt{tc\_win} & 2.5 & 8.0  \\
             \texttt{sp\_win} & 12500 & 45000 \\
             \texttt{d\_win} & 0.2 & 0.45 \\
             \texttt{d\_win\_depth} & 20000 & 50000 \\
             Latitude &                    [34.82$^{\circ}$, 36.55$^{\circ}$] N &                     [-25.0$^{\circ}$, -17.0$^{\circ}$] N \\
             Longitude &                  [-118.33$^{\circ}$, -116.65$^{\circ}$] E &                   [-71.0$^{\circ}$, -66.0$^{\circ}$] E \\
             Depths &                     [-35, 5] km &                       [-250, 5] km \\
\hline
\end{tabular}
\end{table}

\begin{table}
\centering
\caption{Performance comparison using homogeneous (0D) and 1D velocity models}
\label{tab:Versions_battle_chile_associator_stats} \begin{tabular}{rrlllll}
\hline
 Events &  Noise & Associator & Event\_Precision & Event\_Recall & Event\_F1\_Score & Runtime (s) \\
\hline
    100 &     30 &   GaMMA 0D &            0.84 &         0.88 &           0.86 &        2.91 \\
    100 &     30 &   GaMMA 1D &            0.91 &         0.90 &           0.90 &        5.03 \\
    100 &     30 &  PyOcto 0D &            1.00 &         0.99 &           0.99 &        1.29 \\
    100 &     30 &  PyOcto 1D &            1.00 &         0.97 &           0.98 &        3.16 \\
    100 &     30 &    REAL 0D &            1.00 &         0.98 &           0.99 &       60.98 \\
    100 &     30 &    REAL 1D &            0.98 &         0.86 &           0.91 &       72.06 \\
    100 &    100 &   GaMMA 0D &            0.86 &         0.84 &           0.85 &        2.93 \\
    100 &    100 &   GaMMA 1D &            0.88 &         0.84 &           0.86 &        5.28 \\
    100 &    100 &  PyOcto 0D &            1.00 &         1.00 &           1.00 &        0.99 \\
    100 &    100 &  PyOcto 1D &            1.00 &         1.00 &           1.00 &        1.42 \\
    100 &    100 &    REAL 0D &            1.00 &         0.93 &           0.96 &      169.51 \\
    100 &    100 &    REAL 1D &            0.99 &         0.90 &           0.94 &      209.14 \\
    100 &    300 &   GaMMA 0D &            0.92 &         0.93 &           0.93 &        6.10 \\
    100 &    300 &   GaMMA 1D &            0.99 &         0.95 &           0.97 &        6.71 \\
    100 &    300 &  PyOcto 0D &            0.98 &         0.98 &           0.98 &       42.68 \\
    100 &    300 &  PyOcto 1D &            1.00 &         0.99 &           0.99 &        1.19 \\
    100 &    300 &    REAL 0D &            1.00 &         0.94 &           0.97 &      453.73 \\
    100 &    300 &    REAL 1D &            0.96 &         0.92 &           0.94 &      626.48 \\
    500 &     30 &   GaMMA 0D &            0.83 &         0.78 &           0.80 &       41.84 \\
    500 &     30 &   GaMMA 1D &            0.92 &         0.83 &           0.87 &       17.73 \\
    500 &     30 &  PyOcto 0D &            0.99 &         0.97 &           0.98 &       91.96 \\
    500 &     30 &  PyOcto 1D &            1.00 &         0.97 &           0.98 &        6.74 \\
    500 &     30 &    REAL 0D &            1.00 &         0.90 &           0.95 &      306.16 \\
    500 &     30 &    REAL 1D &            0.96 &         0.82 &           0.88 &      424.37 \\
    500 &    100 &   GaMMA 0D &            0.86 &         0.75 &           0.80 &       57.15 \\
    500 &    100 &   GaMMA 1D &            0.90 &         0.80 &           0.85 &       25.74 \\
    500 &    100 &  PyOcto 0D &            0.99 &         0.97 &           0.98 &       45.06 \\
    500 &    100 &  PyOcto 1D &            1.00 &         0.97 &           0.98 &       12.29 \\
    500 &    100 &    REAL 0D &            0.99 &         0.92 &           0.95 &      754.36 \\
    500 &    100 &    REAL 1D &            0.94 &         0.83 &           0.88 &      931.85 \\
    500 &    300 &   GaMMA 0D &            0.53 &         0.47 &           0.50 &      584.25 \\
    500 &    300 &   GaMMA 1D &            0.56 &         0.51 &           0.54 &      145.93 \\
    500 &    300 &  PyOcto 0D &            0.99 &         0.97 &           0.98 &       85.44 \\
    500 &    300 &  PyOcto 1D &            1.00 &         0.97 &           0.98 &       33.46 \\
    500 &    300 &    REAL 0D &            0.98 &         0.91 &           0.95 &     2032.63 \\
    500 &    300 &    REAL 1D &            0.90 &         0.81 &           0.85 &     2226.60 \\
   2000 &     30 &   GaMMA 0D &            0.58 &         0.42 &           0.49 &     1789.12 \\
   2000 &     30 &   GaMMA 1D &            0.68 &         0.48 &           0.57 &      210.15 \\
   2000 &     30 &  PyOcto 0D &            0.96 &         0.89 &           0.93 &      201.67 \\
   2000 &     30 &  PyOcto 1D &            1.00 &         0.91 &           0.95 &       96.97 \\
   2000 &     30 &    REAL 0D &            0.99 &         0.73 &           0.84 &     1300.54 \\
   2000 &     30 &    REAL 1D &            0.93 &         0.64 &           0.76 &     1427.40 \\
   2000 &    100 &   GaMMA 1D &            0.32 &         0.35 &           0.33 &     3598.16 \\
   2000 &    100 &  PyOcto 0D &            0.91 &         0.88 &           0.90 &      417.81 \\
   2000 &    100 &  PyOcto 1D &            0.99 &         0.92 &           0.95 &      131.74 \\
   2000 &    100 &    REAL 0D &            0.95 &         0.70 &           0.81 &     2848.73 \\
   2000 &    100 &    REAL 1D &            0.87 &         0.61 &           0.72 &     2830.46 \\
   2000 &    300 &  PyOcto 0D &            0.16 &         0.44 &           0.24 &     1023.65 \\
   2000 &    300 &  PyOcto 1D &            0.76 &         0.78 &           0.77 &     1790.04 \\
   2000 &    300 &    REAL 0D &            0.54 &         0.54 &           0.54 &     6706.46 \\
   2000 &    300 &    REAL 1D &            0.43 &         0.43 &           0.43 &     5274.96 \\
\hline
\end{tabular}

\end{table}

\begin{table}
\centering
\caption{Subduction Zone Scenario: event-level evaluation of seismic phase associators}
\begin{tabular}{rrlllll}
\hline
 Events &  Noise & Associator & Event\_Precision & Event\_Recall & Event\_F1\_Score & Runtime (s) \\
\hline
    100 &     30 &   GaMMA 1D &            0.91 &         0.90 &           0.90 &        5.03 \\
    100 &     30 &      Genie &            1.00 &         0.99 &           0.99 &      294.52 \\
    100 &     30 &  PhaseLink &            0.93 &         0.82 &           0.87 &        4.99 \\
    100 &     30 &  PyOcto 1D &            1.00 &         0.97 &           0.98 &        3.16 \\
    100 &     30 &    REAL 0D &            1.00 &         0.98 &           0.99 &       60.98 \\
    100 &    100 &   GaMMA 1D &            0.88 &         0.84 &           0.86 &        5.28 \\
    100 &    100 &      Genie &            0.99 &         0.99 &           0.99 &      313.33 \\
    100 &    100 &  PhaseLink &            0.91 &         0.85 &           0.88 &        3.79 \\
    100 &    100 &  PyOcto 1D &            1.00 &         1.00 &           1.00 &        1.42 \\
    100 &    100 &    REAL 0D &            1.00 &         0.93 &           0.96 &      169.51 \\
    100 &    300 &   GaMMA 1D &            0.99 &         0.95 &           0.97 &        6.71 \\
    100 &    300 &      Genie &            1.00 &         0.99 &           0.99 &      325.05 \\
    100 &    300 &  PhaseLink &            0.88 &         0.81 &           0.84 &        4.15 \\
    100 &    300 &  PyOcto 1D &            1.00 &         0.99 &           0.99 &        1.19 \\
    100 &    300 &    REAL 0D &            1.00 &         0.94 &           0.97 &      453.73 \\
    500 &     30 &   GaMMA 1D &            0.92 &         0.83 &           0.87 &       17.73 \\
    500 &     30 &      Genie &            0.99 &         0.97 &           0.98 &      530.98 \\
    500 &     30 &  PhaseLink &            0.85 &         0.60 &           0.71 &        5.11 \\
    500 &     30 &  PyOcto 1D &            1.00 &         0.97 &           0.98 &        6.74 \\
    500 &     30 &    REAL 0D &            1.00 &         0.90 &           0.95 &      306.16 \\
    500 &    100 &   GaMMA 1D &            0.90 &         0.80 &           0.85 &       25.74 \\
    500 &    100 &      Genie &            1.00 &         0.99 &           0.99 &      548.01 \\
    500 &    100 &  PhaseLink &            0.84 &         0.67 &           0.75 &        5.99 \\
    500 &    100 &  PyOcto 1D &            1.00 &         0.97 &           0.98 &       12.29 \\
    500 &    100 &    REAL 0D &            0.99 &         0.92 &           0.95 &      754.36 \\
    500 &    300 &   GaMMA 1D &            0.56 &         0.51 &           0.54 &      145.93 \\
    500 &    300 &      Genie &            0.97 &         0.99 &           0.98 &      593.40 \\
    500 &    300 &  PhaseLink &            0.15 &         0.25 &           0.19 &       11.36 \\
    500 &    300 &  PyOcto 1D &            1.00 &         0.97 &           0.98 &       33.46 \\
    500 &    300 &    REAL 0D &            0.98 &         0.91 &           0.95 &     2032.63 \\
   2000 &     30 &   GaMMA 1D &            0.68 &         0.48 &           0.57 &      210.15 \\
   2000 &     30 &      Genie &            0.98 &         0.92 &           0.95 &     1211.63 \\
   2000 &     30 &  PhaseLink &            0.54 &         0.19 &           0.28 &       14.08 \\
   2000 &     30 &  PyOcto 1D &            1.00 &         0.91 &           0.95 &       96.97 \\
   2000 &     30 &    REAL 0D &            0.99 &         0.73 &           0.84 &     1300.54 \\
   2000 &    100 &   GaMMA 1D &            0.32 &         0.35 &           0.33 &     3598.16 \\
   2000 &    100 &      Genie &            0.95 &         0.89 &           0.92 &     1256.66 \\
   2000 &    100 &  PhaseLink &            0.07 &         0.01 &           0.02 &       16.87 \\
   2000 &    100 &  PyOcto 1D &            0.99 &         0.92 &           0.95 &      131.74 \\
   2000 &    100 &    REAL 0D &            0.95 &         0.70 &           0.81 &     2848.73 \\
   2000 &    300 &      Genie &            0.80 &         0.78 &           0.79 &     1474.35 \\
   2000 &    300 &  PhaseLink &            0.00 &         0.00 &           0.00 &       25.66 \\
   2000 &    300 &  PyOcto 1D &            0.76 &         0.78 &           0.77 &     1790.04 \\
   2000 &    300 &    REAL 0D &            0.54 &         0.54 &           0.54 &     6706.46 \\
\hline
\end{tabular}

\label{tab:sz_stats_event_lvl}
\end{table}

\begin{table}
\centering
\caption{Crustal Scenario: event-level evaluation of seismic phase associators}
\begin{tabular}{rrlllll}
\hline
 Events &  Noise & Associator & Event\_Precision & Event\_Recall & Event\_F1\_Score & Runtime (s) \\
\hline
    100 &     30 &   GaMMA 1D &            1.00 &         1.00 &           1.00 &        9.89 \\
    100 &     30 &      Genie &            1.00 &         1.00 &           1.00 &      319.66 \\
    100 &     30 &  PhaseLink &            0.93 &         0.89 &           0.91 &        5.82 \\
    100 &     30 &  PyOcto 1D &            1.00 &         1.00 &           1.00 &        2.71 \\
    100 &     30 &    REAL 0D &            1.00 &         0.98 &           0.99 &        9.46 \\
    100 &    100 &   GaMMA 1D &            1.00 &         0.99 &           0.99 &       13.22 \\
    100 &    100 &      Genie &            0.99 &         0.99 &           0.99 &      333.67 \\
    100 &    100 &  PhaseLink &            0.99 &         0.96 &           0.97 &        5.10 \\
    100 &    100 &  PyOcto 1D &            1.00 &         0.98 &           0.99 &        3.42 \\
    100 &    100 &    REAL 0D &            1.00 &         0.98 &           0.99 &       31.51 \\
    100 &    300 &   GaMMA 1D &            0.99 &         0.99 &           0.99 &       18.56 \\
    100 &    300 &      Genie &            0.99 &         0.99 &           0.99 &      345.50 \\
    100 &    300 &  PhaseLink &            1.00 &         0.98 &           0.99 &        6.45 \\
    100 &    300 &  PyOcto 1D &            1.00 &         0.99 &           0.99 &        5.84 \\
    100 &    300 &    REAL 0D &            1.00 &         0.99 &           0.99 &       75.32 \\
    500 &     30 &   GaMMA 1D &            0.99 &         0.99 &           0.99 &       46.43 \\
    500 &     30 &      Genie &            0.99 &         0.98 &           0.99 &      582.32 \\
    500 &     30 &  PhaseLink &            0.98 &         0.88 &           0.92 &       10.78 \\
    500 &     30 &  PyOcto 1D &            1.00 &         0.97 &           0.98 &       19.02 \\
    500 &     30 &    REAL 0D &            1.00 &         0.93 &           0.96 &       48.30 \\
    500 &    100 &   GaMMA 1D &            1.00 &         1.00 &           1.00 &       67.81 \\
    500 &    100 &      Genie &            1.00 &         1.00 &           1.00 &      663.37 \\
    500 &    100 &  PhaseLink &            0.99 &         0.88 &           0.93 &       13.31 \\
    500 &    100 &  PyOcto 1D &            1.00 &         0.98 &           0.99 &       10.95 \\
    500 &    100 &    REAL 0D &            0.99 &         0.93 &           0.96 &      145.26 \\
    500 &    300 &   GaMMA 1D &            0.79 &         0.99 &           0.88 &      150.07 \\
    500 &    300 &      Genie &            0.99 &         0.98 &           0.99 &      635.51 \\
    500 &    300 &  PhaseLink &            0.96 &         0.88 &           0.92 &       20.48 \\
    500 &    300 &  PyOcto 1D &            1.00 &         0.98 &           0.99 &       21.26 \\
    500 &    300 &    REAL 0D &            0.98 &         0.93 &           0.96 &      419.16 \\
   2000 &     30 &   GaMMA 1D &            0.98 &         0.99 &           0.98 &      312.27 \\
   2000 &     30 &      Genie &            0.98 &         0.94 &           0.96 &     1445.19 \\
   2000 &     30 &  PhaseLink &            0.93 &         0.61 &           0.74 &       48.07 \\
   2000 &     30 &  PyOcto 1D &            0.99 &         0.93 &           0.96 &       51.07 \\
   2000 &     30 &    REAL 0D &            0.99 &         0.78 &           0.87 &      239.90 \\
   2000 &    100 &   GaMMA 1D &            0.57 &         0.96 &           0.71 &      900.20 \\
   2000 &    100 &      Genie &            0.98 &         0.94 &           0.96 &     1463.43 \\
   2000 &    100 &  PhaseLink &            0.84 &         0.56 &           0.67 &       61.19 \\
   2000 &    100 &  PyOcto 1D &            0.99 &         0.92 &           0.95 &      136.43 \\
   2000 &    100 &    REAL 0D &            0.96 &         0.79 &           0.87 &      551.88 \\
   2000 &    300 &      Genie &            0.95 &         0.86 &           0.90 &     1769.01 \\
   2000 &    300 &  PhaseLink &            0.10 &         0.14 &           0.12 &      143.76 \\
   2000 &    300 &  PyOcto 1D &            0.73 &         0.91 &           0.81 &     1170.82 \\
   2000 &    300 &    REAL 0D &            0.91 &         0.73 &           0.81 &     1985.60 \\
\hline
\end{tabular}

\label{tab:crustal_stats_event_lvl}
\end{table}

\begin{table}
\centering
\caption{Subduction zone scenario: evaluation of seismic phase associators at pick level across different event and noise levels. GT: Ground Truth Picks, Pred: Predicted Picks, CA: Commonly Associated Picks, Missed: Missed Picks, FP: False Picks, WAP: Wrongly Associated Picks.}
\begin{tabular}{lrrlllllllll}
\hline
Associator &  Ev. &  Noise &    GT &  Pred &    CA & Missed &    FP &  WAP & Precision & Recall &   F1 \\
\hline
  GaMMA 1D &  100 &     30 & 28.36 & 26.17 & 26.11 &   2.24 &  0.01 & 0.04 &      1.00 &   0.92 & 0.96 \\
     Genie &  100 &     30 & 27.82 & 27.76 & 27.65 &   0.17 &  0.05 & 0.06 &      1.00 &   0.99 & 1.00 \\
 PhaseLink &  100 &     30 & 28.44 & 24.56 & 23.68 &   4.76 &  0.22 & 0.66 &      0.97 &   0.84 & 0.89 \\
 PyOcto 1D &  100 &     30 & 28.16 & 27.27 & 27.24 &   0.93 &  0.01 & 0.02 &      1.00 &   0.97 & 0.98 \\
   REAL 0D &  100 &     30 & 27.78 & 26.78 & 26.54 &   1.23 &  0.07 & 0.16 &      0.99 &   0.96 & 0.98 \\
  GaMMA 1D &  100 &    100 & 30.14 & 26.77 & 26.75 &   3.39 &  0.02 & 0.00 &      1.00 &   0.90 & 0.94 \\
     Genie &  100 &    100 & 29.01 & 28.91 & 28.79 &   0.22 &  0.10 & 0.02 &      1.00 &   0.99 & 0.99 \\
 PhaseLink &  100 &    100 & 29.79 & 25.75 & 24.58 &   5.21 &  0.67 & 0.51 &      0.96 &   0.83 & 0.89 \\
 PyOcto 1D &  100 &    100 & 29.12 & 28.25 & 28.21 &   0.91 &  0.02 & 0.02 &      1.00 &   0.97 & 0.98 \\
   REAL 0D &  100 &    100 & 29.03 & 28.09 & 27.86 &   1.17 &  0.20 & 0.02 &      0.99 &   0.97 & 0.98 \\
  GaMMA 1D &  100 &    300 & 29.75 & 27.53 & 27.15 &   2.60 &  0.16 & 0.22 &      0.99 &   0.92 & 0.95 \\
     Genie &  100 &    300 & 29.39 & 29.28 & 28.99 &   0.40 &  0.22 & 0.07 &      0.99 &   0.99 & 0.99 \\
 PhaseLink &  100 &    300 & 30.17 & 27.42 & 24.23 &   5.94 &  2.57 & 0.62 &      0.89 &   0.81 & 0.84 \\
 PyOcto 1D &  100 &    300 & 29.66 & 28.92 & 28.78 &   0.88 &  0.07 & 0.07 &      1.00 &   0.97 & 0.98 \\
   REAL 0D &  100 &    300 & 29.02 & 28.16 & 27.40 &   1.62 &  0.46 & 0.30 &      0.98 &   0.95 & 0.96 \\
  GaMMA 1D &  500 &     30 & 29.31 & 26.94 & 26.76 &   2.55 &  0.06 & 0.12 &      0.99 &   0.92 & 0.95 \\
     Genie &  500 &     30 & 28.45 & 28.06 & 27.70 &   0.75 &  0.17 & 0.19 &      0.99 &   0.97 & 0.98 \\
 PhaseLink &  500 &     30 & 29.35 & 28.40 & 24.07 &   5.28 &  1.02 & 3.30 &      0.88 &   0.83 & 0.84 \\
 PyOcto 1D &  500 &     30 & 28.64 & 27.62 & 27.43 &   1.20 &  0.07 & 0.11 &      0.99 &   0.96 & 0.97 \\
   REAL 0D &  500 &     30 & 28.46 & 27.38 & 26.60 &   1.86 &  0.23 & 0.55 &      0.97 &   0.94 & 0.96 \\
  GaMMA 1D &  500 &    100 & 28.86 & 27.12 & 26.77 &   2.09 &  0.23 & 0.11 &      0.99 &   0.93 & 0.95 \\
     Genie &  500 &    100 & 27.85 & 27.33 & 26.81 &   1.04 &  0.39 & 0.12 &      0.98 &   0.96 & 0.97 \\
 PhaseLink &  500 &    100 & 28.89 & 28.81 & 23.08 &   5.81 &  3.96 & 1.76 &      0.81 &   0.80 & 0.80 \\
 PyOcto 1D &  500 &    100 & 28.04 & 27.02 & 26.75 &   1.29 &  0.18 & 0.08 &      0.99 &   0.96 & 0.97 \\
   REAL 0D &  500 &    100 & 28.08 & 27.17 & 25.97 &   2.11 &  0.78 & 0.43 &      0.96 &   0.93 & 0.94 \\
  GaMMA 1D &  500 &    300 & 31.60 & 30.83 & 29.75 &   1.85 &  0.87 & 0.20 &      0.96 &   0.94 & 0.95 \\
     Genie &  500 &    300 & 28.26 & 28.10 & 26.75 &   1.51 &  1.24 & 0.11 &      0.95 &   0.95 & 0.95 \\
 PhaseLink &  500 &    300 & 30.51 & 36.22 & 21.79 &   8.72 & 14.04 & 0.40 &      0.61 &   0.72 & 0.66 \\
 PyOcto 1D &  500 &    300 & 28.45 & 27.59 & 26.92 &   1.53 &  0.58 & 0.09 &      0.97 &   0.95 & 0.96 \\
   REAL 0D &  500 &    300 & 28.43 & 28.50 & 25.49 &   2.94 &  2.57 & 0.45 &      0.89 &   0.90 & 0.89 \\
  GaMMA 1D & 2000 &     30 & 30.92 & 29.94 & 28.88 &   2.04 &  0.33 & 0.73 &      0.96 &   0.94 & 0.95 \\
     Genie & 2000 &     30 & 28.38 & 26.79 & 25.76 &   2.62 &  0.44 & 0.59 &      0.96 &   0.91 & 0.93 \\
 PhaseLink & 2000 &     30 & 29.17 & 32.04 & 22.48 &   6.69 &  4.95 & 4.61 &      0.72 &   0.78 & 0.74 \\
 PyOcto 1D & 2000 &     30 & 28.52 & 27.13 & 26.56 &   1.96 &  0.20 & 0.37 &      0.98 &   0.93 & 0.95 \\
   REAL 0D & 2000 &     30 & 29.10 & 28.35 & 25.39 &   3.72 &  0.93 & 2.04 &      0.90 &   0.88 & 0.88 \\
  GaMMA 1D & 2000 &    100 & 32.25 & 32.54 & 30.35 &   1.90 &  1.20 & 1.00 &      0.93 &   0.94 & 0.93 \\
     Genie & 2000 &    100 & 28.49 & 26.79 & 24.97 &   3.52 &  1.30 & 0.53 &      0.93 &   0.88 & 0.90 \\
 PhaseLink & 2000 &    100 & 29.78 & 35.85 & 20.41 &   9.37 & 14.00 & 1.44 &      0.58 &   0.69 & 0.63 \\
 PyOcto 1D & 2000 &    100 & 28.32 & 27.01 & 25.87 &   2.44 &  0.73 & 0.40 &      0.96 &   0.91 & 0.93 \\
   REAL 0D & 2000 &    100 & 29.26 & 29.60 & 24.40 &   4.85 &  3.22 & 1.98 &      0.82 &   0.84 & 0.83 \\
     Genie & 2000 &    300 & 28.78 & 25.07 & 21.99 &   6.79 &  2.68 & 0.41 &      0.88 &   0.76 & 0.81 \\
 PhaseLink & 2000 &    300 &  0.00 &  0.00 &  0.00 &   0.00 &  0.00 & 0.00 &       nan &    nan &  nan \\
 PyOcto 1D & 2000 &    300 & 28.84 & 26.59 & 23.82 &   5.02 &  2.32 & 0.45 &      0.89 &   0.82 & 0.84 \\
   REAL 0D & 2000 &    300 & 30.46 & 33.26 & 22.93 &   7.53 &  8.58 & 1.75 &      0.69 &   0.76 & 0.72 \\
\hline
\end{tabular}

\label{tab:sz_pick_lvl_table}
\end{table}

\begin{table}
\centering
\caption{Crustal scenario: evaluation of seismic phase associators at the pick level across different event densities and noise conditions.}
\begin{tabular}{lrrlllllllll}
\hline
Associator &  Ev. &  Noise &    GT &  Pred &    CA & Missed &    FP &  WAP & Precision & Recall &   F1 \\
\hline
  GaMMA 1D &  100 &     30 & 81.78 & 67.74 & 67.69 &  14.09 &  0.01 & 0.04 &      1.00 &   0.83 & 0.91 \\
     Genie &  100 &     30 & 81.78 & 81.62 & 81.51 &   0.27 &  0.03 & 0.08 &      1.00 &   1.00 & 1.00 \\
 PhaseLink &  100 &     30 & 80.97 & 70.52 & 69.70 &  11.27 &  0.13 & 0.69 &      0.99 &   0.86 & 0.92 \\
 PyOcto 1D &  100 &     30 & 81.78 & 81.69 & 81.60 &   0.18 &  0.01 & 0.08 &      1.00 &   1.00 & 1.00 \\
   REAL 0D &  100 &     30 & 81.80 & 81.93 & 81.54 &   0.26 &  0.11 & 0.28 &      1.00 &   1.00 & 1.00 \\
  GaMMA 1D &  100 &    100 & 83.47 & 68.96 & 68.87 &  14.61 &  0.04 & 0.05 &      1.00 &   0.83 & 0.91 \\
     Genie &  100 &    100 & 83.15 & 82.94 & 82.67 &   0.48 &  0.11 & 0.16 &      1.00 &   0.99 & 0.99 \\
 PhaseLink &  100 &    100 & 83.46 & 77.60 & 75.83 &   7.62 &  0.56 & 1.21 &      0.98 &   0.91 & 0.94 \\
 PyOcto 1D &  100 &    100 & 83.33 & 83.30 & 83.09 &   0.23 &  0.10 & 0.10 &      1.00 &   1.00 & 1.00 \\
   REAL 0D &  100 &    100 & 83.33 & 83.46 & 82.56 &   0.77 &  0.45 & 0.45 &      0.99 &   0.99 & 0.99 \\
  GaMMA 1D &  100 &    300 & 81.78 & 68.81 & 68.66 &  13.12 &  0.10 & 0.05 &      1.00 &   0.84 & 0.91 \\
     Genie &  100 &    300 & 81.25 & 81.28 & 80.81 &   0.44 &  0.39 & 0.08 &      0.99 &   0.99 & 0.99 \\
 PhaseLink &  100 &    300 & 81.20 & 79.08 & 77.21 &   3.99 &  1.56 & 0.31 &      0.98 &   0.95 & 0.96 \\
 PyOcto 1D &  100 &    300 & 81.27 & 81.34 & 80.86 &   0.41 &  0.37 & 0.11 &      0.99 &   0.99 & 0.99 \\
   REAL 0D &  100 &    300 & 81.25 & 81.88 & 80.14 &   1.11 &  1.40 & 0.33 &      0.98 &   0.99 & 0.98 \\
  GaMMA 1D &  500 &     30 & 81.38 & 68.95 & 68.79 &  12.58 &  0.07 & 0.09 &      1.00 &   0.85 & 0.92 \\
     Genie &  500 &     30 & 81.18 & 80.66 & 80.26 &   0.92 &  0.16 & 0.24 &      0.99 &   0.99 & 0.99 \\
 PhaseLink &  500 &     30 & 81.15 & 79.85 & 76.29 &   4.86 &  0.91 & 2.65 &      0.96 &   0.94 & 0.95 \\
 PyOcto 1D &  500 &     30 & 81.26 & 80.81 & 80.35 &   0.91 &  0.14 & 0.32 &      0.99 &   0.99 & 0.99 \\
   REAL 0D &  500 &     30 & 81.59 & 82.05 & 80.01 &   1.59 &  0.68 & 1.37 &      0.98 &   0.98 & 0.98 \\
  GaMMA 1D &  500 &    100 & 81.90 & 70.19 & 69.90 &  12.00 &  0.24 & 0.05 &      1.00 &   0.86 & 0.92 \\
     Genie &  500 &    100 & 81.70 & 81.33 & 80.58 &   1.12 &  0.60 & 0.14 &      0.99 &   0.99 & 0.99 \\
 PhaseLink &  500 &    100 & 82.10 & 82.20 & 75.35 &   6.75 &  3.16 & 3.69 &      0.93 &   0.92 & 0.92 \\
 PyOcto 1D &  500 &    100 & 81.72 & 81.71 & 80.99 &   0.73 &  0.53 & 0.19 &      0.99 &   0.99 & 0.99 \\
   REAL 0D &  500 &    100 & 82.17 & 83.24 & 79.45 &   2.72 &  2.43 & 1.35 &      0.95 &   0.97 & 0.96 \\
  GaMMA 1D &  500 &    300 & 82.03 & 72.53 & 71.76 &  10.27 &  0.70 & 0.07 &      0.99 &   0.88 & 0.93 \\
     Genie &  500 &    300 & 81.83 & 82.00 & 80.06 &   1.78 &  1.78 & 0.17 &      0.98 &   0.98 & 0.98 \\
 PhaseLink &  500 &    300 & 81.89 & 84.70 & 70.74 &  11.15 & 11.50 & 2.45 &      0.84 &   0.87 & 0.85 \\
 PyOcto 1D &  500 &    300 & 81.72 & 81.93 & 80.18 &   1.55 &  1.48 & 0.27 &      0.98 &   0.98 & 0.98 \\
   REAL 0D &  500 &    300 & 82.02 & 84.81 & 76.67 &   5.35 &  6.99 & 1.16 &      0.90 &   0.94 & 0.92 \\
  GaMMA 1D & 2000 &     30 & 81.91 & 71.86 & 71.21 &  10.70 &  0.33 & 0.32 &      0.99 &   0.87 & 0.93 \\
     Genie & 2000 &     30 & 82.07 & 80.11 & 78.61 &   3.46 &  0.70 & 0.80 &      0.98 &   0.96 & 0.97 \\
 PhaseLink & 2000 &     30 & 82.66 & 87.21 & 73.25 &   9.41 &  4.51 & 9.45 &      0.86 &   0.89 & 0.87 \\
 PyOcto 1D & 2000 &     30 & 82.10 & 81.12 & 79.58 &   2.52 &  0.62 & 0.92 &      0.98 &   0.97 & 0.97 \\
   REAL 0D & 2000 &     30 & 83.56 & 85.44 & 77.77 &   5.79 &  2.68 & 4.99 &      0.91 &   0.93 & 0.92 \\
  GaMMA 1D & 2000 &    100 & 82.41 & 74.76 & 73.41 &   9.01 &  1.06 & 0.29 &      0.98 &   0.89 & 0.93 \\
     Genie & 2000 &    100 & 82.40 & 80.91 & 77.97 &   4.44 &  2.14 & 0.80 &      0.96 &   0.95 & 0.95 \\
 PhaseLink & 2000 &    100 & 82.83 & 90.16 & 67.90 &  14.93 & 15.49 & 6.78 &      0.77 &   0.82 & 0.79 \\
 PyOcto 1D & 2000 &    100 & 82.30 & 81.88 & 78.97 &   3.33 &  2.00 & 0.91 &      0.96 &   0.96 & 0.96 \\
   REAL 0D & 2000 &    100 & 83.76 & 87.35 & 74.36 &   9.40 &  8.66 & 4.33 &      0.85 &   0.89 & 0.87 \\
     Genie & 2000 &    300 & 82.05 & 79.62 & 73.52 &   8.53 &  5.37 & 0.74 &      0.92 &   0.89 & 0.91 \\
 PhaseLink & 2000 &    300 & 84.95 & 98.24 & 56.00 &  28.95 & 41.26 & 0.98 &      0.57 &   0.66 & 0.61 \\
 PyOcto 1D & 2000 &    300 & 81.58 & 81.63 & 74.69 &   6.90 &  5.99 & 0.95 &      0.91 &   0.91 & 0.91 \\
   REAL 0D & 2000 &    300 & 83.99 & 91.09 & 64.99 &  18.99 & 22.72 & 3.37 &      0.71 &   0.77 & 0.74 \\
\hline
\end{tabular}

\label{tab:crustal_stats_pick_lvl}
\end{table}

%\begin{table}
%\centering
%\caption{Associator statistics of high station density configuration. }
%\label{tab:high_sta_density_associator_stats}
%\input{figures/sta_density_analysis/high_density/associator_stats}
%\end{table}

%\begin{table}
%\centering
%\caption{Associator pick statistics of high station density configuration. }
%\label{tab:high_sta_density_pick_stats}
%\input{figures/sta_density_analysis/high_density/pick_stats_table}
%\end{table}

%Low station density tables

%\begin{table}
%\centering
%\caption{Associator statistics of low station density configuration. }
%\label{tab:low_sta_density_associator_stats}
%\input{figures/sta_density_analysis/low_density/associator_stats}
%\end{table}

%\begin{table}
%\centering
%\caption{Associator pick statistics of low station density configuration. }
%\label{tab:low_sta_density_pick_stats}
%\input{figures/sta_density_analysis/low_density/pick_stats_table}
%\end{table}